\documentclass[a4paper,11pt]{article}
\pdfoutput=1 

\usepackage[margin=1.0in]{geometry}

\usepackage[utf8]{inputenc}
\usepackage{graphics}
\usepackage{graphicx}
\usepackage{url}
\usepackage{caption} 
\usepackage{amsmath}
\usepackage[merge,numbers,compress]{natbib}
\usepackage{hyperref}
\usepackage{placeins}
\usepackage{morefloats}
\usepackage{slashed}
\usepackage[colorinlistoftodos, shadow]{todonotes}
\usepackage{titlesec}
\usepackage{gensymb}
\usepackage{textcomp}
\usepackage{amssymb}
\usepackage{verbatim}
\usepackage{xspace}

\allowdisplaybreaks

\allowdisplaybreaks[1]

\def\refeq#1{\mbox{(\ref{#1})}}
\def\reffi#1{\mbox{Fig.~\ref{#1}}}
\def\reffis#1{\mbox{Figs.~\ref{#1}}}
\def\refta#1{\mbox{Table~\ref{#1}}}
\def\reftas#1{\mbox{Tables~\ref{#1}}}
\def\refse#1{\mbox{Section~\ref{#1}}}
\def\refses#1{\mbox{Sections~\ref{#1}}}

\def\citere#1{\mbox{Ref.~\cite{#1}}}
\def\citeres#1{\mbox{Refs.~\cite{#1}}}

\newcommand{\GeV}{\unskip\,\mathrm{GeV}}
\newcommand{\MeV}{\unskip\,\mathrm{MeV}}
\newcommand{\TeV}{\unskip\,\mathrm{TeV}}

\def\mathswitch#1{\relax\ifmmode#1\else$#1$\fi}
\def\mathswitchr#1{\relax\ifmmode{\mathrm{#1}}\else$\mathrm{#1}$\fi}
\def\mathswitchit#1{\relax\ifmmode{#1}\else$#1$\fi}

\newcommand{\Pb}{\mathswitchr b}

\newcommand{\Pt}{\mathswitchr t}
\newcommand{\Pp}{\mathswitchr p}

\newcommand{\Pe}{\mathswitchr e}
\newcommand{\Pep}{\mathswitchr {e^+}}
\newcommand{\Pem}{\mathswitchr {e^-}}

\newcommand{\Pmum}{\mathswitchr {\mu^-}}

\newcommand{\PW}{\mathswitchr W}
\newcommand{\PZ}{\mathswitchr Z}

\newcommand{\Pg}{\mathswitchr g}
\newcommand{\PH}{\mathswitchr H}

\newcommand{\Pne}{\mathswitch \nu_{\mathrm{e}}}

\newcommand{\MW}{\mathswitch {M_\PW}}

\newcommand{\MZ}{\mathswitch {M_\PZ}}

\newcommand{\sw}{\mathswitch {s_{\mathrm{w}}}}
\newcommand{\cw}{\mathswitch {c_{\mathrm{w}}}}

\newcommand{\GF}{\mathswitch {G_\mu}}

\newcommand{\cgw}{ g_{\rm w}}

\def\bfi{\begin{figure}}
\def\efi{\end{figure}}

\def\ie{i.e.\ }

\newcommand{\im}{{\rm  i}}

\newcommand{\rT}{{\mathrm{T}}}

\newcommand{\OS}{{\mathrm{OS}}}

\newcommand{\recola}{{\sc Recola}\xspace}
\newcommand{\recolatwo}{{\sc Recola2}\xspace}
\newcommand{\collier}{{\sc Collier}\xspace}

\newcommand{\Reptil}{{\sc REPT1L}\xspace}
\newcommand{\FeynRules}{{\sc FeynRules}\xspace}

\newcommand{\FortranNinety}{{\sc Fortran95}\xspace}
\newcommand{\UFO}{{\sc UFO}\xspace}
\newcommand{\Python}{{\sc Python}\xspace}

\def\draftdate{\relax}
\def\mda{\relax}
\def\mua{\relax}
\def\mla{\relax}
\def\draft{
  \def\thtystars{******************************}
  \def\sixtystars{\thtystars\thtystars}
  \typeout{}
  \typeout{\sixtystars**}
  \typeout{* Draft mode!
    For final version remove \protect\draft\space in source file *}
  \typeout{\sixtystars**}
  \typeout{}
  \def\draftdate{\today}
  \def\mua{\marginpar[\boldmath\hfil$\uparrow$]%
    {\boldmath$\uparrow$\hfil}%
    \typeout{marginpar: $\uparrow$}\ignorespaces}
  \def\mda{\marginpar[\boldmath\hfil$\downarrow$]%
    {\boldmath$\downarrow$\hfil}%
    \typeout{marginpar: $\downarrow$}\ignorespaces}
  \def\mla{\marginpar[\boldmath\hfil$\rightarrow$]%
    {\boldmath$\leftarrow $\hfil}%
    \typeout{marginpar: $\leftrightarrow$}\ignorespaces}
  \def\Mua{\marginpar[\boldmath\hfil$\Uparrow$]%
    {\boldmath$\Uparrow$\hfil}%
    \typeout{marginpar: $\uparrow$}\ignorespaces}
  \def\Mda{\marginpar[\boldmath\hfil$\Downarrow$]%
    {\boldmath$\Downarrow$\hfil}%
    \typeout{marginpar: $\downarrow$}\ignorespaces}
  \def\Mla{\marginpar[\boldmath\hfil$\Rightarrow$]%
    {\boldmath$\Leftarrow $\hfil}%
    \typeout{marginpar: $\leftrightarrow$}\ignorespaces}
  \overfullrule 5pt
  \oddsidemargin -15mm
  \marginparwidth 29mm
}

\numberwithin{equation}{section}

\begin{document} 

\thispagestyle{empty}
\def\thefootnote{\fnsymbol{footnote}}
\setcounter{footnote}{1}
\null

\vfill
\begin{center}
  {\Large {\boldmath\bf {Anomalous triple-gauge-boson interactions in vector-boson pair
  production with \recolatwo}
      \par} \vskip 2.5em
    {\large
      {\sc Mauro Chiesa$^{1}$, Ansgar Denner$^{1}$, Jean-Nicolas Lang$^{2}$
      }\\[2ex]
      {\normalsize \it
        $^1$Julius-Maximilians-Universit\"at W\"urzburg,
        Institut f\"ur Theoretische Physik und Astrophysik, \\
        Emil-Hilb-Weg 22, D-97074 W\"urzburg, Germany
      }\\[2ex]
      {\normalsize \it
        $^2$Universit\"at Z\"urich, Physik-Institut, CH-8057 Z\"urich,
        Switzerland}
    }
  }
  \par \vskip 1em
\end{center}\par
\vskip .0cm \vfill {\bf Abstract:}
\par
Diboson production at the LHC is a process of great importance both in
the context of tests of the SM and for direct searches for new
physics. In this paper we present a phenomenological study of
$\PW\PW$~($\to \Pep \Pne \Pmum \bar{\nu}_{\mu}$), $\PW\PZ$~($\to \Pem
\overline{\nu}_{\rm e} \mu^+ \mu^-$), and $\PZ\PZ$~($\to \Pep \Pem
\mu^+ \mu^-$) production considering event selections of interest for
the anomalous triple-gauge-boson-coupling searches at the LHC: we
provide theoretical predictions within the Standard Model at NLO QCD
and NLO EW accuracy and study the effect of the anomalous
triple-gauge-boson interactions at NLO QCD. For $\PW\PW$ and $\PZ\PZ$
the contribution of the loop-induced $\Pg\Pg\to\PW^+\PW^-$ and
$\Pg\Pg\to\PZ\PZ$ processes is included. Anomalous triple-gauge-boson
interactions are parametrized in the EFT framework. This paper is the
first application of \recolatwo in the EFT context.
\par
\vskip 1cm
\noindent
\par
\null \setcounter{page}{0} \clearpage
\def\thefootnote{\arabic{footnote}} \setcounter{footnote}{0}

\tableofcontents


\section{Introduction}

Diboson production processes are of great importance in high-energy
physics.  On one hand, they are sensitive to the gauge-boson self
interaction so that their measurement provides a crucial test of the
Standard Model (SM) description of the gauge-boson dynamics. On the
other hand, diboson production at the LHC is a source of background
for other SM processes like for instance Higgs production as well as
for direct searches of new physics.  Therefore, a precise theoretical
knowledge of these processes is mandatory not only in view of
precision tests of the SM but also regarding new-physics searches.

The production of leptonically decaying electroweak boson pairs has
been intensively studied at the Tevatron
~\cite{Aaltonen:2009aa,Aaltonen:2012vu,Abazov:2012ze,D0:2013rca,Aaltonen:2014yfa}
and in Run I of the LHC~\cite{Aad:2016ett,
  Aad:2016wpd,Aaboud:2016urj,Khachatryan:2015sga,
  CMS:2014xja,Khachatryan:2015pba,Khachatryan:2016poo}, searching for
deviations from the SM predictions and setting limits on the strength
of possible non-SM triple-gauge-boson interactions (anomalous
triple-gauge-boson couplings, aTGCs in the following). Recently, the
first results for $\PW\PW$, $\PZ\PZ$ and $\PW\PZ$ production at
$13\TeV$ have been presented in
\citeres{Aaboud:2017qkn,Aad:2015zqe,Aaboud:2016yus}
and~\cite{CMS:2016vww,Khachatryan:2016txa,Khachatryan:2016tgp} by the
ATLAS and CMS collaborations, respectively.

Together with Higgs and Drell--Yan production, diboson production is
one of the LHC processes known with highest theoretical precision.
Theoretical predictions for $q \overline{q}^{(')} \to V V^{(')}$
($V=\PW$, $\PZ$) with stable external vector bosons have been computed
in \citeres{Brown:1978mq,Brown:1979ux} at leading order (LO) and in
\citeres{Ohnemus:1991gb,Ohnemus:1991kk,Ohnemus:1990za,Mele:1990bq,Frixione:1992pj,Frixione:1993yp}
at NLO (next-to-leading) QCD accuracy.  The NNLO QCD corrections for
on shell $V$ and $V^{(')}$ have been presented in
\citeres{Cascioli:2014yka,Gehrmann:2014fva}. The leptonic decays of
the vector bosons have been included in the LO calculation of
\citere{Gunion:1985mc}, while higher-order QCD corrections including
leptonic decays of $V$ and $V^{(')}$ have been computed in
\citeres{Ohnemus:1994ff,Dixon:1999di,Campbell:1999ah,Campbell:2011bn}
at NLO and in
\citeres{Grazzini:2015hta,Grazzini:2016swo,Grazzini:2016ctr,Grazzini:2017mhc,Heinrich:2017bvg} at NNLO.
The NLO QCD corrections to $q \overline{q}^{(')} \to V V^{(')}+1$~jet
($V=\PW$, $\PZ$) have been evaluated in
\citeres{Dittmaier:2007th,Campbell:2007ev,Dittmaier:2009un,Binoth:2009wk,Campanario:2010hp,Yong:2016njr}.
Besides fixed-order calculations, diboson production processes have
been studied at NLO QCD accuracy matched with Parton Shower (NLOPS) in
the MC@NLO~\cite{Frixione:2002ik} and in the {\tt
  POWHEG}~\cite{Nason:2004rx,Frixione:2007vw} framework in
\citeres{Frixione:2002ik}
and~\cite{Nason:2006hfa,Hamilton:2010mb,Hoche:2010pf,Melia:2011tj,Nason:2013ydw},
respectively.  NLOPS predictions for $\PW\PW+$~jets with NLO merging
of zero and one jet multiplicities have been investigated in
\citere{Cascioli:2013gfa} in the {\sc Sherpa+OpenLoops}
framework~\cite{Gleisberg:2008ta,Cascioli:2011va}.  The NNLO QCD
corrections have been matched to resummation of the transverse
momentum of the diboson~\cite{Grazzini:2015wpa} and of the hardest
jet~\cite{Dawson:2016ysj}.

Formally, the loop-induced processes $\Pg\Pg \to V V$ ($V=\PW$, $\PZ$)
contribute to $\PW\PW$ and $\PZ\PZ$ production at the same
perturbative order as the NNLO QCD corrections to $q
\overline{q}^{(')} \to VV$, however, their contributions are relatively
large because of the gluon luminosity.  LO predictions for the
$\Pg\Pg$ channel have been computed in
\citeres{Dicus:1987dj,Glover:1988fe,Glover:1988rg} for stable $V$s
and in \citeres{Binoth:2005ua,Zecher:1994kb,Binoth:2006mf}
including leptonic vector-boson decays, while the NLO QCD
corrections have been published in
\citeres{Caola:2015rqy,Caola:2015psa} where the interference with
the Higgs-mediated process $\Pg\Pg \to \PH \to V V$ has been
neglected.  In the same approximation, the process $\Pg\Pg \to \PZ\PZ$
has been considered at NLOPS accuracy in \citere{Alioli:2016xab}.
The interference with $\Pg\Pg \to \PH \to V V$ has been studied at LO
in
\citeres{Kauer:2013qba,Campbell:2011cu,Campbell:2013una,Kauer:2012hd,Kauer:2015dma},
the LOPS predictions have been presented in \citere{Bellm:2016cks},
the universal soft--collinear terms of the QCD corrections have been
included in \citere{Bonvini:2013jha}, and the full NLO calculation
has been published in \citere{Caola:2016trd}.

Diboson production via quark--anti-quark annihilation is sensitive to
the gauge-boson self-interaction. The impact of potential non-SM
triple-gauge-boson interactions has been considered at LO in
\citeres{Zeppenfeld:1987ip,Hagiwara:1989mx,Nuss:1996dg,Kuss:1997mf,Baur:2000ae,Baur:2000cx}
and at NLO QCD in \citeres{Baur:1994aj,Baur:1995uv,Dixon:1999di}.  In
\citere{Baglio:2017bfe} the effect of anomalous triple-gauge-boson and
fermion couplings on $q\bar{q}\to\PW^+\PW^-$ (with on-shell $\PW$s)
has been studied at NLO QCD accuracy and compared to the one-loop
electroweak corrections.  The anomalous triple-gauge-boson couplings
for $\PW\PW$ and $\PW\PZ$ production have been included in the QCD NLO
Monte Carlo integrators {\sc MCFM}~\cite{Campbell:2010ff} and {\sc
  VBF@NLO}~\cite{Arnold:2008rz,Arnold:2011wj,Baglio:2014uba}. The
event generators {\sc
  MC@NLO}~\cite{Frixione:2010wd,Alwall:2011uj,Alwall:2014hca} and {\tt
  POWHEG} allow to simulate $\PW\PW$ and $\PW\PZ$ production at NLOPS
accuracy including the effect of the anomalous $W^+W^-V$ ($V=Z$,
$\gamma$) couplings, while both charged and neutral anomalous
triple-gauge-boson couplings are included in {\sc Sherpa} at LO.
Theoretical predictions for $\PW\PZ$ production including aTGCs in the
EFT framework have been presented in \citere{Franceschini:2017xkh} at
NLO QCD plus parton-shower merging.

One-loop electroweak (EW)\footnote{EW corrections or $\mathcal{O}(
  \alpha )$ corrections in the following.}  corrections are usually
small at the level of integrated cross sections, however, they can
have a significant effect on the shape of the distributions of
interest. On one hand, photonic corrections can lead to pronounced
radiative tails near resonances or kinematical thresholds and, on the
other hand, the size of the EW corrections can reach the order of
several tens of percent in the high $p_{\rT}$ or invariant-mass tails
of distributions because of the so-called Sudakov
logarithms~\cite{Beenakker:1993tt,Beccaria:1998qe,
  Ciafaloni:1998xg,Kuhn:1999de,Ciafaloni:2000df,Denner:2000jv}. This
in particular implies that the EW corrections have a large impact in
those regions of the phase space of interest for the searches for
physics beyond the SM. As far as diboson production is concerned, the
logarithmic part~\cite{Denner:2000jv,Denner:2001gw} of the
$\mathcal{O}( \alpha )$ corrections to the process $q
\overline{q}^{(')} \to V V^{(')}$ ($V=\PW$, $\PZ$) has been computed
in \citeres{Accomando:2001fn,Accomando:2004de} and in
\citere{Accomando:2005xp} in the context of the searches for aTGCs.
The full one-loop EW corrections have been studied in
\citeres{Bierweiler:2012kw,Bierweiler:2013dja,Baglio:2013toa} for
stable external $V$ and $V^{(')}$, while the leptonic vector-boson
decays were included in the form of a consistent expansion about the
resonances for $\PW\PW$ production in~\citere{Billoni:2013aba}, and in
an approximate variant via the {\sc Herwig++}~\cite{Bellm:2015jjp}
Monte Carlo generator for $\PW\PW$, $\PZ\PZ$ and $\PW\PZ$ production
in \citere{Gieseke:2014gka}. The full $\mathcal{O}( \alpha ) $
calculations based on full $2 \to 4$ particle amplitudes, including
all off-shell effects, have been presented for
$\PW$-pair~\cite{Biedermann:2016guo},
$\PZ$-pair~\cite{Biedermann:2016yvs,Biedermann:2016lvg} and $\PZ
\PW$~\cite{Biedermann:2017oae} production.  The one-loop EW
corrections to the process $\Pp\Pp\to 2l2\nu$ have been computed in
\citere{Kallweit:2017khh}.

The aim of this paper is on the one hand to compare the effects of
anomalous couplings including QCD corrections with SM electroweak
corrections for typical experimental event selections.  On the other
hand, this paper documents the first application of
\recolatwo~\cite{Denner:2017vms,Denner:2017wsf} for a Lagrangian with
anomalous couplings. To this end \recolatwo\ model files have been
constructed with \Reptil~\cite{Denner:2017wsf} and verified by
comparisons with calculations in the literature.

This article is organized as follows. In \refse{sec:calculation},
the details of the calculation are described together with the
cross-checks that have been performed.  In \refse{sec:EFT}, we
present our treatment of the anomalous triple-gauge-boson interactions
in diboson production and collect the conversion rules between the EFT
description of these interactions and the one based on \mbox{aTGCs}. The
input parameters and event selections considered in our
phenomenological studies are collected in \refse{sect:params}. In
\refse{sect:numres}, numerical results are presented for
integrated cross sections and differential distributions for $\PW\PW$,
$\PW\PZ$, and $\PZ\PZ$ production.

\section{Technical details of the calculation}
\label{sec:calculation}

We compute the NLO QCD corrections to the
four-lepton\footnote{Four-lepton stands for four charged leptons, two
  leptons plus two neutrinos or three charged leptons plus neutrino.}
production processes at the LHC including the effect of the anomalous
triple-gauge-boson interactions.

We consider as LO the processes $q\bar{q}'\to V_1 V_2 \to$
$l_1\overline{l}'_1l_2\overline{l}'_2$, where $V_{1(2)}=\PW$, $\PZ$
and $\gamma$. In addition to the SM ${\cal O}(\alpha^4)$ contribution,
we include the effect of the anomalous triple-gauge-boson interactions
corresponding to the higher-dimensional operators described in
\refse{sec:EFT}. We study the impact of dimension-6 operators for
$\PW\PW$ and $\PW\PZ$ production, and of dimension-8 operators for
$\PZ\PZ$ production which is insensitive to dimension-6 operators.

The NLO QCD corrections to $q\bar{q}'\to V_1 V_2 \to$
$l_1\overline{l}'_1l_2\overline{l}'_2$ are of order
$\alpha^4 \alpha_{\rm s}$ in the SM. As for the LO calculation,
in addition to the SM contribution, we include the effect of
the anomalous triple-gauge-boson interaction corresponding
to dimension-6 operators  (dimension-8 operators if both $V_1$
and $V_2$ are neutral gauge bosons) at NLO QCD. For the SM processes
$q\bar{q}'\to V_1 V_2 \to$ we also compute the corresponding NLO EW corrections. 

Another contribution to $\PW\PW$ and $\PZ\PZ$ production in the SM is the
loop-induced process $\Pg\Pg \to l_1\overline{l}'_1l_2\overline{l}'_2$:
though this occurs at $\mathcal{O}( \alpha_{\rm s}^2 \alpha^4 )$, it
can be phenomenologically relevant because of the gluon luminosity. 
The $\Pg\Pg$ channel is not sensitive to the aTGCs; however, we
compute the $\Pg\Pg$ diagrams at LO accuracy and include their
contribution in our phenomenological studies.

{\sloppypar
Our calculation relies on tools like {\FeynRules}~\cite{Christensen:2008py,Alloul:2013bka},
\Reptil~\cite{Denner:2017wsf}
and \recolatwo \cite{Denner:2017vms,Denner:2017wsf}  
together with an efficient Monte Carlo integrator.
}

We used the {\sc Mathematica} package {\FeynRules} to implement the
SM Lagrangian (according to the conventions of
\citere{Denner:1991kt}) and the dimension-6 and -8 operators
relevant for the anomalous triple-gauge-boson interaction, as
described in \refse{sec:EFT}.

The \UFO model file~\cite{Degrande:2011ua} generated by {\FeynRules}
is then converted into a model file for \recolatwo by means of the
{\Python} library \Reptil (Recola's rEnormalization Procedure Tool at
1 loop): besides deriving the tree-level as well as the one-loop
\recolatwo model files from the \UFO format, \Reptil performs in a
fully automated way the counterterm expansion of the vertices, sets up
and solves the renormalization conditions and computes the rational
terms of type R2 for the model under consideration.

\recolatwo is used for the automated generation and the numerical evaluation
of the tree-level and one-loop amplitudes starting from the model file generated
by \Reptil. \recolatwo is an enhanced version of the \FortranNinety code
\recola~\cite{Actis:2016mpe},
designed for the computation of tree-level and one-loop amplitudes
in general gauge theories and using the tensor-integral library \collier\cite{Denner:2016kdg}.

The phase-space integration is carried out with a multi-channel Monte Carlo
integrator that is a further development of the one described in
\citeres{Berends:1994pv,Dittmaier:2002ap}.

As a cross check, \recolatwo has been interfaced to the {\tt
  POWHEG-BOX-V2} generator
~\cite{Nason:2004rx,Frixione:2007vw,Alioli:2010xd}, and the results at
NLO QCD in the SM have been compared for the processes
$q\overline{q}^{(')} \to \PW\PW$, $\PW\PZ$ and
$\PZ\PZ$~\cite{Melia:2011tj,Nason:2013ydw}. In order to validate the
implementation of the non-SM $W^+W^-V$ ($V=Z$, $\gamma$) interaction,
we compared our results for the LO matrix-element squared computed
with \recolatwo
with the ones obtained with the {\sc VBF@NLO} program for
the CP-even dimension-6 operators.
Moreover, the NLO QCD corrections to the diagrams involving the
anomalous triple-gauge-boson interactions have been computed
analytically and the results have been used to cross check the
predictions from \recolatwo at the amplitude level. We also used the
matrix elements coded in the {\tt Wgamma} package~\cite{Barze:2014zba}
of {\tt POWHEG-BOX-V2} to validate the implementation of the CP-even
anomalous triple-gauge-boson interaction.

As a further validation, we implemented another model into \recolatwo
where the anomalous gauge-boson-interaction is parametrized in terms
of anomalous couplings rather than Wilson coefficients. We verified
that this model reproduces the results of
\citeres{Falkowski:2016cxu} and~\cite{Baur:2000ae} for $\PW\PW$,
$\PW\PZ$, and $\PZ\PZ$ production within the accuracy of the plots
presented there. The two models have been compared
at the matrix element level by using the conversion formulas of
\refse{sec:EFT} and we found perfect agreement when the
gauge-boson widths are set to zero.%
\footnote{In the complex-mass scheme, the electroweak mixing angle and
  thus the relations between Wilson coefficients and anomalous
  couplings, Eqs.~\refeq{eq:TGCEFTconv} and
  \refeq{eq:NTGCEFTconv}, become complex, while we keep the anomalous
  couplings and Wilson coefficients real.}

\section{EFT framework for triple-gauge-boson interaction}
\label{sec:EFT}

Beyond Standard Model (BSM) effects can be parametrized in a
model-independent way by means of an effective field theory (EFT). In
the Standard Model EFT, the SM Lagrangian is generalized by adding
non-renormalizable gauge-invariant operators with canonical dimension
$D>4$:
\begin{equation}
  {\cal L}^{\rm eff.}={\cal L}^{\rm SM} + \sum_i \frac{c^i_6}{\Lambda^2}{\cal O}^i_6
  + \sum_i \frac{c^i_8}{\Lambda^4}{\cal O}^i_8 + \cdots .
  \label{eq:eftgen}
\end{equation}
In Eq.~(\ref{eq:eftgen}) the operators ${ \cal O }^i_D$ represent the
effect of new physics with a mass scale $\Lambda$ much larger than the
electroweak scale and are multiplied by the corresponding Wilson
coefficients $c^i_D$.

In the EFT language, the anomalous $W^+W^-V$ ($V=Z$, $\gamma$)
interaction can be parametrized in terms of the following set of
dimension-6
operators~\cite{Hagiwara:1992eh,Hagiwara:1993ck,Degrande:2012wf,Degrande:2013rea}
\begin{equation}
  \begin{aligned}
    {\cal O}_{WWW}&= - \frac{\cgw^3}{4} \epsilon_{ijk} W_{\mu\nu}^i W^{\nu\rho\;j}W_{\rho}^{~\mu\;k},\\
    {\cal O}_W&    = - \im \cgw (D_\mu\Phi)^\dagger \frac{\tau_k}{2} W^{\mu\nu\;k}(D_\nu\Phi), \\
    {\cal O}_B&    = + \im \frac{g_1}{2} (D_\mu\Phi)^\dagger B^{\mu\nu}(D_\nu\Phi), \\
    {\cal O}_{\widetilde{W}WW}&=  + \frac{\cgw^3}{4} \epsilon_{ijk} {\widetilde{W}}_{\mu\nu}^i W^{\nu\rho\;j}W_{\rho}^{~\mu\;k},\\
    {\cal O}_{\widetilde{W}}&  =  + \im \cgw (D_\mu\Phi)^\dagger \frac{\tau_k}{2} {\widetilde{W}}^{\mu\nu\; k}(D_\nu\Phi) , 
  \end{aligned}
  \label{eq:tgcWWv}
\end{equation}
where $\cgw=e/\sw$ and $g_1=e/\cw$ correspond to the ${\rm SU}(2)_{\rm
  w}$ and ${\rm U}(1)_{\rm Y}$ gauge couplings, respectively, $\tau$
are the Pauli matrices (twice the ${\rm SU}(2)_{\rm w}$ generators)
and $\Phi$ stands for the Higgs doublet.\footnote{Note that our
  definitions of the dimension-6 and dimension-8 operators in
  Eqs.~(\ref{eq:tgcWWv}) and~(\ref{eq:ntgcop}) differ from the ones in
  \citeres{Degrande:2012wf} and~\cite{Degrande:2013kka} in order to
  match the conventions of \citere{Denner:1991kt} for the SM vertices
  while preserving the relations~(\ref{eq:TGCEFTconv})
  and~(\ref{eq:NTGCEFTconv}). The field-strength tensor in Eq.~(6) of
  \citere{Degrande:2012wf} should be replaced by $W_{\mu\nu}=\im
  g\tau^I (\partial_\mu W^I_{\nu}- \partial_\nu
  W^I_{\mu}-g\epsilon_{IJK} W^J_{\mu} W^K_{\nu})/2$ for internal
  consistency of that paper (see also \citere{Hagiwara:1993ck}). With
  this definition we find the following conversion rules between
  \citere{Degrande:2012wf} and \citere{Denner:1991kt}: $W^i_{\mu} \to
  -W^i_{\mu}$, $W^\pm_{\mu} \to -W^\pm_{\mu}$ and $Z_{\mu} \to
  -Z_{\mu}$. The conversion rules between \citere{Degrande:2013kka}
  and \citere{Denner:1991kt} read: $B_{\mu} \to -B_{\mu}$ and
  $Z_{\mu} \to -Z_{\mu}$.  Furthermore we assume that the definition
  of the $\epsilon$ tensor is $\epsilon_{0123}=+1$ and
  $\epsilon^{0123}=+1$ in \citeres{Degrande:2012wf}
  and~\cite{Degrande:2013kka}, respectively. This is suggested by a
  comparison with results in \citeres{Hagiwara:1986vm} and
  \cite{Gounaris:1999kf}, respectively.}  We use the definitions:
\begin{equation}
  \begin{aligned}
   D_{\mu}\Phi & =\left(\partial_{\mu}-\im \cgw \frac{\tau^k}{2} W^k_{\mu}+\im \frac{1}{2}g_1 B_{\mu}\right)\Phi,\\
    W_{\mu\nu}^i & = \partial_\mu W^i_\nu - \partial_\nu W^i_\mu
    + \cgw \epsilon_{ijk} W^j_\mu W^k_\nu , \\ 
    B_{\mu \nu} & = \partial_\mu B_\nu - \partial_\nu B_\mu  \; ,\\
    {\widetilde{W}}_{\mu\nu}^i & = \frac{1}{2} \epsilon_{\mu\nu\rho\sigma} W^{\rho\sigma \; i}, \quad {\rm with } \quad \epsilon^{0123}=+1.
  \end{aligned}
  \label{eq:fields}
\end{equation}
In the literature, the anomalous $W^+W^-V$ ($V=Z$, $\gamma$)
interaction is often parametrized in terms of the phenomenological
Lagrangian~\cite{Hagiwara:1986vm,DeRujula:1991ufe,Zeppenfeld:1987ip}
($V=\gamma,Z$):
\begin{equation}
 \begin{aligned}
  \frac{\cal L}{g_{WWV}}=& \im   \left(g_1^V(W_{\mu\nu}^+W^{-\mu}-W^{+\mu}W_{\mu\nu}^-)V^\nu
+\kappa_VW_\mu^+W_\nu^-V^{\mu\nu}
+\frac{\lambda_V}{M_{\PW}^2}W^{+\mu\nu}W_\nu^{-\rho}V_{\rho\mu}
\right.\\&\left.
+\im g_4^VW_\mu^+W^-_\nu(\partial^\mu V^\nu+\partial^\nu V^\mu)
+\im g_5^V\epsilon^{\mu\nu\rho\sigma}(W_\mu^+\partial_\rho W^-_\nu-\partial_\rho W_\mu^+W^-_\nu)V_\sigma
\right.\\&\left.
-\tilde{\kappa}_VW_\mu^+W_\nu^-\tilde{V}^{\mu\nu}
-\frac{\tilde{\lambda}_V}{M_{\PW}^2}W^{+\mu\nu}W_\nu^{-\rho}\tilde{V}_{\rho\mu}
\right) \, ,
 \end{aligned}
 \label{eq:tgcpheno}
\end{equation}
with $X_{\mu\nu} = \partial_\mu X_\nu -\partial_\nu X_\mu$ and
$g_{WWV}$ is the $WWV$ coupling in the SM
($g_{WW\gamma}=-e$, $g_{WWZ}=e\cw/\sw$).  It is possible to relate
Eq.~(\ref{eq:tgcpheno}) to the EFT framework of Eqs.~(\ref{eq:tgcWWv})
according to the relations
\begin{align}
    g_1^Z &= 1+c_W\frac{M_{\PZ}^2}{2\Lambda^2},\notag\\
    \kappa_\gamma &= 1+(c_W+c_B)\frac{M_{\PW}^2}{2\Lambda^2},\notag\\
    \kappa_Z &= 1+\left(c_W-c_B\frac{\sw^2}{\cw^2}\right)\frac{M_{\PW}^2}{2\Lambda^2},\notag\\
    \lambda_\gamma &= \lambda_Z = c_{WWW} \cgw^2 \frac{3 M_{\PW}^2}{2\Lambda^2},\notag\\
    g_4^V &= g_5^V=0,\notag\\
    \tilde{\kappa}_\gamma &=
    c_{\tilde{W}}\frac{M_{\PW}^2}{2\Lambda^2},\notag\\
    \tilde{\kappa}_Z &=
    -c_{\tilde{W}}\frac{\sw^2}{\cw^2}\frac{M_{\PW}^2}{2\Lambda^2},\notag\\
    \tilde{\lambda}_\gamma &= \tilde{\lambda}_Z = c_{\tilde{W}WW} \cgw^2 \frac{3 M_{\PW}^2}{2\Lambda^2}.
  \label{eq:TGCEFTconv}
\end{align}
At tree-level, there is no triple-gauge-boson interaction in the
neutral sector in the SM.  However, this kind of interaction can arise
in some extensions of the SM and can be described in the EFT
framework. We follow the approach of \citere{Degrande:2013kka} and
consider the set of dimension-8 operators,\footnote{There is
  no dimension-6 contribution to neutral triple-gauge-boson
  interactions~\cite{Buchmuller:1985jz,Grzadkowski:2010es}.}
\begin{equation}
  \begin{aligned}
  \mathcal{O}_{BW}&= - \im \, \Phi^\dagger   B_{\mu\nu}
  \frac{\tau_i}{2} W^{\mu\rho\; i} \left\{D_\rho,D^\nu\right\} \Phi + \mathrm{h.c.}, \\
  \mathcal{O}_{WW}&= \im \, \Phi^\dagger \frac{\tau_i}{2} \frac{\tau_j}{2} W_{\mu\nu}^iW^{\mu\rho \;j} \left\{D_\rho,D^\nu\right\} \Phi + \mathrm{h.c.} , \\
  \mathcal{O}_{BB}&= \im \, \Phi^\dagger  B_{\mu\nu}B^{\mu\rho} \left\{D_\rho,D^\nu\right\} \Phi + \mathrm{h.c.} , \\
  \mathcal{O}_{\widetilde{B}W}&= - \im \, \Phi^\dagger  \widetilde{B}_{\mu\nu} \frac{\tau_i}{2} W^{\mu\rho \; i} \left\{D_\rho,D^\nu\right\} \Phi + \mathrm{h.c.},
  \end{aligned}
  \label{eq:ntgcop}
\end{equation}
which are added to the SM Lagrangian (h.c. denotes the  hermitian conjugate).
In Eq.~(\ref{eq:ntgcop}), $D_{\mu}$ represents the ${\rm SU}(2)_{\rm
  w} \times {\rm U}(1)_{Y}$ covariant derivative and
$\left\{D_\mu,D^\nu\right\}=D_{\mu}D^{\nu}+D^{\nu}D_{\mu}$.  As for
the case of the anomalous $W^+W^-V$ interaction, in the literature the
neutral triple-gauge-boson interaction has been described in terms of
phenomenological Lagrangians~\cite{Gaemers:1978hg,Hagiwara:1986vm,Gounaris:1999kf,Baur:2000cx,Gounaris:2000dn,Gounaris:2000tb}:
\begin{eqnarray}
  \label{eq:neutralTGC}
        {\cal{L}}_{VVV} &=& \frac{e}{\MZ^{2}} \Bigg [
          -[f_4^\gamma (\partial_\mu A^{\mu \beta})-
            f_4^Z (\partial_\mu Z^{\mu \beta}) ] Z_\alpha
          ( \partial^\alpha Z_\beta)\nonumber\\
          & & \qquad\;{}+[f_5^\gamma (\partial^\sigma A_{\sigma \mu})-
            f_5^Z (\partial^\sigma Z_{\sigma \mu}) ] \widetilde{Z}^{\mu \beta} Z_\beta
          \nonumber \\
          &&\qquad\;{}+  [h_1^\gamma (\partial^\sigma A_{\sigma \mu})
            -h_1^Z (\partial^\sigma Z_{\sigma \mu})] Z_\beta A^{\mu \beta}
          +[h_3^\gamma  (\partial_\sigma A^{\sigma \rho})
            - h_3^Z  (\partial_\sigma Z^{\sigma \rho})] Z^\alpha
          \widetilde{A}_{\rho \alpha}
          \nonumber \\
          &&\qquad\;{}+ \left \{\frac{h_2^\gamma}{\MZ^{2}} [\partial_\alpha \partial_\beta
            \partial^\rho A_{\rho \mu} ]
          -\frac{h_2^Z}{\MZ^{2}} [\partial_\alpha \partial_\beta
            (\square +\MZ^{2}) Z_\mu] \right \} Z^\alpha A^{\mu \beta}
          \nonumber \\
          &&\qquad\;{}- \left \{
          \frac{h_4^\gamma}{2\MZ^{2}}[\square \partial^\sigma
            A^{\rho \alpha}] -
          \frac{h_4^Z}{2 \MZ^{2}} [(\square +\MZ^{2}) \partial^\sigma
            Z^{\rho \alpha}] \right \} Z_\sigma \widetilde{A}_{\rho \alpha }
          \Bigg ] ~ .
\end{eqnarray}
Note that our conventions differ from those
of~\citere{Gounaris:1999kf} by a minus sign in the $\PZ$-boson field.
The constants can be expressed in terms of Wilson coefficients via the
relations:%
\footnote{Note that in \citere{Degrande:2013kka} the coefficient $f_4^Z$ is wrong by a factor 2} 
\allowdisplaybreaks[4]
 \begin{align}
    f_4^\gamma &=  \frac{ {\rm vev}^2 M_{\PZ}^2}{4 \cw \sw\Lambda^4} 
    \left(\cw\sw c_{WW}- \left(\cw^2-\sw^2\right)c_{BW}-4 \cw \sw
      {c_{BB}}\right),\notag\\
    f_4^Z &=  \frac{ M_{\PZ}^2 {\rm vev}^2}{4 \cw \sw\Lambda ^4} 
    \left(\cw^2 {c_{WW}}+2 \cw \sw {c_{BW}}+4 \sw^2
      {c_{BB}}\right),\notag\\
    f_5^\gamma &=  \frac{ {\rm vev}^2 M_{\PZ}^2 }{4 \cw \sw} \frac{c_{\widetilde{B}W}}{\Lambda^4},\notag\\
    f_5^Z&=0,\notag\\
    h_1^\gamma&=-\frac{
      {\rm vev}^2 M_{\PZ}^2}{4 \cw \sw\Lambda ^4}  \left(\sw^2 c_{WW}-2 \cw \sw {c_{BW}}+4 \cw^2 {c_{BB}}\right),\notag\\
    h_1^Z&= \frac{ {\rm vev}^2 M_{\PZ}^2}{4 \cw \sw\Lambda^4} 
    \left(-\cw\sw c_{WW}+ \left(\cw^2-\sw^2\right)c_{BW}+4 \cw \sw {c_{BB}}\right),\notag\\
    h_2^\gamma&=0,\notag\\
    h_2^Z&=0,\notag\\
    h_3^\gamma&=0,\notag\\
    h_3^Z&=\frac{  {\rm vev}^2 M_{\PZ}^2 }{4 \cw \sw} \frac{c_{\widetilde{B}W}}{\Lambda^4},\notag\\
    h_4^\gamma&=  0, \notag\\
    h_4^Z&=0.
  \label{eq:NTGCEFTconv}
\end{align}

In Eqs.~(\ref{eq:tgcpheno})--(\ref{eq:NTGCEFTconv}), $M_{V}$ ($V=\PW$,
$\PZ$) represent the gauge-boson masses, $\cw=\MW/\MZ$ and
$\sw=\sqrt{1-\cw^2}$ are the cosine and sine of the weak-mixing angle,%
\footnote{Be careful to discriminate the Wilson coefficient $c_W$ and
  the cosine of the weak mixing angle $\cw$.}
$e$ is the electric charge, and ${\rm vev}=2 M_\PW \sw/e$ represents
the vacuum expectation value of the Higgs-doublet field $\Phi$.

Cross sections and/or differential distributions obtained from the Lagrangian in Eq.~(\ref{eq:eftgen}) have the form
\begin{equation}
  \sigma = \sigma_{{\rm SM}^2} + \sigma_{{\rm SM}\times {\rm EFT6}} +
  \sigma_{{\rm EFT6}^2} + \sigma_{{\rm SM}\times {\rm EFT8}}  +
  \sigma_{{\rm EFT8}^2} + \dots ,  
  \label{eq:EFTobs}
\end{equation}
with
\begin{equation}
  \sigma_{{\rm SM}\times {\rm EFT6}} \propto \frac{c_6}{\Lambda^2} ,
  \quad  
\sigma_{{\rm EFT6}^2} \propto \frac{c_6^2}{\Lambda^4} \quad 
  \sigma_{{\rm SM}\times {\rm EFT8}} \propto \frac{c_8}{\Lambda^4}, \quad 
\sigma_{{\rm EFT8}^2} \propto \frac{c_8^2}{\Lambda^8} 
.  
  \label{eq:EFTobs2}
\end{equation}
It is clear from Eqs.~(\ref{eq:EFTobs})--(\ref{eq:EFTobs2}) that the $\sigma_{{\rm EFT6}^2}$ and $\sigma_{{\rm SM}\times {\rm EFT8}}$ are
of the same order in the $1/ \Lambda$ expansion. This means that for a generic EFT model a consistent $1/ \Lambda$ expansion
at the lowest order should only include the $\sigma_{{\rm SM}\times {\rm EFT6}}$ term. On the other hand, a wide range of strongly
interacting BSM models exists where the  $\sigma_{{\rm SM}\times {\rm EFT8}}$ term is subleading with respect to $\sigma_{{\rm EFT6}^2}$ 
terms without invalidating the EFT expansion~\cite{Biekoetter:2014jwa,Contino:2016jqw,Falkowski:2016cxu}. For these reasons in
\refse{sect:numres} we show our numerical results for the impact of the anomalous triple-gauge-boson interaction to $\PW\PW$ and
$\PW \PZ$ production both with and without the contribution of the $\sigma_{{\rm EFT6}^2}$ terms.  Similar considerations hold for
dimension-8 operators in $\PZ\PZ$ production.

In the phenomenological analysis of
\refses{subsect:wwres}--\ref{subsect:wzres} for $\PW\PW$ and $\PW\PZ$
production we consider the following values for the Wilson
coefficients corresponding to the dimension-6 operators in
Eq.~(\ref{eq:tgcWWv}) that are consistent with experimental limits of
\citere{Sirunyan:2017bey}:
\begin{equation}\arraycolsep 2pt
  \begin{array}[b]{lcllcllcl}
    \frac{c_{W}^{+}}{\Lambda^2}           & = &  3 \times 10^{-6} \GeV^{-2}, &
    \frac{c_{W}^{-}}{\Lambda^2}           & = & -3 \times
    10^{-6}\GeV^{-2}, \\
    \frac{c_{B}^{+}}{\Lambda^2}           & = &  1.5 \times
    10^{-5} \GeV^{-2}, \quad  &
    \frac{c_{B}^{-}}{\Lambda^2}           & = & -1.5 \times 10^{-5} \GeV^{-2}, \\
    \frac{c_{WWW}^{+}}{\Lambda^2}         & = &  3 \times 10^{-6} \GeV^{-2}, &
    \frac{c_{WWW}^{-}}{\Lambda^2}         & = & -3 \times 10^{-6} \GeV^{-2}, \\
    \frac{\tilde{c}_{W}^{+}}{\Lambda^2}   & = &  1 \times 10^{-6} \GeV^{-2}, &
    \frac{\tilde{c}_{W}^{-}}{\Lambda^2}   & = & -1 \times 10^{-6} \GeV^{-2}, \\
    \frac{\tilde{c}_{WWW}^{+}}{\Lambda^2} & = &  3 \times 10^{-6} \GeV^{-2}, &
    \frac{\tilde{c}_{WWW}^{-}}{\Lambda^2} & = & -3 \times 10^{-6} \GeV^{-2}.
  \end{array}
  \label{eq:wwwilsoncoeffs}
\end{equation}
For the Wilson coefficients of the dimension-8 operators in Eq.~(\ref{eq:ntgcop}) we use the values
\begin{equation}\arraycolsep 2pt
  \begin{array}[b]{lcllcllcl}
    \frac{c_{BB}^{+}}{\Lambda^4}           & = &  2 \times
    10^{-12} \GeV^{-4}, \quad &
    \frac{c_{BB}^{-}}{\Lambda^4}           & = & -2 \times 10^{-12} \GeV^{-4}, \\
    \frac{c_{WW}^{+}}{\Lambda^4}           & = &  3.5 \times 10^{-12} \GeV^{-4}, &
    \frac{c_{WW}^{-}}{\Lambda^4}           & = & -3.5 \times 10^{-12} \GeV^{-4}, \\
    \frac{c_{BW}^{+}}{\Lambda^4}           & = &  2 \times 10^{-12} \GeV^{-4}, &
    \frac{c_{BW}^{-}}{\Lambda^4}           & = & -2 \times 10^{-12} \GeV^{-4}, \\
    \frac{c_{\tilde{B}W}^{+}}{\Lambda^4}   & = &  2 \times 10^{-12} \GeV^{-4}, &
    \frac{c_{\tilde{B}W}^{-}}{\Lambda^4}   & = & -2 \times 10^{-12} \GeV^{-4},
  \end{array}
  \label{eq:zzwilsoncoeffs}
\end{equation}
which are consistent with the experimental bounds of \citere{Aaboud:2016urj}.

\section{Input parameters and cuts}
\label{sect:params}

We study the impact of the anomalous triple-gauge-boson interactions 
at the LHC with a centre-of-mass energy of $13\TeV$. Our numerical 
predictions are obtained using the $\GF$ scheme, where the electromagnetic 
coupling $\alpha$ is derived from $\GF$ with the relation 
\begin{align}
 \alpha_{G_\mu}=\frac{\sqrt{2}}{\pi}G_\mu 
 M_\PW^2\left(1-\frac{M_\PW^2}{M_\PZ^2}\right).
\end{align}
 
The relevant SM input parameters are~\cite{Patrignani:2016xqp}:
\begin{equation}\arraycolsep 2pt
  \begin{array}[b]{lcllcllcl}
    \GF & = & 1.1663787 \times 10^{-5} \GeV^{-2}, 
    \\
    \MW^{\OS} & = & 80.385\GeV, &
    \quad \Gamma_\PW^{\OS} & = & 2.085\GeV, \\
    \MZ^{\OS} & = & 91.1876\GeV, &
    \quad \Gamma_\PZ^{\OS} & = & 2.4952\GeV, \\
     M_\PH & = & 125\GeV, &
    \quad \Gamma_\PH & = & 4.097\MeV,\\
    m_\Pt & = & 173.2\GeV, &
    \quad \Gamma_\Pt & = & 1.369\GeV.
  \end{array}
  \label{eq:SMpar}
\end{equation}
Except for the top quark, all the other fermions are considered as
massless, and we use a diagonal CKM matrix. The on-shell $\PW$ and
$\PZ$ masses are converted to the corresponding pole values
as~\cite{Bardin:1988xt}:
\begin{align}
  M_V= \frac{M^{\OS}_V}{\sqrt{1+(\Gamma^{\OS}_V/M_V^{\OS})^2 } }, \qquad \Gamma_V = \frac{\Gamma_V^{\OS}}{\sqrt{1+(\Gamma_V^{\OS}/M_V^{\OS})^2 } }, \quad V=\PW,\PZ.
\end{align}

The complex-mass scheme
(CMS)~\cite{Denner:1999gp,Denner:2005fg,Denner:2006ic} is used in
order to deal with the presence of resonances. In the CMS the weak
mixing angle is derived from the ratio $\mu_{\PW}/\mu_{\PZ}$ with
$\mu_V^2=M_V^2-\im \Gamma_V M_V$.  Besides the $\PW$ and $\PZ$
resonances, also top resonances appear in the real QCD corrections to
$\PW\PW$ production with initial-state $\Pb$~quarks. In the
loop-induced processes $\Pg\Pg \to \PW \PW$ and $\Pg\Pg \to \PZ \PZ$
also Higgs resonances are present.

For the parton distribution functions (PDFs), the {\sc LHAPDF6.1.6}
package~\cite{Buckley:2014ana} is used and the {\tt
  NNPDF23\_nlo\_as\_0118\_qed} PDF
set~\cite{Ball:2012cx,Ball:2013hta,Ball:2014uwa} is employed. The same
PDF set is used to compute both the LO and NLO results. The
factorization and renormalization scales for the processes $\Pp \Pp
\to V V'$ are set to $(M_V^{\OS}+M_{V'}^{\OS})/2$.  The corresponding
value of $\alpha_{\rm s}$ is taken from the used PDFs.

For $\PW\PW$ production we consider the cuts of \citeres{Aad:2016wpd} and~\cite{Khachatryan:2015sga}
for ATLAS and CMS, respectively. The ATLAS setup can be summarized as follows:
\begin{equation}
  \begin{aligned} 
    & p_{\rT,l}> 20\GeV, \quad | \eta_l | < 2.5, \quad
       p_{\rT,l}^{\rm max} > 25\GeV, \\
    & M^{\rm inv}_{ll} > 10\GeV, \quad 
       E_{\rT}^{\rm miss} > 20\GeV, \quad
       E_{\rT}^{\rm rel} > 15\GeV, \\
    & 0 \; {\rm jets } \; {\rm with } \; p_{\rT,\rm jet} > 25\GeV, \quad  | \eta_{\rm jet} | < 4.5.
  \end{aligned}
  \label{eq:wwatlascut}
\end{equation}     
The CMS setup is:
\begin{equation}
  \begin{aligned}
    & p_{\rT,l}> 20\GeV, \quad | \eta_l | < 2.5, \\
    & M^{\rm inv}_{ll} > 12\GeV, \quad  E_{\rT}^{\rm miss} > 20\GeV, \quad  p_{\rT,ll} > 30\GeV,  \\
    & 0 \; {\rm jets } \; {\rm with } \; p_{\rT,\rm jet } > 30\GeV,
    \quad | \eta_{\rm jet} | < 5 .
  \end{aligned}
  \label{eq:wwcmscut}
\end{equation}     
In Eqs.~(\ref{eq:wwatlascut})--(\ref{eq:wwcmscut}), $l$ stands for a
charged lepton, $p_{\rT,ll}$ and $M^{\rm inv}_{ll}$ are the transverse
momentum and the invariant mass of the charged lepton pair,
$p_{\rT,l}^{\rm max}$ is the transverse momentum of the hardest
lepton, \ie the lepton with highest $p_{\rT}$, and $E_T^{\rm miss}$ is
the missing momentum in the transverse plane obtained from the sum of
the momenta of the two neutrinos.  Finally, $E_{\rT}^{\rm rel}$ is
defined as
\begin{equation}
 E_{\rT}^{\rm rel} =  \left\{ \begin{array}{ll}
  E_{\rT}^{\rm miss} {\rm sin } \Delta \phi_l \quad  &
    {\rm if } \; \Delta \phi_l \; \in \left[ -\frac{\pi}{2}, \frac{\pi}{2} \right], \\
   E_{\rT}^{\rm miss} \quad & 
    {\rm if } \; \Delta \phi_l \; \notin \left[ -\frac{\pi}{2}, \frac{\pi}{2} \right], 
  \end{array}\right.
  \label{eq:etrel}
\end{equation}
where $\Delta \phi_l$ is the difference in azimuthal angle between the
direction of the missing-momentum vector $\vec{E}_{\rT}^{\rm miss}$
and the momentum of the charged lepton closest to $\vec{E}_{\rT}^{\rm
  miss}$.  

At NLO QCD the large-$p_{\rT,l}$ region of $\PW\PW$ production is
dominated by kinematical configurations where a $\PW$~boson is
recoiling against a hard quark that radiates a soft $\PW$~boson. Since
this kind of process does not depend on aTGCs~\cite{Baur:1995uv}, the
sensitivity to aTGCs is largely lost when moving from LO to NLO.
Therefore, both ATLAS and CMS impose a jet veto for the search of
aTGCs in the $\PW\PW$ channel.  We define jets according to the
anti-$k_t$
algorithm~\cite{Cacciari:2005hq,Cacciari:2008gp,Cacciari:2011ma} with
$R$ parameter $0.4$ and $0.5$ for the ATLAS and the CMS event
selection, respectively.

The ATLAS analysis setup for the process $\Pp\Pp \to \PW\PZ$ reads~\cite{Aad:2016ett}:
\begin{equation}
  \begin{aligned}
    & p_{\rT,l_{i}} > 15\GeV, \quad | \eta_{l_{i}} | < 2.5,\quad
      p_{\rT,l_{\PW}} > 20\GeV , \quad | \eta_{l_{\PW}} | < 2.5,\\
    & | M^{\rm inv}_{l_{1}l_{2}} - M_{\PZ} | < 10\GeV, \quad M_{\rT,\PW}  > 30\GeV ,\\
    & \Delta R_{l_{i},l_{\PW}} >0.3, \quad \Delta R_{l_{1},l_{2}} >0.2, \quad
      p_{\rT,l}^{{\rm max}}  > 25\GeV ,
  \end{aligned}
  \label{eq:wzatlascut}
\end{equation}     
while the CMS one reads~\cite{Khachatryan:2016poo}:
\begin{equation}
  \begin{aligned}
    & p_{\rT,l} > 20\GeV, \quad | \eta_{l} | < 2.5, \quad
       E_{\rT}^{\rm miss} > 30\GeV, \quad M^{\rm inv}_{3l} > 100\GeV ,\\
    & M^{\rm inv}_{l_{1}l_{2}} \in \left[ 71,111\right] \GeV, \quad \Delta R_{l_{i},l_{\PW}}>0.1 . 
  \end{aligned}
  \label{eq:wzcmscut}
\end{equation}
In Eqs.~(\ref{eq:wzatlascut})--(\ref{eq:wzcmscut}) $l_i$, $i=1,2$, are
the two leptons coming from the $\PZ$ decay, $l_W$ is the charged
lepton from the $\PW$ decay, $M_{\rm T}^{\PW}$ is the transverse mass
of the $\PW$~boson defined as
\begin{equation}
  M_{\rm T}^{\PW} = \sqrt{2p_{\rT,l} p_{\rT,\nu}(1- {\rm cos} \Delta \phi_{l\nu})},
  \label{eq:mtendef}
\end{equation}
$M^{\rm inv}_{3l}$ is the invariant mass of the three charged leptons, and
$M^{\rm inv}_{l_{1}l_{2}}$ is the invariant mass of charged-lepton pair coming from the $\PZ$ decay. If more than one $l^+l^-$
pair can be assigned to the $\PZ$~boson, the $\PZ$-boson candidate with invariant mass $M^{\rm inv}_{l_{1}l_{2}}$ closest to the nominal
$\PZ$-boson mass is selected. Finally, 
\begin{equation}
\Delta R_{l_il_j}=\sqrt{(\eta_{l_i}-\eta_{l_j})^2+\Delta\phi_{l_il_j}^2}
\end{equation}
is the rapidity--azimuthal-angle separation of the leptons $l_i$ and $l_j$.

The ATLAS and the CMS cuts for the four-charged-lepton analysis of
\citeres{Aaboud:2016urj,CMS:2014xja} are very similar. In our
phenomenological studies we consider the ATLAS event selection:
\begin{equation}
  \begin{aligned}
    & p_{\rT,l} > 7\GeV, \quad | \eta_{l} | < 2.5, \quad
      p_{\rT,l}^{\mathrm{max}} > 25\GeV, \quad \Delta R_{l_{i},l_{j}} > 0.2, \\
    & M^{\rm inv}_{\PZ_{1}} \in \left[ 66,116\right] \GeV, \quad
      M^{\rm inv}_{\PZ_{2}} \in \left[ 66,116\right] \GeV,
  \end{aligned}
  \label{eq:zz4latlascut}
\end{equation}
In Eq.~(\ref{eq:zz4latlascut}), 
$\PZ_{1}$ and $\PZ_{2}$ stand for the two $\PZ$~bosons reconstructed from 
pairs of same flavour and opposite-charge leptons (in the following we will only
consider the process $\Pp\Pp\to\Pep\Pem\mu^+\mu^-$ where only one pairing is
possible).

In all the setups described above, when computing NLO EW corrections
we recombine final-state charged leptons and photons if $\Delta
R_{l\gamma} < 0.1$.

\section{Phenomenological results}
\label{sect:numres}
\subsection{WW production}
\label{subsect:wwres}

Our predictions for the integrated cross sections for the process
$\Pp\Pp \to \Pep \Pne \Pmum \bar{\nu}_{\mu}$ are collected in
\refta{tab:ww-xsec}. The LO results are compared to the ones at NLO
QCD and NLO EW accuracy. The contribution of the loop-induced
$\Pg\Pg\to\PW\PW$ process is also shown. The ATLAS and CMS setups in
\refta{tab:ww-xsec} correspond to the event selection in
Eqs.~(\ref{eq:wwatlascut}) and~(\ref{eq:wwcmscut}), respectively. For
both setups we present our results both with and without the
contribution of the processes with initial-state $\Pb$~quarks. The
impact of these processes is of order $2\%$ for LO and NLO EW, but
becomes of order $11{-}17\%$ for NLO QCD, because the
$\Pg\Pb\to\PW\PW\Pb$ channel is enhanced by the presence of top
resonances.  On one hand, this channel is absent in the four-flavour
scheme, on the other hand, the $t$-channel single-top contribution is
usually subtracted in experimental analyses. We prefer to use the same
PDF set {\tt NNPDF23\_nlo\_as\_0118\_qed} and the 5-flavour scheme for
all diboson production processes and simply discard the contribution
of initial-state $\Pb$~quarks (at the integrated-cross-section level
the two approaches lead to very similar results as shown in
\refta{tab:ww-ptjcut}).  If not otherwise stated, the effect of
initial-state $\Pb$~quarks is included in the following.
\begin{table}
  \begin{center}
\renewcommand{\arraystretch}{1.4}
    \begin{tabular}{|l|l|l|l|l|l}
      \hline
      Setup & LO [fb] & NLO QCD [fb] & NLO EW [fb] & $\Pg\Pg$  [fb] \\
      \hline
      ATLAS no \Pb   & $281.13(3)^{+6.1\%}_{-7.4\%}$ & $262.1(1)^{+2.9\%}_{-2.5\%}$ & $272.66(8)^{+6.3\%}_{-7.6\%}$ & $29.5(5)^{+25\%}_{-18\%}$\\
      \hline
      ATLAS with \Pb & $285.61(3)^{+6.5\%}_{-7.8\%}$ & $291.4(1)^{+2.6\%}_{-3.3\%}$ & $276.98(8)^{+6.6\%}_{-7.9\%}$ & $29.5(5)^{+25\%}_{-18\%}$\\
      \hline
      CMS no \Pb   & $239.77(2)^{+6.0\%}_{-7.3\%}$ & $238.3(1)^{+2.3\%}_{-2.4\%}$ & $231.58(7)^{+6.1\%}_{-7.4\%}$ & $27.3(3)^{+25\%}_{-18\%}$ \\
      \hline
        CMS with \Pb & $243.84(2)^{+6.3\%}_{-7.7\%}$ & $279.1(1)^{+2.8\%}_{-3.9\%}$ & $235.49(7)^{+6.5\%}_{-7.8\%}$ & $27.3(3)^{+25\%}_{-18\%}$ \\
      \hline
    \end{tabular}
  \end{center} 
  \caption{Integrated cross section for the process $\Pp\Pp \to \Pep \Pne \Pmum \bar{\nu}_{\mu}$ at $\sqrt{s}=13\TeV$ in the ATLAS and CMS setups of Eqs.~(\ref{eq:wwatlascut})
    and~(\ref{eq:wwcmscut}), respectively. The numbers in parentheses correspond to the statistical error on the last digit. The uncertainties are estimated from the 
    scale dependence, as explained in the text.}
  \label{tab:ww-xsec}
\end{table}

In \refta{tab:ww-xsec} and following tables the numbers in parentheses
correspond to the statistical integration error, while the
uncertainties are estimated from scale variation: we set the
factorization and renormalization scales to $\mu_{\rm F}=K_{\rm
  F}\mu_0$ and $\mu_{\rm R}=K_{\rm R}\mu_0$ ($\mu_0$ being the central
scale choice described in \refse{sect:params}) and we evaluate the
cross sections for the following combinations of $(K_{\rm F},K_{\rm
  R})$:
\begin{equation}
  (K_{\rm F},K_{\rm R})=\left\{ \Big( \frac{1}{2},\frac{1}{2} \Big),\Big(\frac{1}{2},1\Big),\Big(1,\frac{1}{2}\Big),\Big(1,1\Big),\Big(1,2\Big),\Big(2,1\Big),\Big(2,2\Big)\right\}.
  \label{eq:scalevar}
\end{equation}     
The upper and lower values of the cross sections in
\refta{tab:ww-xsec} correspond to the upper and lower limits of the
so-obtained scale variations. At LO and NLO EW, scale uncertainties
only result from variation of the factorization scale and are of the
same order. In contrast, the scale dependence at NLO QCD results from
variation of both factorization and renormalization scales and is
smaller than the LO one.

The $\Pg\Pg\to\PW\PW$ channel gives a positive contribution of order
$10\%$ with respect to the LO results. The NLO EW corrections are of
order $-3\%$. If the initial-state $\Pb$-quark contribution is not
included, also the NLO QCD corrections turn out to be negative: this
is a consequence of the jet veto in Eqs.~(\ref{eq:wwatlascut})
and~(\ref{eq:wwcmscut}) that basically removes the real QCD
corrections to $\PW\PW$ production. 

The dependence of the NLO QCD corrections to the fiducial cross
section on the jet veto is shown in \refta{tab:ww-ptjcut}, for
different setups, where we set the maximum jet-$p_{\rm T}$ cut to
$25$,~$50$,~$100$,~$200$ and~$1000\GeV$.  Besides results based on
5-flavour PDFs with and without initial-state bottom contributions we
also provide  results based on 4-flavour PDFs. While the
cross sections in the 5-flavour scheme agree well with those in the
4-flavour scheme when omitting the b-induced contributions, the latter
give a sizable extra contribution that grows with increasing jet veto.
\begin{table}
  \begin{center}
\renewcommand{\arraystretch}{1.4}
    \begin{tabular}{|l|l|l|l|l|l|l|}
\hline
ATLAS & \multicolumn{5}{|c|}{$\sigma^\mathrm{NLO\;QCD}$  [fb] } \\
\hline
  $p_{\rm T, \, jet}\leq $ & $25\GeV$ & $50\GeV$  & $100\GeV$ & $200\GeV$  &  $1000\GeV$ \\
      \hline
      5f with \Pb & $291.4(1)^{+2.6\%}_{-3.3\%}$ & $458.0(1)^{+4.5\%}_{-5.5\%}$ & $699.5(2)^{+7.1\%}_{-7.5\%}$ & $796.3(2)^{+7.7\%}_{-7.5\%}$ & $817.2(2)^{+7.8\%}_{-7.2\%}$\\
      \hline
      5f no \Pb & $262.1(1)^{+2.9\%}_{-2.5\%}$ & $334.4(1)^{+2.0\%}_{-2.6\%}$ & $389.2(1)^{+3.3\%}_{-2.7\%}$ & $423.0(1)^{+4.0\%}_{-3.2\%}$ & $439.0(1)^{+4.3\%}_{-3.5\%}$\\
      \hline
      4f & $260.1(1)^{+2.8\%}_{-2.4\%}$ & $330.3(1)^{+1.9\%}_{-2.6\%}$ & $383.7(1)^{+3.4\%}_{-2.7\%}$ & $416.9(1)^{+4.1\%}_{-3.3\%}$ & $432.6(1)^{+4.5\%}_{-3.5\%}$\\
      \hline
    \end{tabular}
  \end{center} 
  \caption{Integrated cross section at NLO QCD for the process $\Pp\Pp
    \to \Pep \Pne \Pmum \bar{\nu}_{\mu}$ at $\sqrt{s}=13\TeV$ in the
    ATLAS setup of Eq.~(\ref{eq:wwatlascut})   for different values of
    the jet veto. The results in the third line (5f with \Pb) are
    computed using the 5-flavour PDF set 
    {\tt NNPDF23\_nlo\_as\_0118\_qed} with initial-state $\Pb$-quark
    contribution included. These contributions are omitted in the
    results in the fourth line (5f no \Pb). In the last line
    the 4-flavour PDF set {\tt NNPDF30\_nlo\_as\_0118\_nf\_4} is used. 
    Same notation and conventions as in \refta{tab:ww-xsec}.}
  \label{tab:ww-ptjcut}
\end{table}
In \reffi{fig:wwqcdcut} the dependence of the NLO QCD corrections on
the jet veto is illustrated for the distributions in the transverse
momenta of the hardest lepton ($p_{\rT,l}$) and of the
charged-lepton-pair ($p_{\rT,ll}$). The real QCD corrections become
more and more important as the jet veto is loosened, leading to large
positive QCD corrections in particular in the high-$p_{\rm T}$ region.
Even for strict jet veto, for both the setups of
Eqs.~(\ref{eq:wwatlascut}) and~(\ref{eq:wwcmscut}) the NLO QCD
corrections become positive once the processes with initial-state
$\Pb$~quarks are taken into account (not shown).
\bfi
\begin{center}
  \begin{minipage}{0.40\textwidth}
    \includegraphics[width=\textwidth]{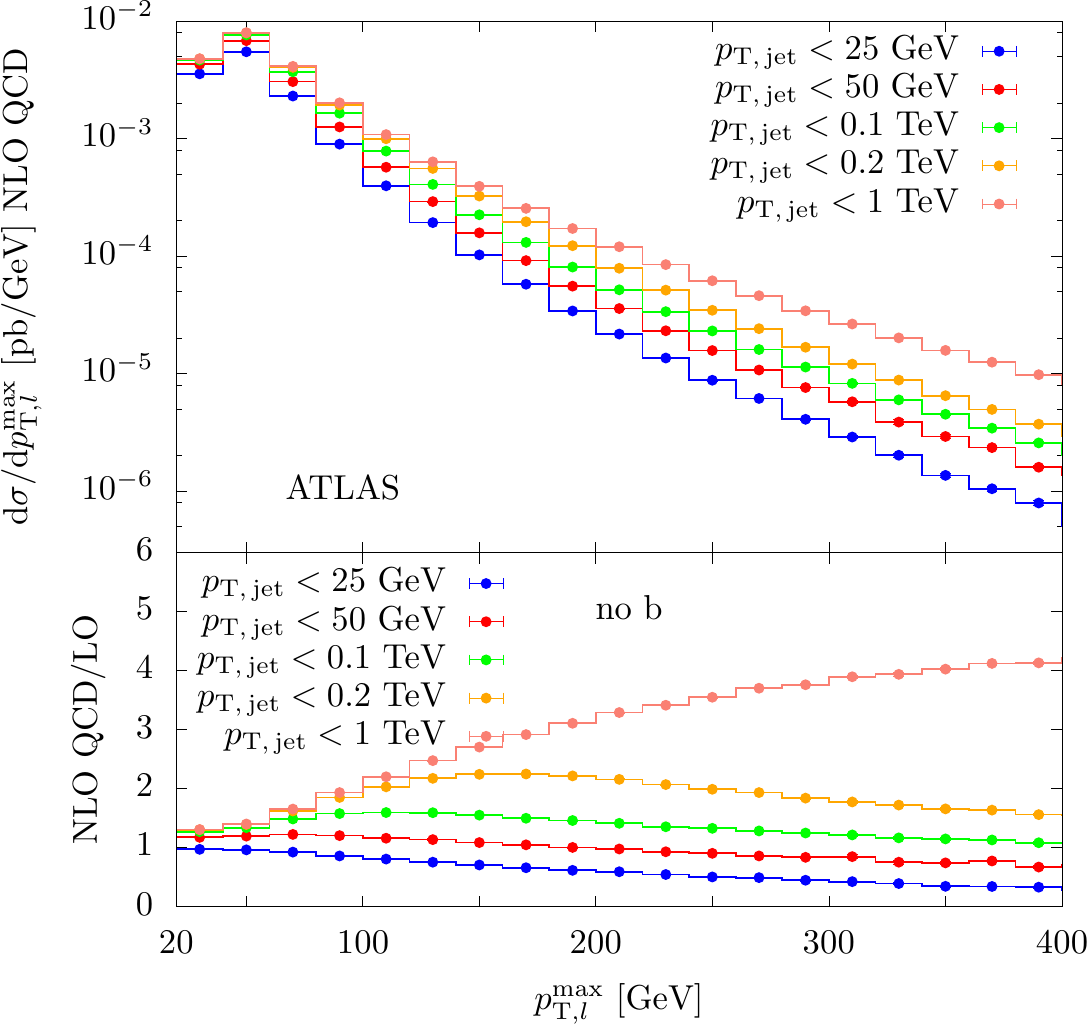}
  \end{minipage}
  \begin{minipage}{0.40\textwidth}
    \includegraphics[width=\textwidth]{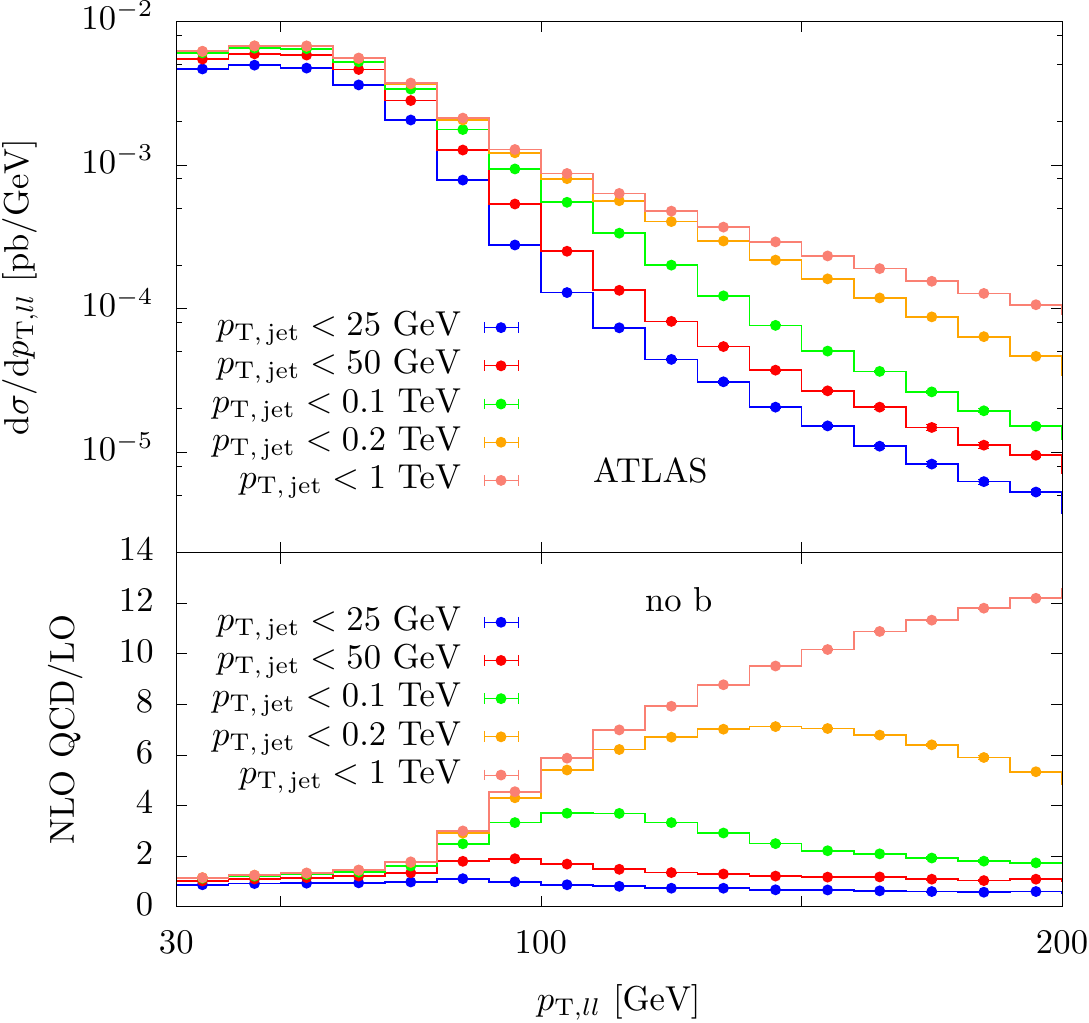}
  \end{minipage}
\end{center}
\caption{Differential distribution in the transverse momentum of the hardest lepton ($p_{{\rm T},l}^{\rm max}$) and in the $p_{\rm T}$ of the charged-lepton pair
  ($p_{{\rm T},ll}$) for the process $\Pp\Pp \to \Pep \Pne \Pmum
  \bar{\nu}_{\mu}$ at $\sqrt{s}=13\TeV$ under the event selections of
  Eq.~(\ref{eq:wwatlascut}) for different values of the jet $p_{\rm T}$
  entering the veto  condition in Eq.~(\ref{eq:wwatlascut}).
  Lower panels: ratio between the NLO QCD and the LO predictions (the
  same PDF set is used at LO and NLO QCD). The contribution from
  initial-state~$\Pb$ quarks is not included.}
\label{fig:wwqcdcut}
\efi

Our predictions at the differential-distribution level are collected
in \reffis{fig:wwpte}--\ref{fig:wwy} for some sample observables.
Figures~\ref{fig:wwpte}--\ref{fig:wwy} confirm the pattern described
above at the cross-section level: both the EW and QCD corrections are
small in those bins that give the largest contribution to the
integrated cross section, while their size increases in the
high-$p_{\rm T}$ and invariant-mass regions.
\bfi
\begin{center}
  \begin{minipage}{0.40\textwidth}
    \includegraphics[width=\textwidth]{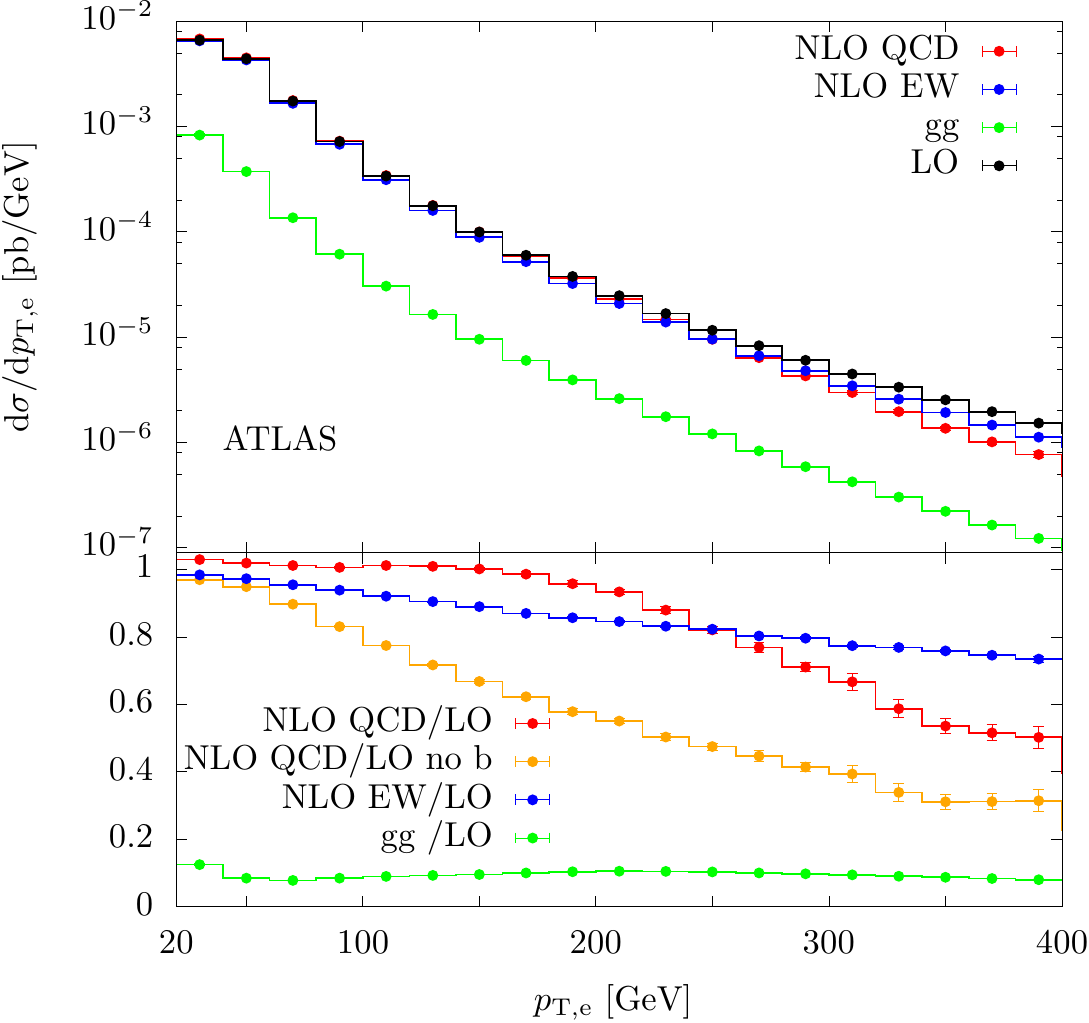}
  \end{minipage}
  \begin{minipage}{0.40\textwidth}
    \includegraphics[width=\textwidth]{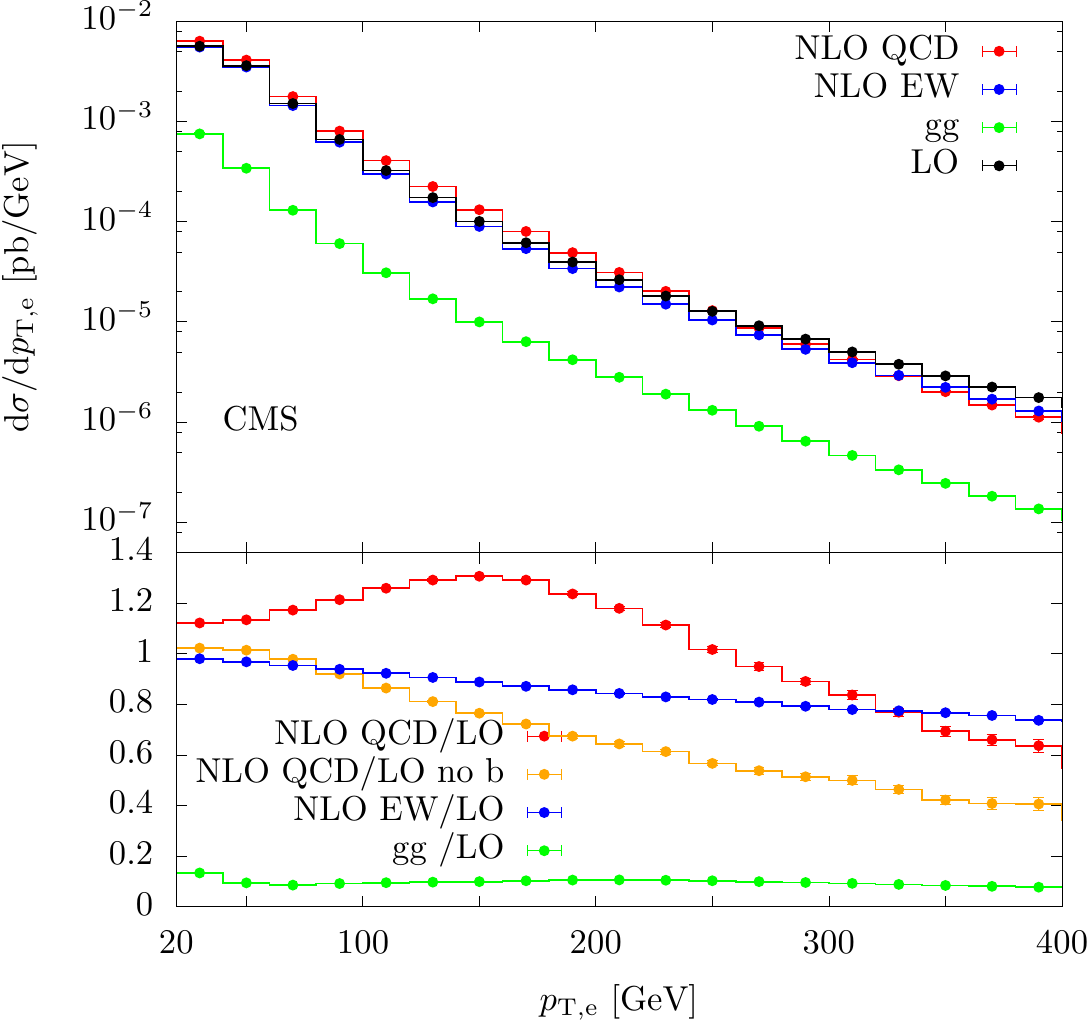}
  \end{minipage}
\end{center}
\caption{Differential distributions in the transverse
  momentum of the positron ($p_{{\rm T},\Pe}$) for the process
  $\Pp\Pp \to \Pep \Pne \Pmum \bar{\nu}_{\mu}$
  at $\sqrt{s}=13\TeV$ for the ATLAS and CMS event
  selections of Eqs.~(\ref{eq:wwatlascut}) and~(\ref{eq:wwcmscut}),
  respectively. The LO results (black lines) are compared to the ones
  at NLO QCD (red lines) and NLO EW (blue lines). The $\Pg\Pg$
  contribution is also shown (green lines).  Lower panels: ratio of
  the NLO QCD, NLO EW and $\Pg\Pg$ contributions with respect to the
  LO (red, blue and green lines, respectively). The orange lines
  correspond to the ratio of the NLO QCD corrections and the LO
  predictions if the processes with initial-state $\Pb$~quarks are not
  included (see text for details). For all curves the central value of
  the factorization and renormalization scales is used and the error
  bars correspond to the statistical integration uncertainties. Note
  that the same PDF set is employed for both the LO and NLO
  predictions.}
\label{fig:wwpte}
\efi

\bfi
\begin{center}
  \begin{minipage}{0.40\textwidth}
    \includegraphics[width=\textwidth]{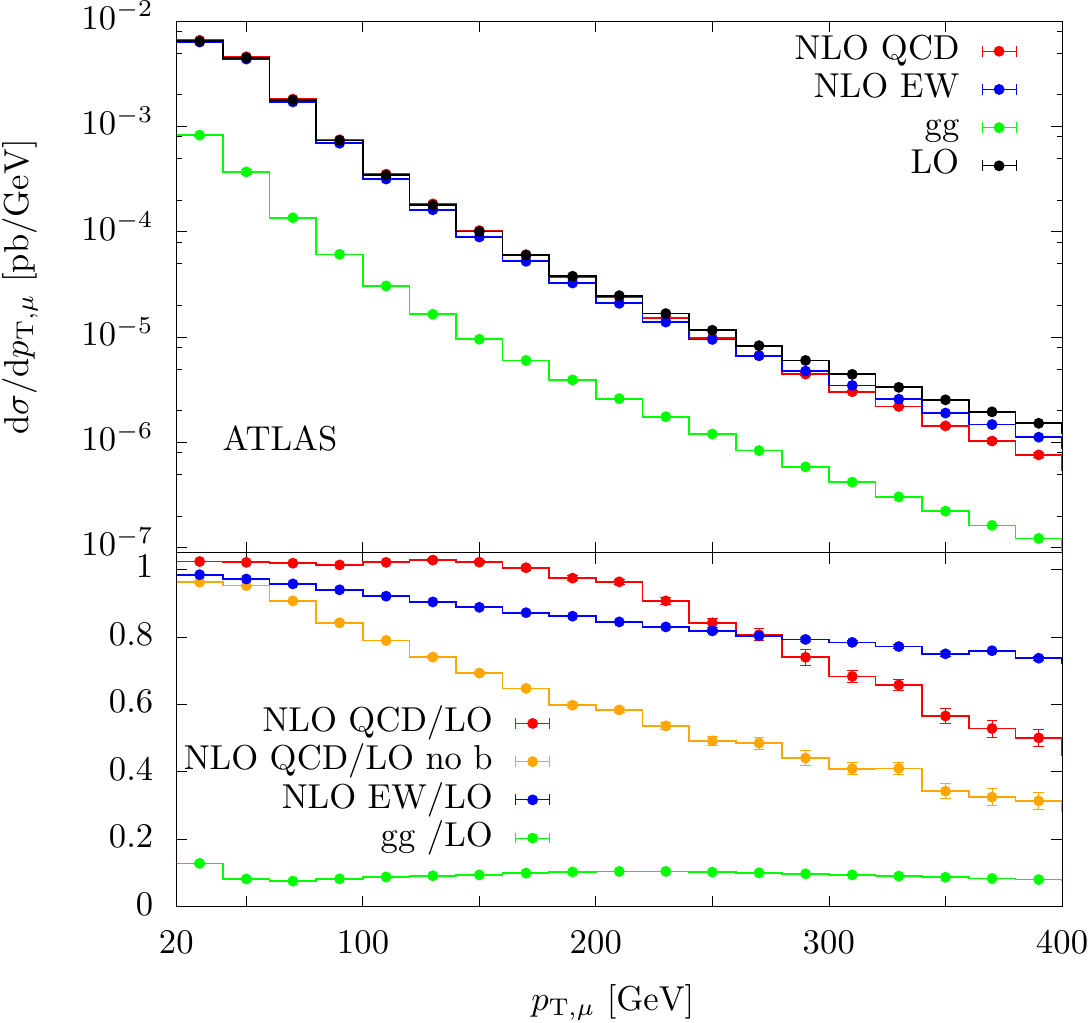}
  \end{minipage}
  \begin{minipage}{0.40\textwidth}
    \includegraphics[width=\textwidth]{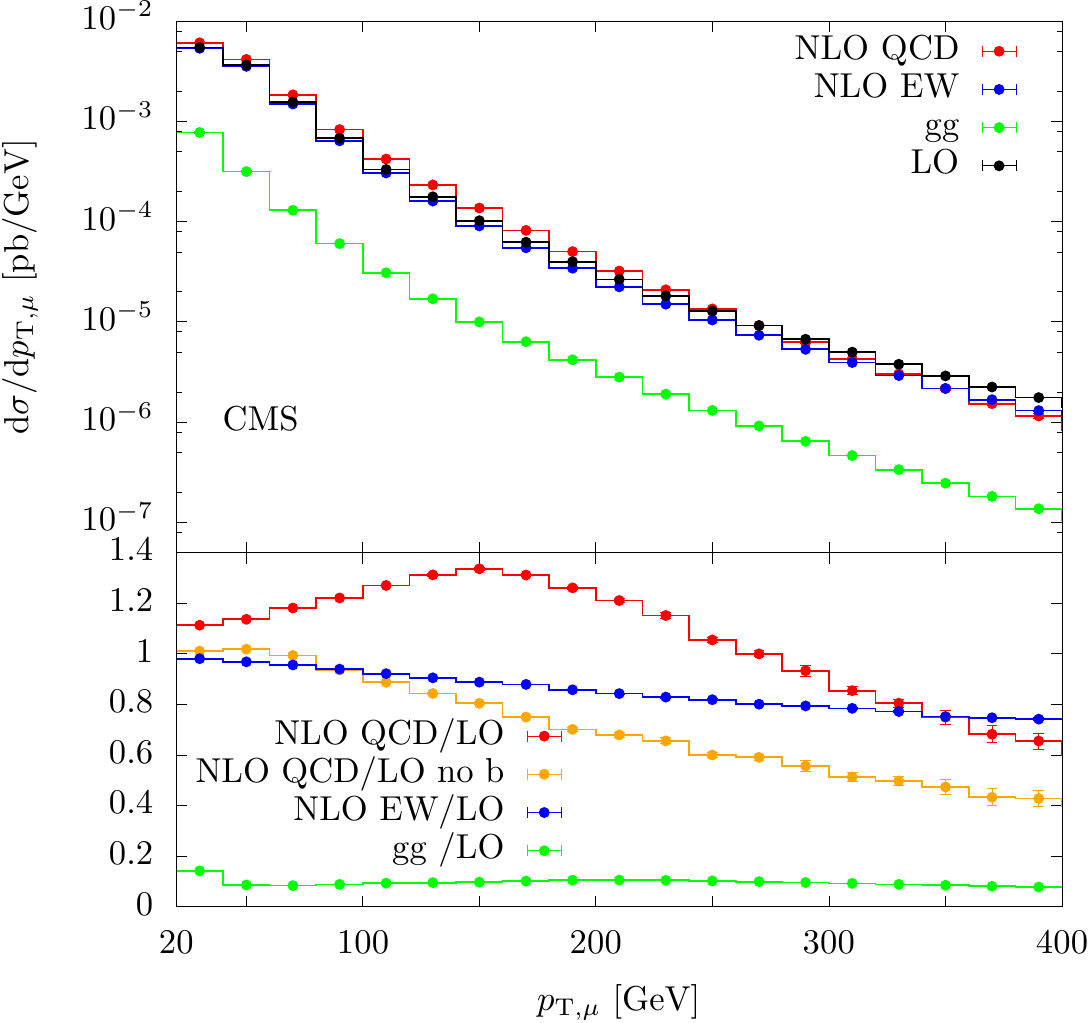}
  \end{minipage}
\end{center}
\caption{Differential distributions in the transverse
  momentum of the muon  ($p_{{\rm T},\mu}$) for the process
  $\Pp\Pp \to \Pep \Pne \Pmum \bar{\nu}_{\mu}$
  at $\sqrt{s}=13\TeV$ for the ATLAS and CMS event
  selections of Eqs.~(\ref{eq:wwatlascut}) and~(\ref{eq:wwcmscut}),
  respectively. Same notations and conventions as in \reffi{fig:wwpte}.}
\label{fig:wwptm}
\efi
\bfi
\begin{center}
  \begin{minipage}{0.40\textwidth}
    \includegraphics[width=\textwidth]{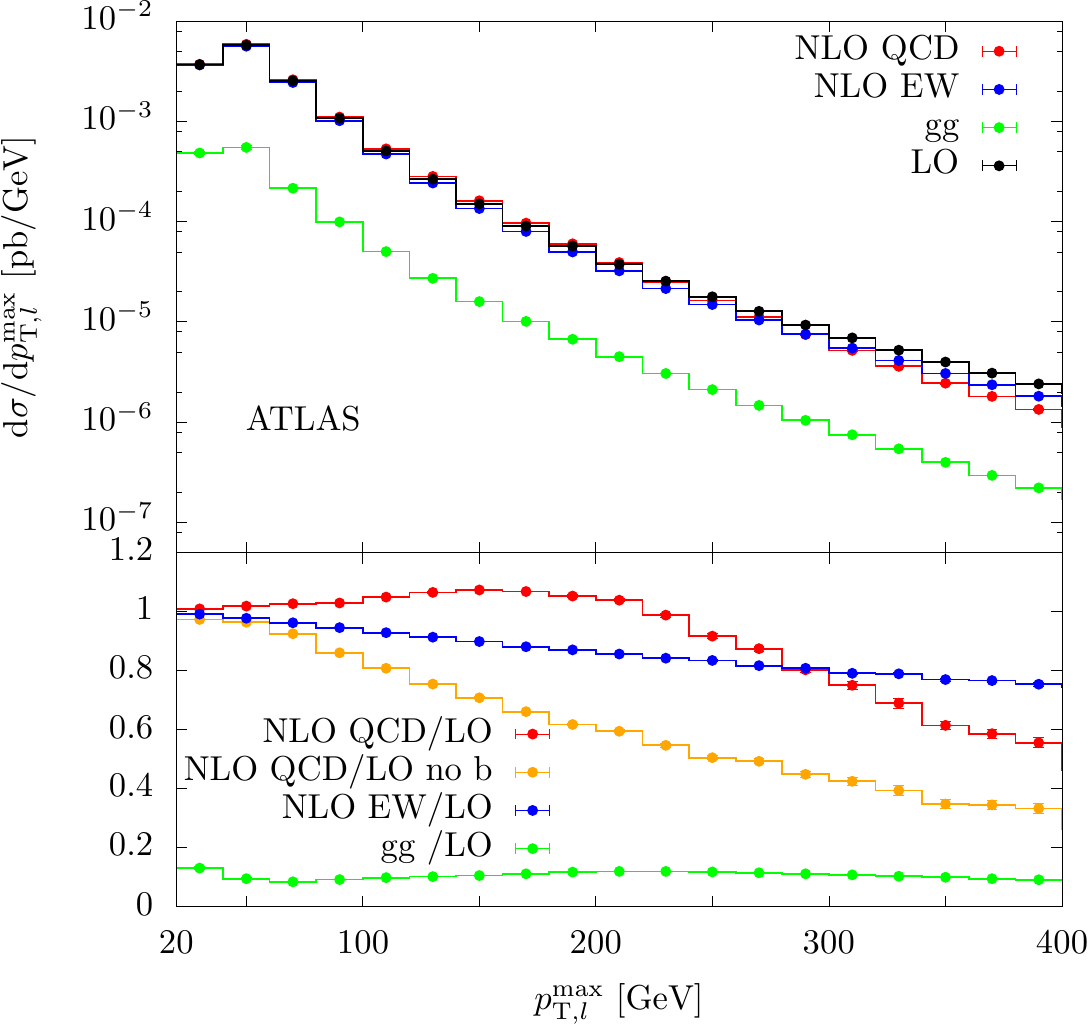}
  \end{minipage}
  \begin{minipage}{0.40\textwidth}
    \includegraphics[width=\textwidth]{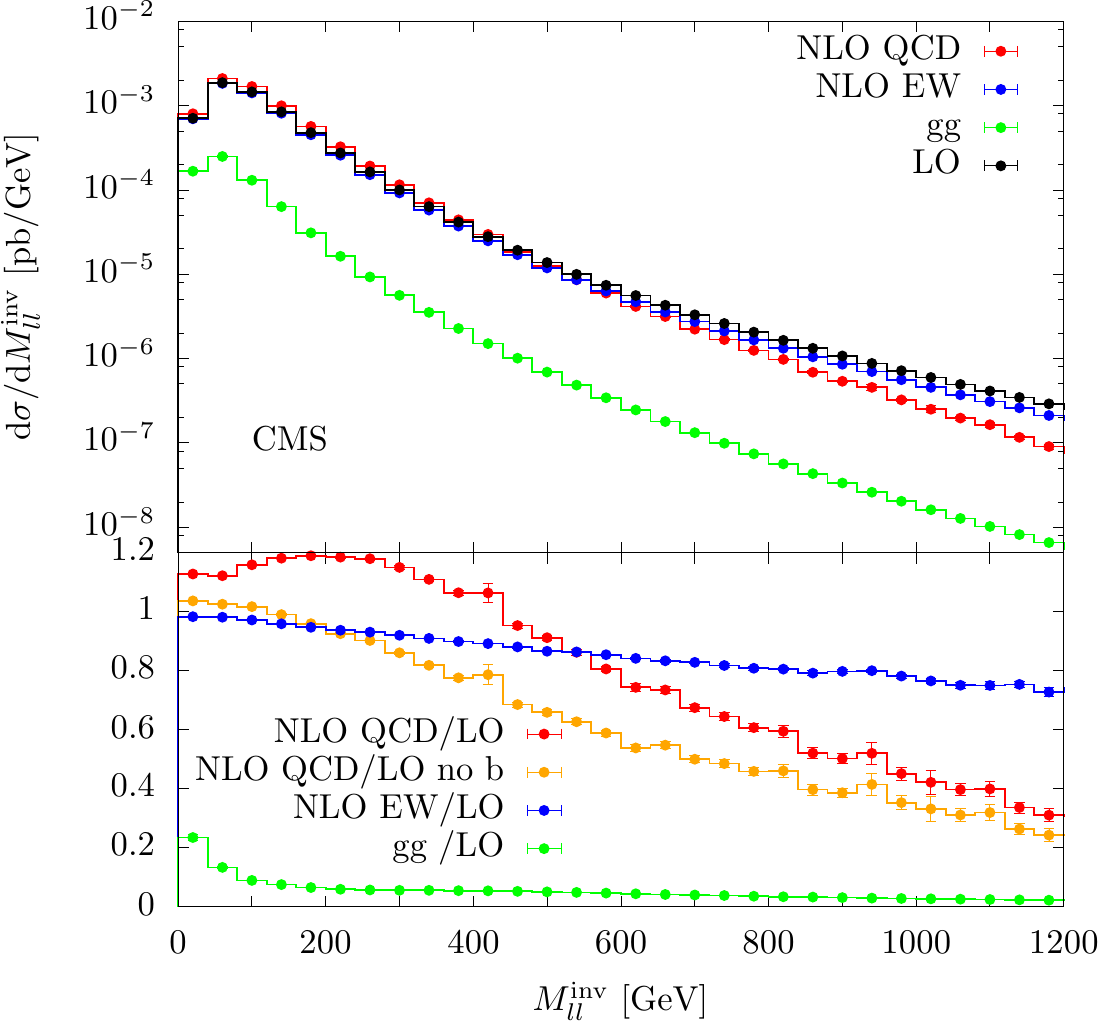}
  \end{minipage}
\end{center}
\caption{Differential distribution in the transverse momentum of the hardest lepton ($p_{{\rm T},l}^{\rm max}$) and in the invariant mass 
  of the charged-lepton pair ($M^{\rm inv}_{ll}$) for the process $\Pp\Pp \to \Pep \Pne \Pmum \bar{\nu}_{\mu}$ at $\sqrt{s}=13\TeV$ under the event
  selections of Eqs.~(\ref{eq:wwatlascut}) and~(\ref{eq:wwcmscut}). Same notations and conventions as in \reffi{fig:wwpte}.}
\label{fig:wwpthardmll}
\efi
\bfi
\begin{center}
  \begin{minipage}{0.40\textwidth}
    \includegraphics[width=\textwidth]{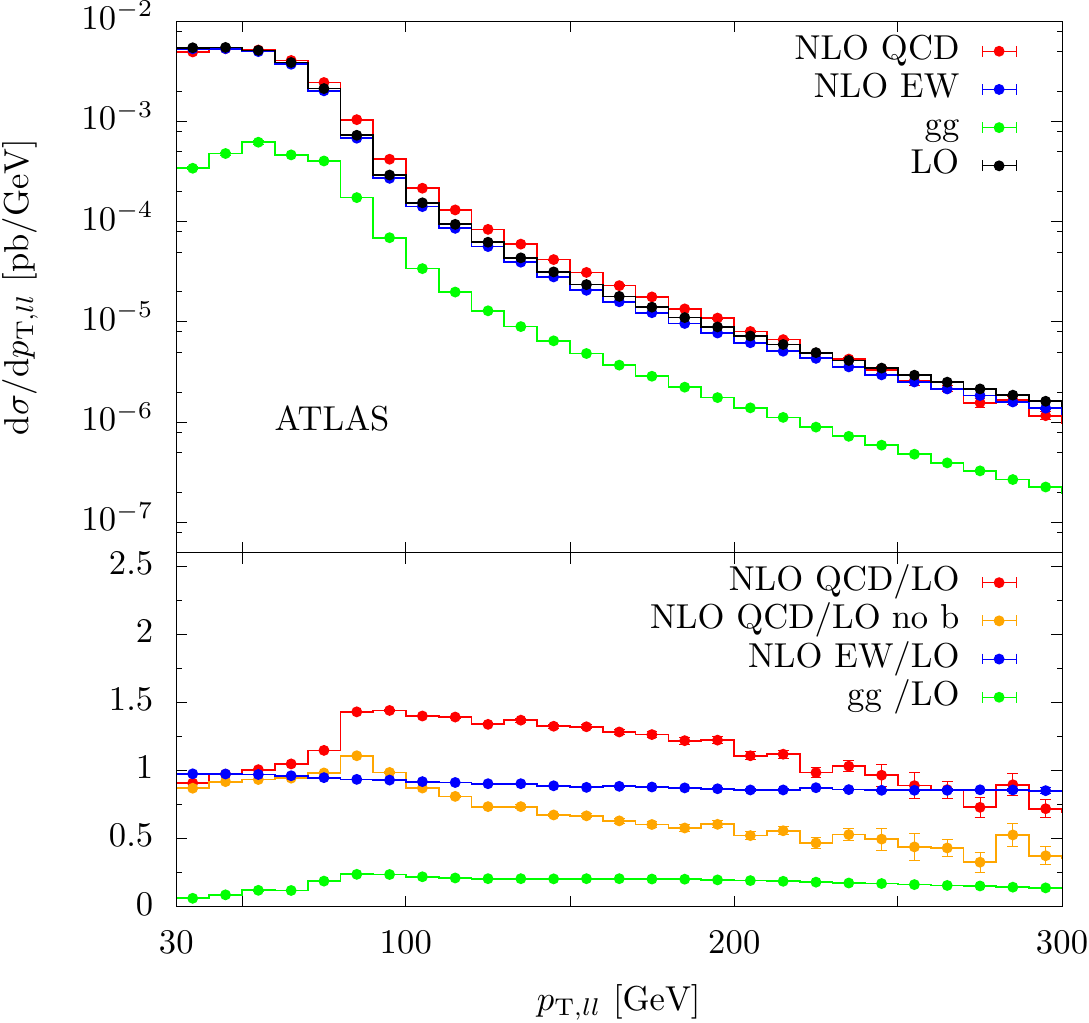}
  \end{minipage}
  \begin{minipage}{0.40\textwidth}
    \includegraphics[width=\textwidth]{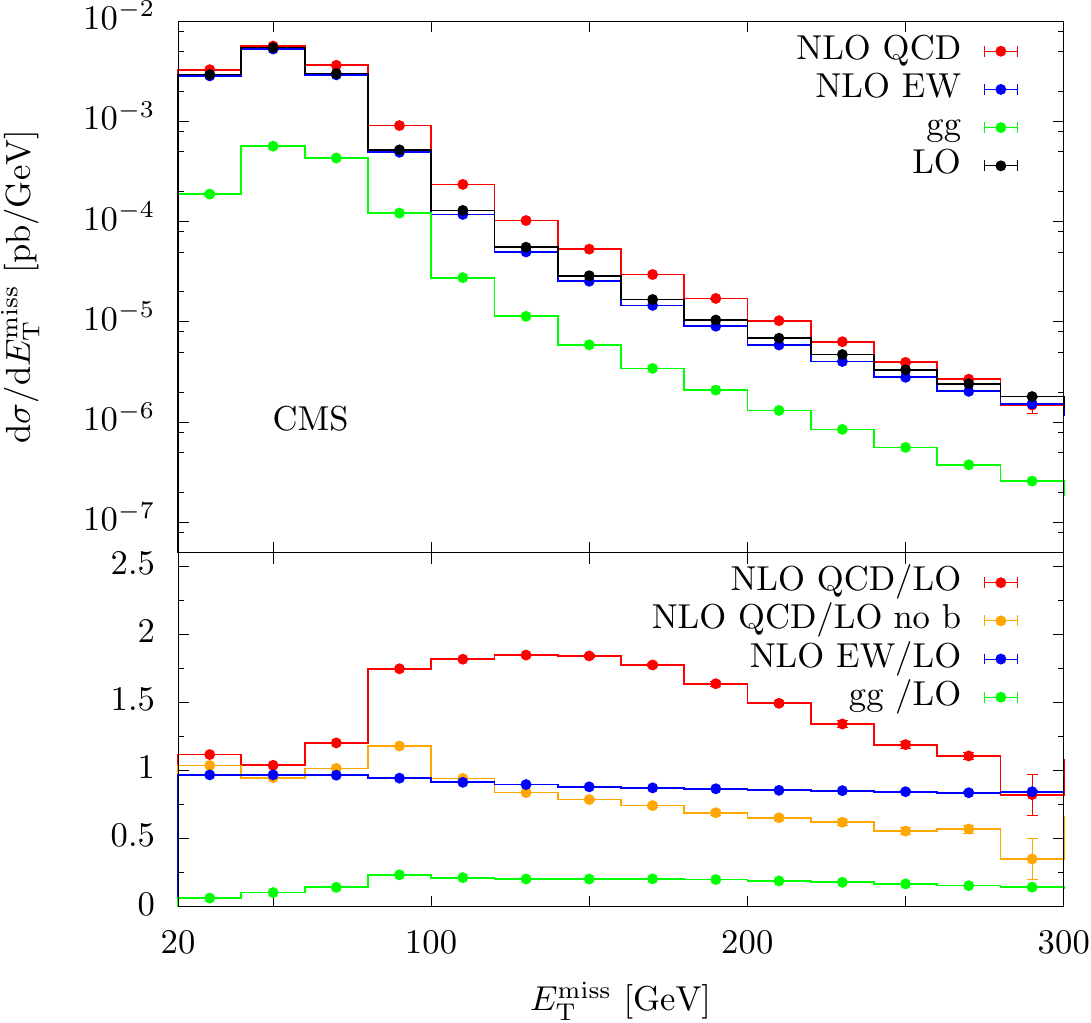}
  \end{minipage}
\end{center}
\caption{Differential distribution in the transverse momentum of the charged-lepton pair ($p_{{\rm T},ll}$) and in the missing transverse
  momentum ($E_{\rT}^{\rm miss}$) for the process $\Pp\Pp \to \Pep \Pne \Pmum \bar{\nu}_{\mu}$ at $\sqrt{s}=13\TeV$ under the event
  selections of Eqs.~(\ref{eq:wwatlascut}) and~(\ref{eq:wwcmscut}). Same notations and conventions as in \reffi{fig:wwpte}.}
\label{fig:wwptnptll}
\efi
\bfi
\begin{center}
  \begin{minipage}{0.40\textwidth}
    \includegraphics[width=\textwidth]{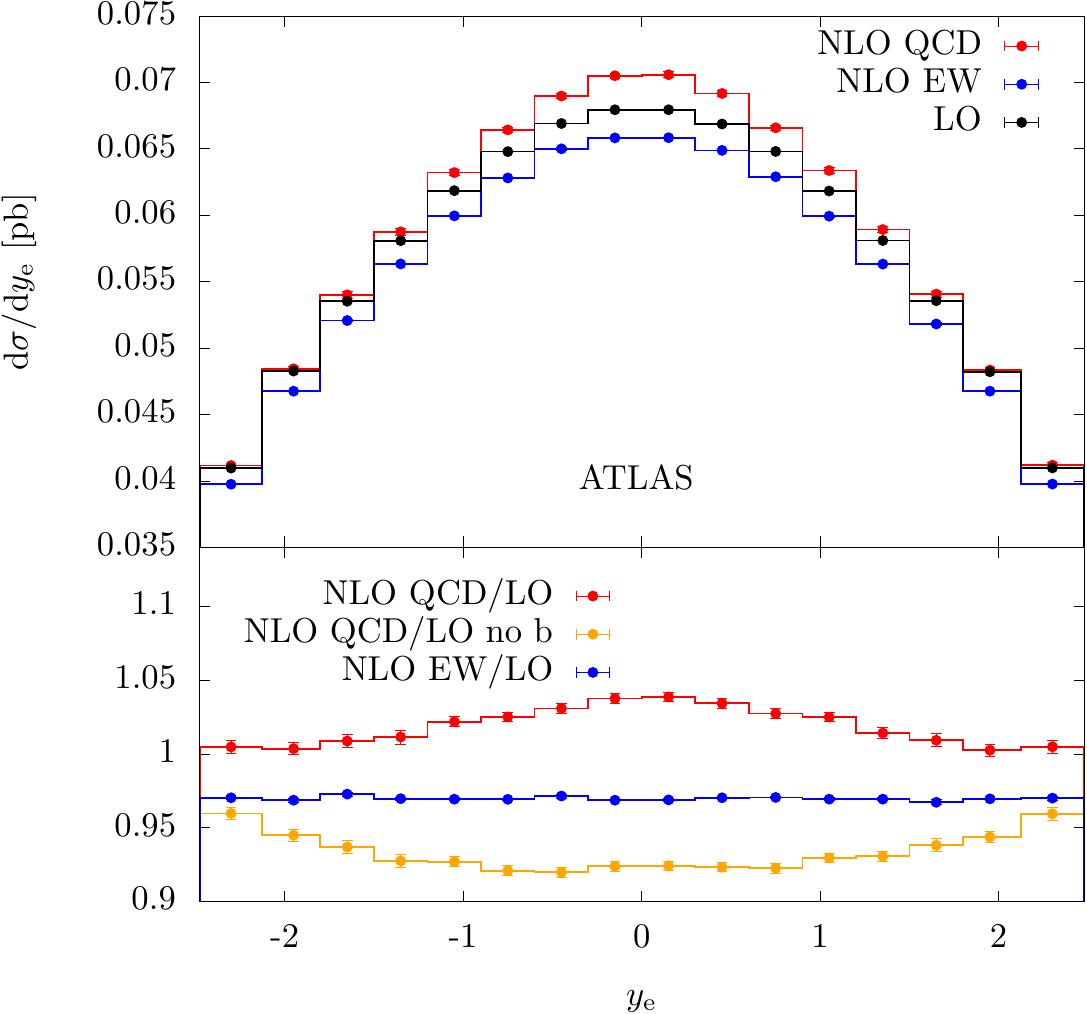}
  \end{minipage}
  \begin{minipage}{0.40\textwidth}
    \includegraphics[width=\textwidth]{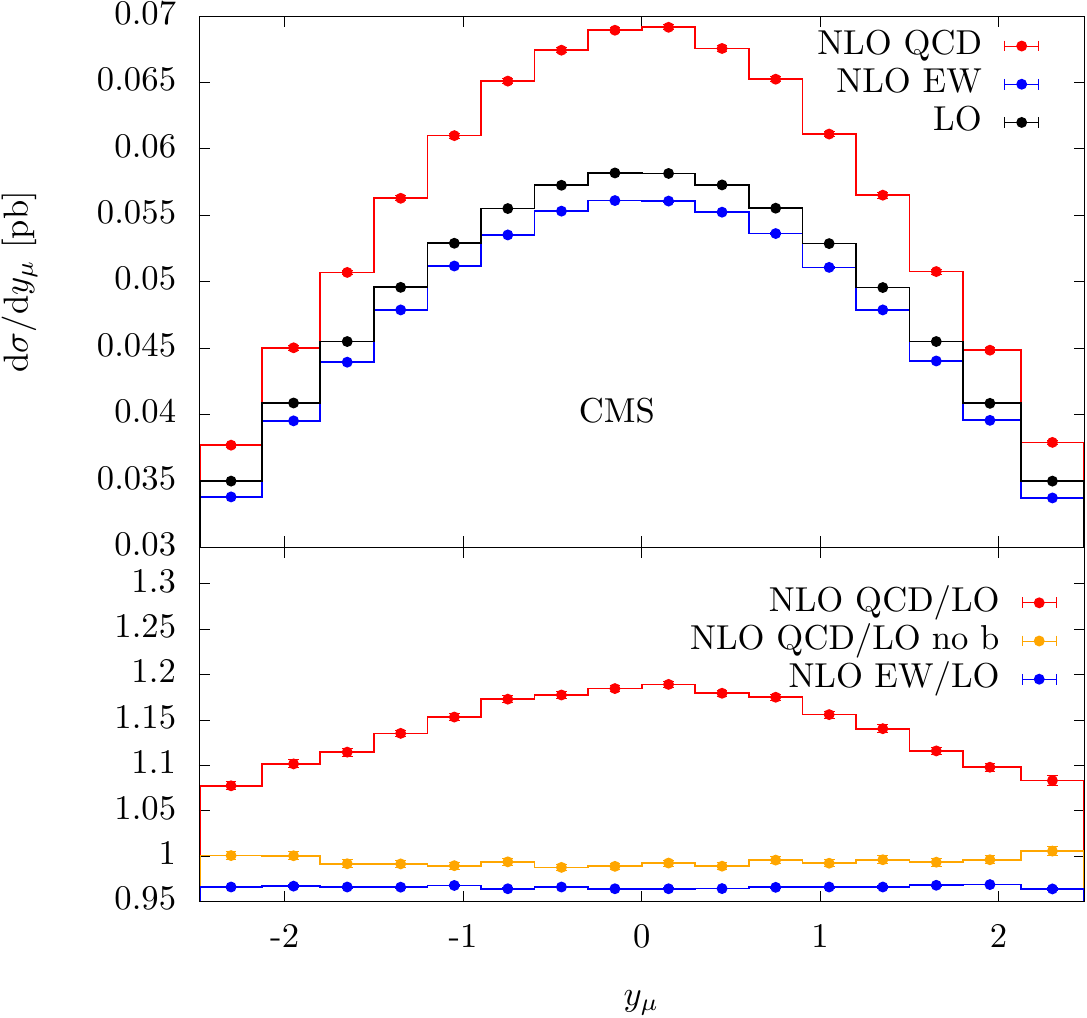}
  \end{minipage}
\end{center}
\caption{Differential distribution in the positron and muon rapidities ($y_{ \rm e }$ and $y_{ \mu }$, respectively) for the process
  $\Pp\Pp \to \Pep \Pne \Pmum \bar{\nu}_{\mu}$ at $\sqrt{s}=13\TeV$
  under the event selections of Eqs.~(\ref{eq:wwatlascut})
  and~(\ref{eq:wwcmscut}). Same notations and conventions as in
  \reffi{fig:wwpte}.  The $\Pg\Pg$ channel gives a flat contribution
  of a few percent and is not shown to improve the plot
  readability.}
\label{fig:wwy}
\efi

The distributions in the transverse momenta of the positron, muon, and
hardest lepton ($p_{\rm T , \Pe}$, $p_{\rm T , \mu}$, and $p_{\rm T
  ,l}^{\mathrm{max}}$, respectively), as well as in the invariant mass
of the charged lepton pair $M^{\rm inv}_{ll}$ are shown in
\reffis{fig:wwpte}--\ref{fig:wwpthardmll}. For these observables the
NLO EW corrections decrease monotonically and become of order $-20\%$
in the tails of the distributions. If the processes with initial-state
$\Pb$~quarks are not considered, the NLO QCD corrections behave
similar as the NLO EW ones and reach $-60\%$ to $-70\%$ for $p_{\rm T}
\simeq 400\GeV$ and $M^{\rm inv}_{ll} \simeq 1.2\TeV$. The
contribution of initial-state $\Pb$~quarks is positive and mainly
concentrated in the region between $100$ and $300\GeV$ in the $p_{\rm
  T}$ distribution ($100$ and $600\GeV$ in the $M^{\rm inv}_{ll}$
distribution).  By looking at \reffis{fig:wwpte} and~\ref{fig:wwptm}
we find a large difference in the contribution of the processes with
initial-state $\Pb$~quarks to the $p_{\rm T , \Pe}$ (and $p_{\rm T ,
  \mu}$) distributions for the ATLAS and CMS setups: we verified that
this effect mainly comes from the difference in the jet veto threshold
in Eqs.~(\ref{eq:wwatlascut}) and~(\ref{eq:wwcmscut}).  In the $p_{\rm
  T}$ and $M^{\rm inv}_{ll}$ range considered in
\reffis{fig:wwpte}--\ref{fig:wwpthardmll}, the impact of the $\Pg\Pg$
channel is basically one order of magnitude smaller than the LO
prediction (except for the first few bins of the $M^{\rm inv}_{ll}$
distribution, where this contribution is of order $20{-}30\%$).

The differential distributions of the charged-lepton-pair transverse
momentum ($p_{{\rm T}, ll}$) and the missing transverse
energy\footnote{In our calculation $E_{\rT}^{\rm miss}$ corresponds to
  the transverse momentum of the two neutrinos.}  ($E_{\rT}^{\rm miss}$)
are shown in \reffi{fig:wwptnptll}.  Since at LO $E_{\rT}^{\rm
  miss}=p_{{\rm T}, ll}$, these two observables are closely related,
and the corresponding NLO corrections are similar. As in the case of
the lepton-$p_{\rm T}$ distributions in
\reffis{fig:wwpte}--\ref{fig:wwpthardmll}, the NLO EW corrections are
negative and their size increases with $p_{{\rm T}, ll}$
($E_{\rT}^{\rm miss}$) reaching the value of $-15\%$ for $p_{{\rm T},
  ll},E_{\rT}^{\rm miss} \simeq 300\GeV$. If the processes with
initial-state $\Pb$~quarks are not included, the NLO QCD corrections
become negative and large (of order $-50\%$) in the tail of the
distributions.  The peak in the NLO QCD corrections around $90\GeV$ is
a consequence of the jet veto in
Eqs.~(\ref{eq:wwatlascut})--(\ref{eq:wwcmscut}): as can be seen in
\reffi{fig:wwqcdcut}, the position of the peak is shifted to larger
$p_{{\rm T}, ll}$ values as the jet-$p_{\rm T}$ threshold is
increased. A similar feature is there for the initial-state
$\Pb$-quark contribution, where the peak is much more pronounced.

Figure~\ref{fig:wwy} shows the differential distributions in the
positron and muon rapidities ($y_{\Pe}$ and $y_{\mu}$, respectively).
Both EW and QCD corrections are basically flat as a function of the
lepton rapidities, while the processes with initial-state $\Pb$~quarks
give a larger contribution in the central region.

\bfi
\begin{center}
  \begin{minipage}{0.40\textwidth}
    \includegraphics[width=\textwidth]{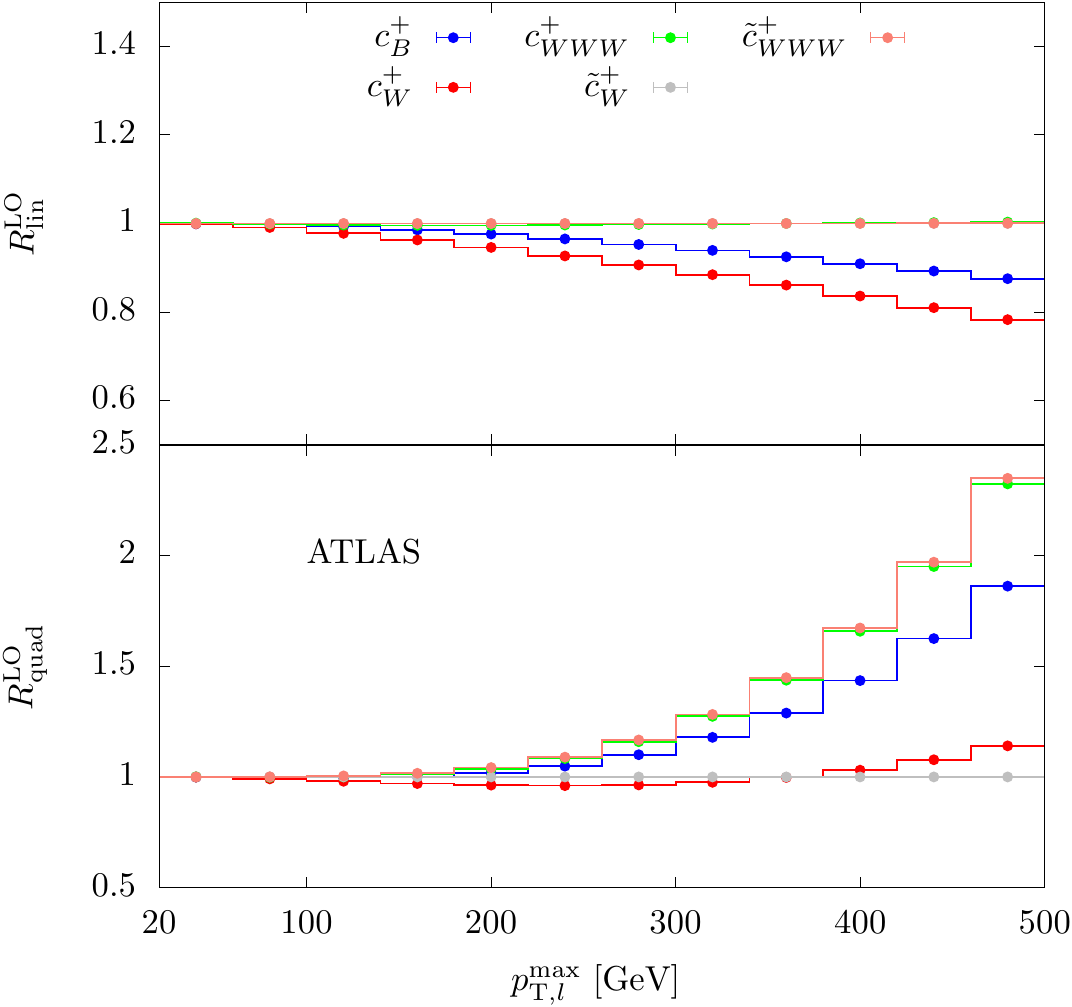}
  \end{minipage}
  \begin{minipage}{0.40\textwidth}
    \includegraphics[width=\textwidth]{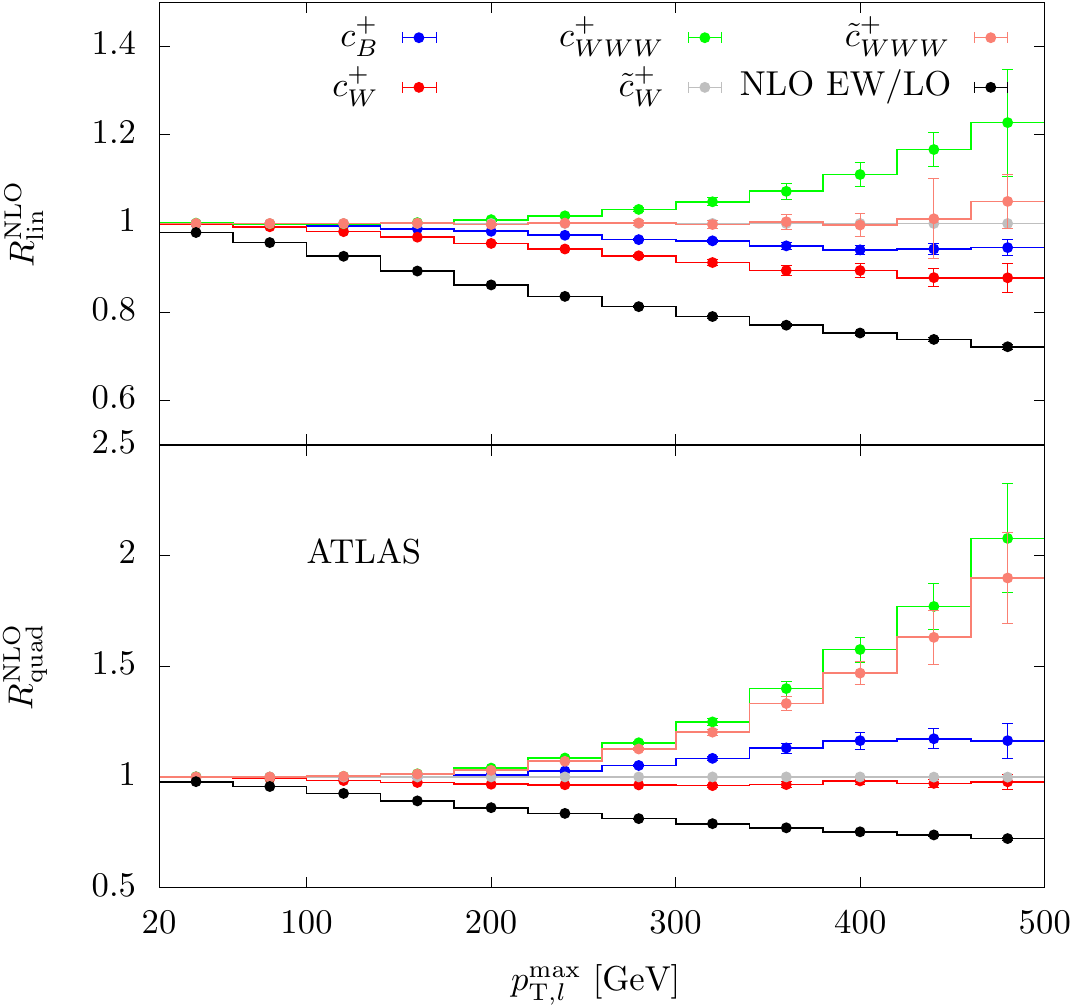}
  \end{minipage}\\[3ex]
  \begin{minipage}{0.40\textwidth}
    \includegraphics[width=\textwidth]{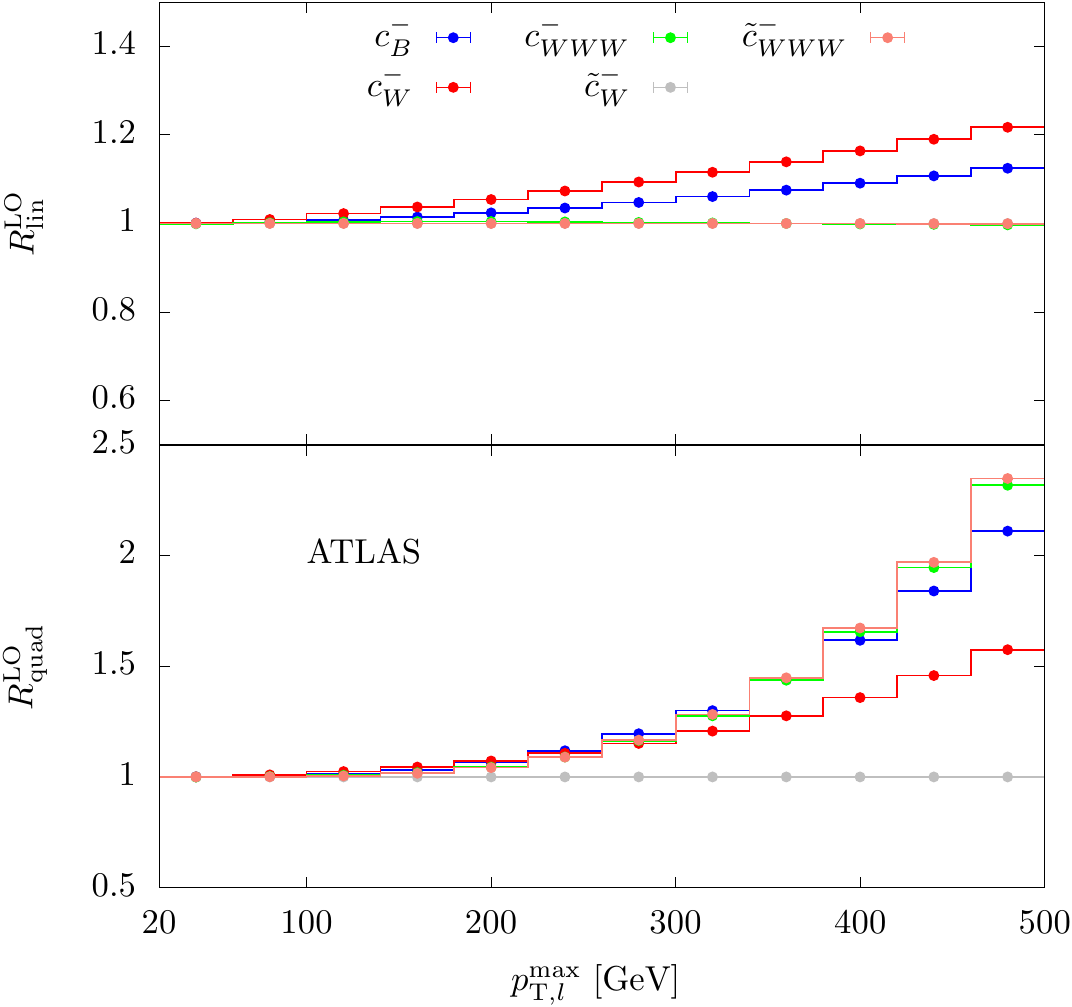}
  \end{minipage}
  \begin{minipage}{0.40\textwidth}
    \includegraphics[width=\textwidth]{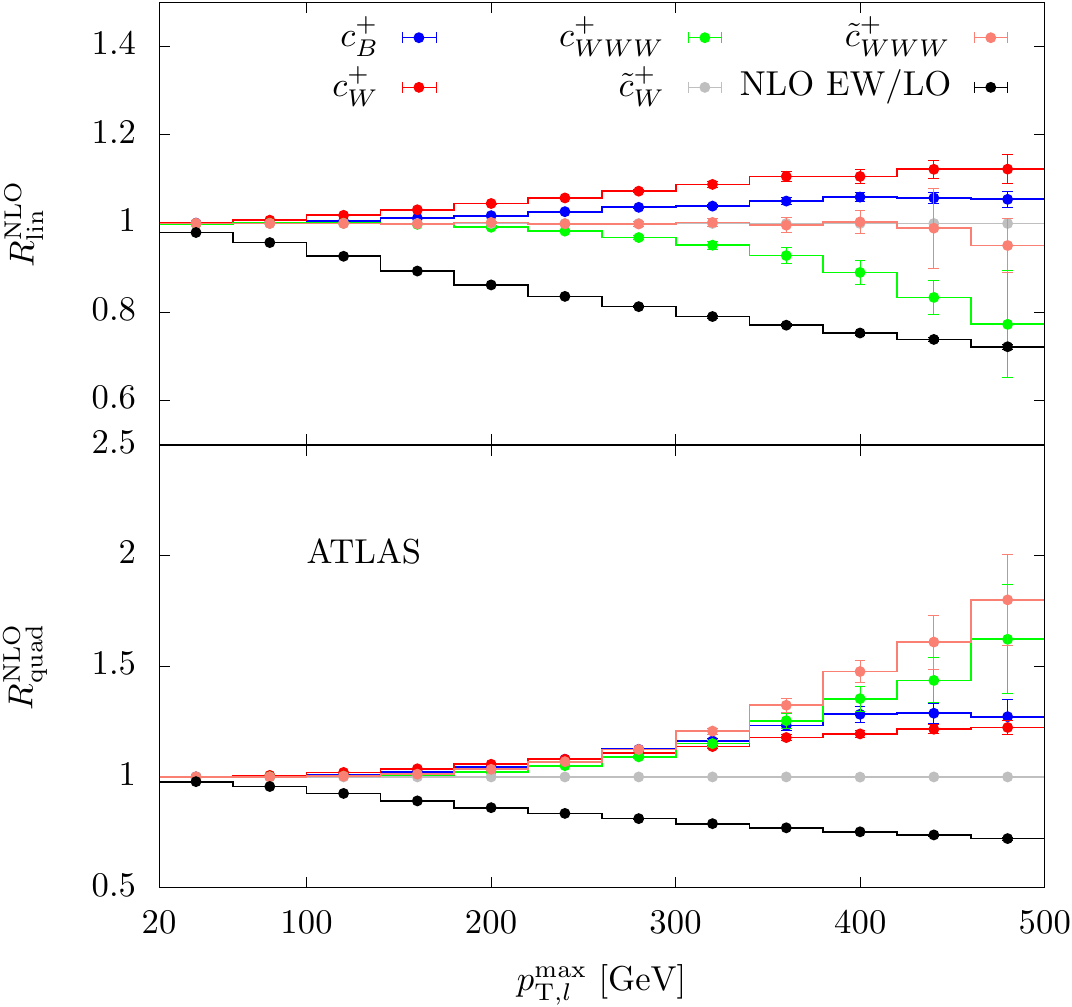}
  \end{minipage}
\end{center}
\caption{Ratio $R^{\rm LO(NLO)}_{\rm lin(quad)}$ as a function of the 
  transverse momentum of the hardest lepton for the process $\Pp\Pp
  \to \Pep \Pne \Pmum \bar{\nu}_{\mu}$ in the ATLAS setup of
  Eq.~(\ref{eq:wwatlascut}).  The ratios between the theoretical
  predictions including both the aTGCs and the SM results at LO (left
  plots) are compared to the ratios at NLO QCD accuracy (right
  plots). In each plot the upper panel corresponds to the ratio where
  the non-SM contributions are included only up to the interference
  terms $\sigma^{\rm LO(NLO)}_{{\rm SM}\times {\rm EFT6}}$, while in
  the lower panel both the $\sigma^{\rm LO(NLO)}_{{\rm SM}\times {\rm
      EFT6}}$ and the $\sigma^{\rm LO(NLO)}_{{\rm EFT6}^2}$ terms are
  considered.  For all curves the central value of the factorization
  and renormalization scales is used and the error bars correspond to
  the statistical integration uncertainties.}
\label{fig:wwtgcpth}
\efi
\bfi
\begin{center}
  \begin{minipage}{0.40\textwidth}
    \includegraphics[width=\textwidth]{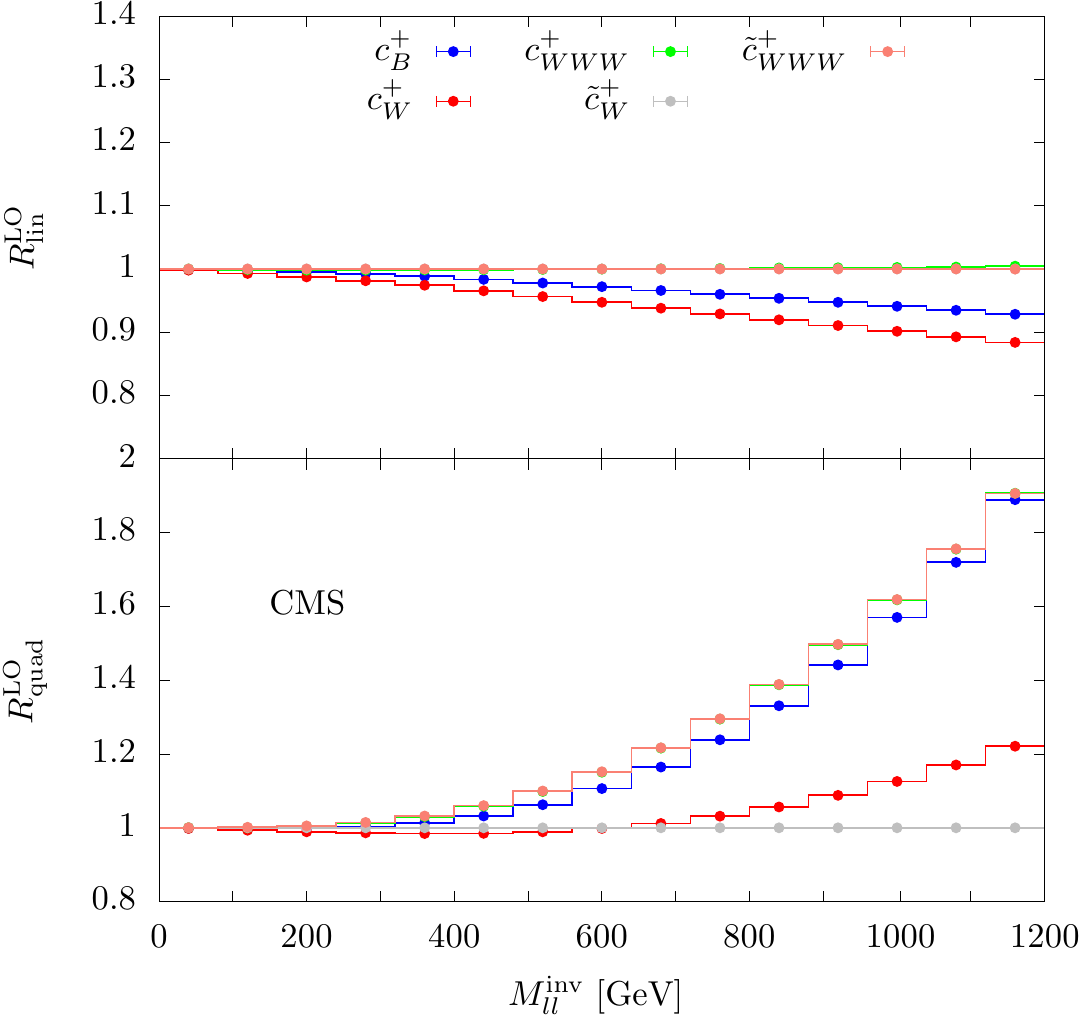}
  \end{minipage}
  \begin{minipage}{0.40\textwidth}
    \includegraphics[width=\textwidth]{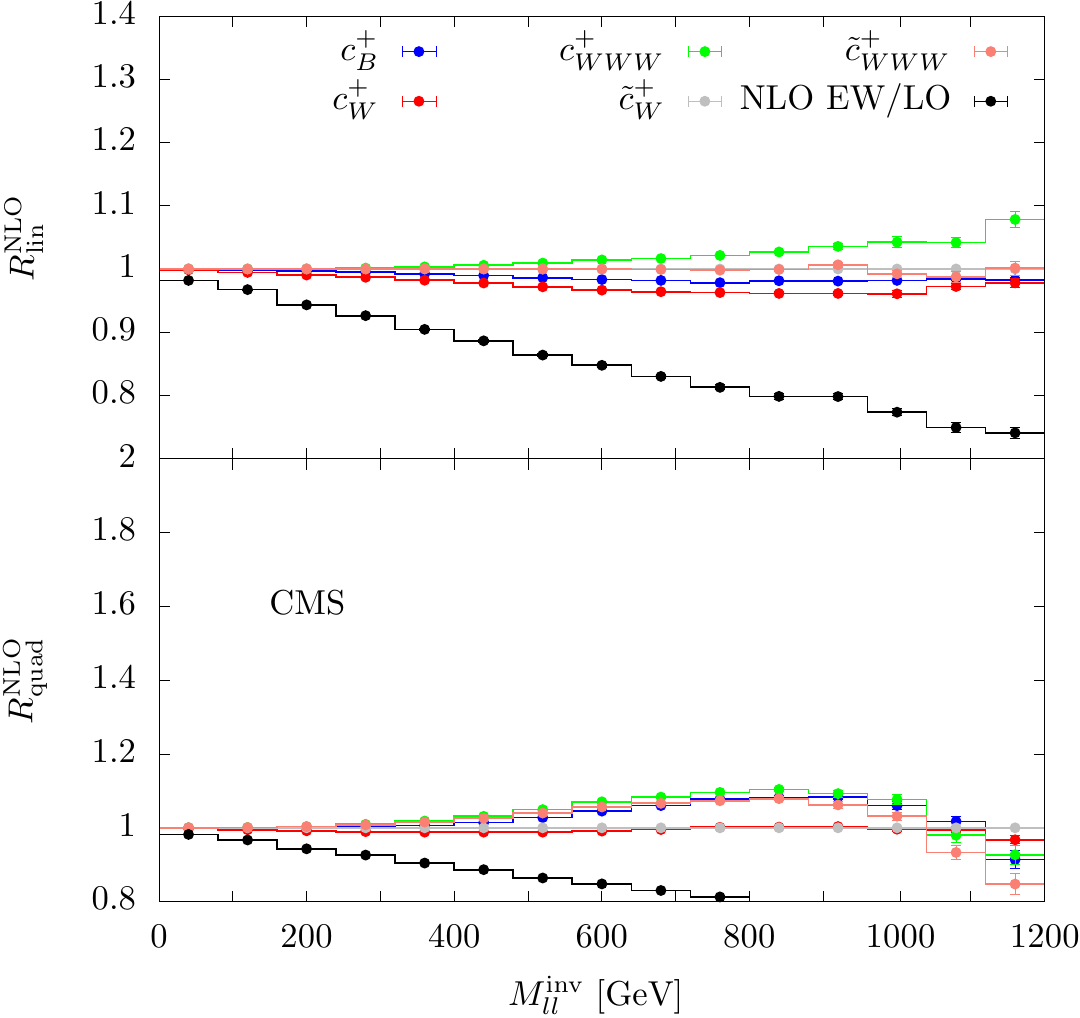}
  \end{minipage}\\[3ex]
  \begin{minipage}{0.40\textwidth}
    \includegraphics[width=\textwidth]{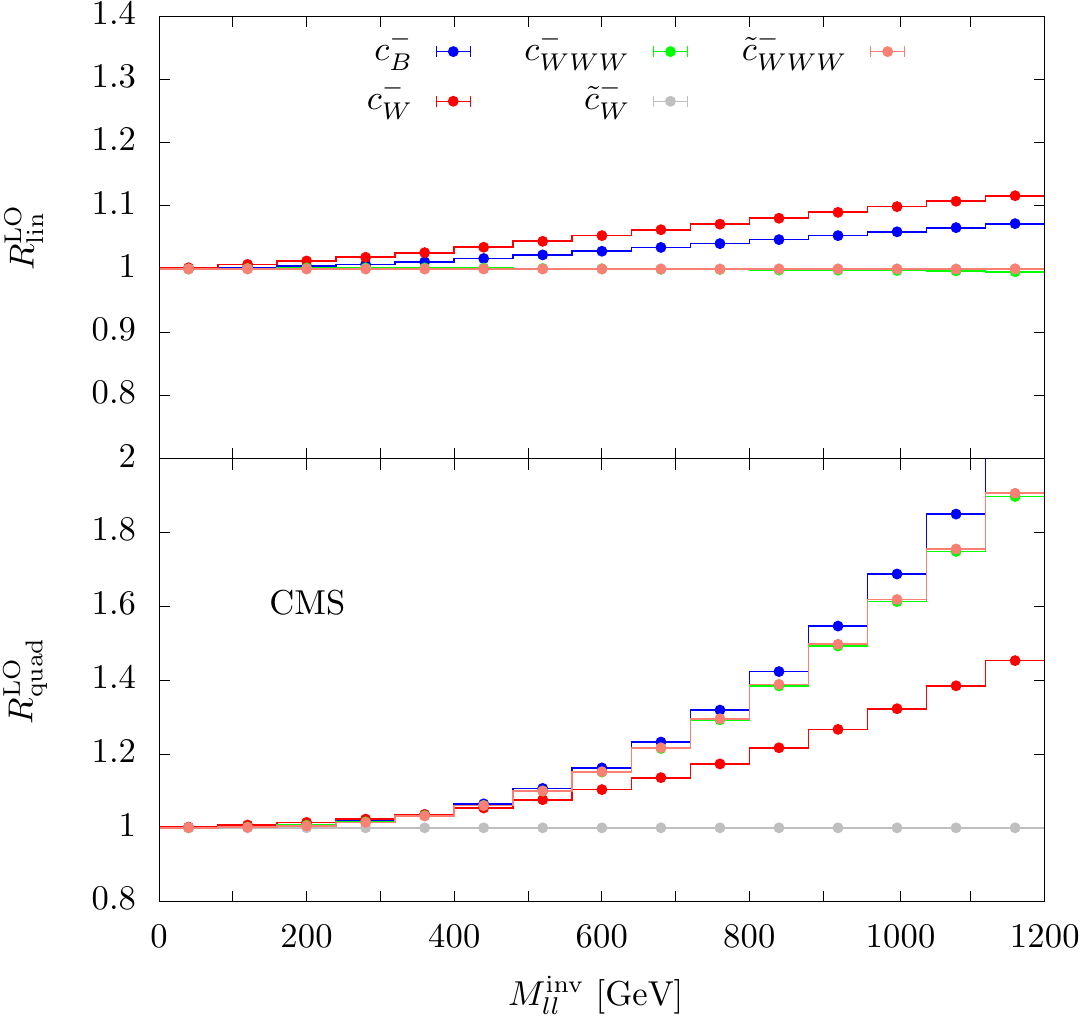}
  \end{minipage}
  \begin{minipage}{0.40\textwidth}
    \includegraphics[width=\textwidth]{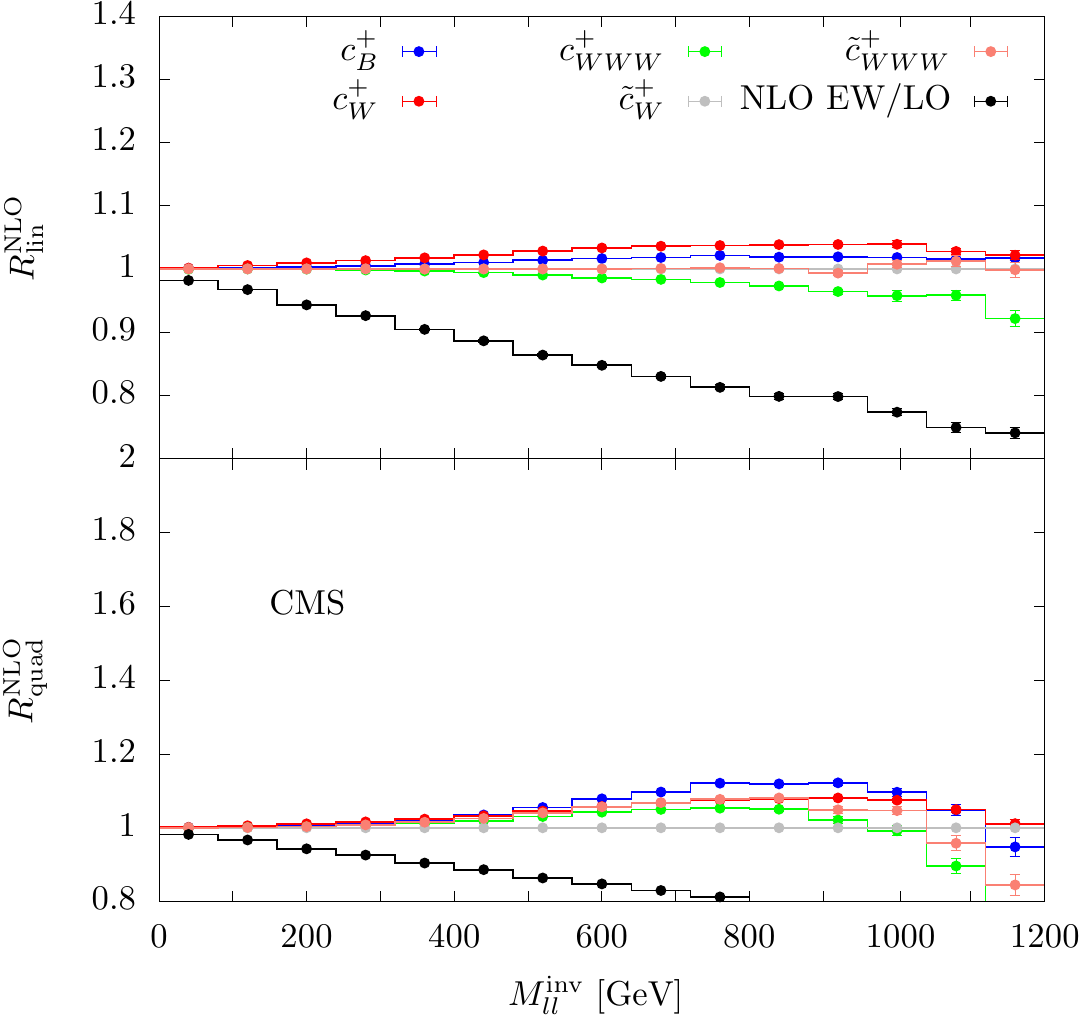}
  \end{minipage}
\end{center}
\caption{Ratio $R^{\rm LO(NLO)}_{\rm lin(quad)}$ as a function of the 
  invariant mass of the charged-lepton pair for the process $\Pp\Pp
  \to \Pep \Pne \Pmum \bar{\nu}_{\mu}$ in the CMS setup of
  Eq.~(\ref{eq:wwcmscut}). Same notation and conventions as in
  \reffi{fig:wwtgcpth}.}
\label{fig:wwtgcmll}
\efi
Concerning the impact of the anomalous triple-gauge-boson interactions, we consider the ratios
\begin{equation}
  \begin{aligned}
  R^{\rm LO (NLO)}_{\rm lin } &=\frac{{\rm d}  \Big( \sigma_{{\rm SM}^2} + \sigma_{{\rm SM}\times {\rm EFT}} \Big)^{{\rm LO(NLO)}\,{\rm QCD}}/{\rm d}X}
  { {\rm d}  \sigma_{{\rm SM}^2}^{{\rm LO(NLO)}\,{\rm QCD}} /{\rm d} X }, \\
  R^{\rm LO (NLO)}_{\rm quad}&=\frac{{\rm d}  \Big( \sigma_{{\rm SM}^2} + \sigma_{{\rm SM}\times {\rm EFT}} + \sigma_{{\rm EFT}^2}\Big)^{{\rm LO(NLO)}\,{\rm QCD}}/{\rm d}X}
  { {\rm d}  \sigma_{{\rm SM}^2}^{{\rm LO(NLO)}\,{\rm QCD}} /{\rm d} X },
  \end{aligned}
  \label{eq:defratio}
\end{equation}
(with $X=p_{{\rm T}, l}^{\rm max}\, , M^{\rm inv}_{ll}$) at LO and NLO
QCD accuracy. In the $R^{\rm NLO}_{\rm lin \, (quad)}$ ratios the NLO
QCD corrections to the diagrams involving the aTGCs are included as
well. For the Wilson coefficients we use the values listed in
Eq.~(\ref{eq:wwwilsoncoeffs}). In
\reffis{fig:wwtgcpth}--\ref{fig:wwtgcmll} only one Wilson coefficient
is different from zero for each curve.  Since the $t$-channel
single-top contribution is subtracted in the experimental searches for
aTGCs, in \reffis{fig:wwtgcpth}--\ref{fig:wwtgcmll} the contribution
of the processes with initial-state $\Pb$~quarks is not included.
Without these processes the NLO QCD corrections become of order $-60\%$
and $-70\%$ for $p_{{\rm T}, l}^{\rm max} \simeq 500\GeV$ and
$M^{\rm inv}_{ll} \simeq 1.2\TeV$: we thus limit our analysis to the $p_{{\rm
    T}, l}^{\rm max}$ and $M^{\rm inv}_{ll}$ values below $500\GeV$
and $1.2\TeV$, respectively, since for larger transverse momenta or
invariant masses  the differential distributions
are strongly suppressed by the NLO QCD corrections.

The impact of aTGCs on the distributions in the transverse momentum of
the leading lepton ($p_{{\rm T}, l}^{\rm max}$) and the invariant mass
of the charged lepton pair ($M^{\rm inv}_{ll}$) at LO is shown in the
left plots of \reffis{fig:wwtgcpth}--\ref{fig:wwtgcmll}.  Comparing
the predictions for $R^{\rm LO}_{\rm lin}$ and $R^{\rm LO}_{\rm quad}$
reveals that the $\sigma^{\rm LO}_{{\rm EFT6}^2}$ terms give the
largest contribution in the high $p_{\rm T}$ and/or invariant-mass
regions. Since the $\sigma^{\rm LO}_{{\rm EFT6}^2}$ terms are
positive, the results for $R^{\rm LO}_{\rm quad}$ using the two sets
$c^{-}$ and $c^{+}$ of Wilson coefficients in
Eq.~(\ref{eq:wwwilsoncoeffs}) are very similar. With the numerical
values in Eq.~(\ref{eq:wwwilsoncoeffs}), the largest deviation from
the SM predictions come from the $c_{B}$ and $c_{W}$ coefficients as
far as only the $\sigma^{\rm LO}_{{\rm SM}\times {\rm EFT6}}$
interferences are considered, while also the $c_{WWW}$ and
$\tilde{c}_{WWW}$ coefficients give a sizable contribution when the
$\sigma^{\rm LO}_{{\rm EFT6}^2}$ terms are included (the contributions
of the $c_{WWW}$ and $\tilde{c}_{WWW}$ coefficients to $R^{\rm LO}_{\rm lin}$
and $R^{\rm LO}_{\rm quad}$ are very similar and the corresponding curves
basically overlap in \reffis{fig:wwtgcpth}--\ref{fig:wwtgcmll}).

The results for $R^{\rm NLO}_{\rm lin }$ and $R^{\rm NLO}_{\rm quad}$
are shown in the right plots of \reffis{fig:wwtgcpth}--\ref{fig:wwtgcmll}.  
\bfi
\begin{center}
  \begin{minipage}{0.40\textwidth}
    \includegraphics[width=\textwidth]{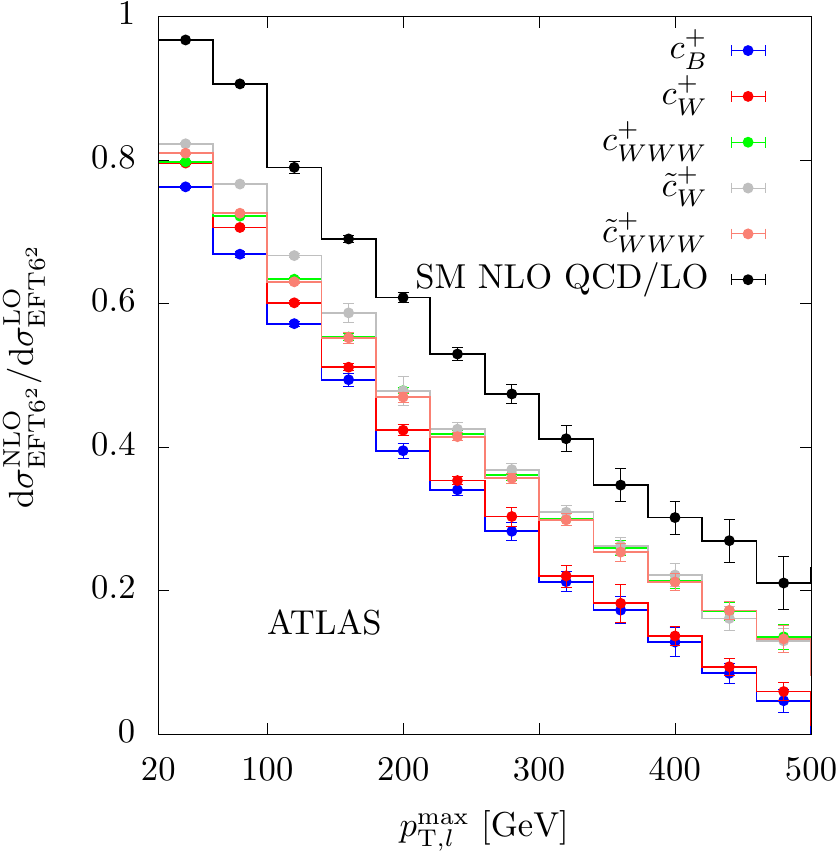}
  \end{minipage}
  \begin{minipage}{0.40\textwidth}
    \includegraphics[width=\textwidth]{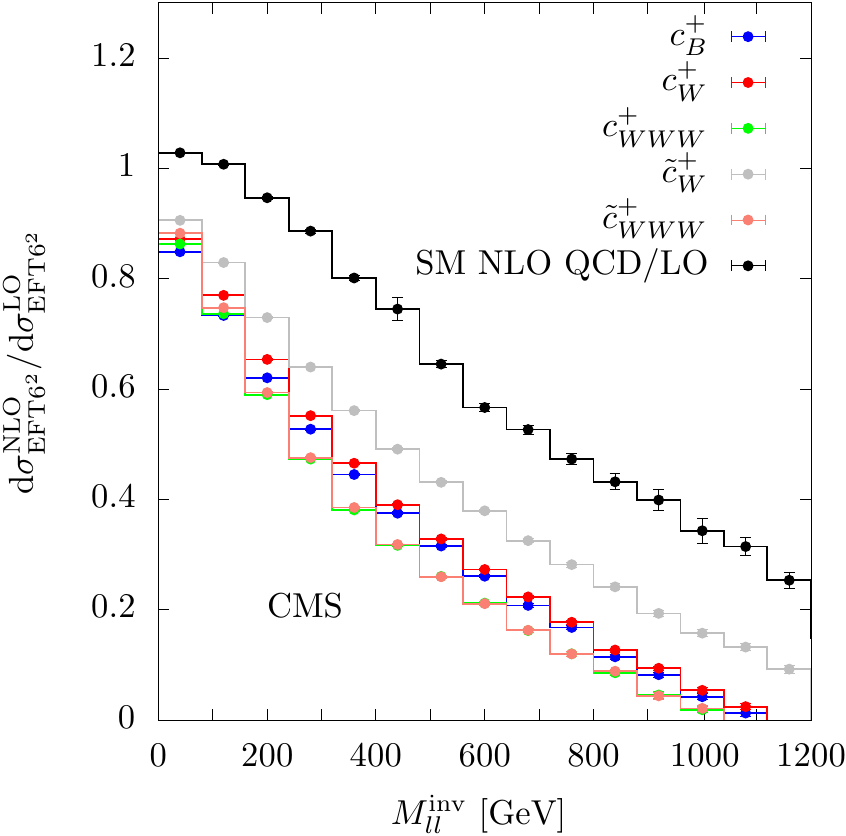}
  \end{minipage}
\end{center}
\caption{Ratio between the ${\rm EFT6}^2$ contribution to the process
  $\Pp\Pp \to \Pep \Pne \Pmum \bar{\nu}_{\mu}$ computed at NLO QCD
  ($\sigma_{{\rm EFT6}^2}^{\rm NLO}$) and LO ($\sigma_{{\rm EFT6}^2}^{\rm LO}$)
  accuracy  as a function of the hardest lepton $p_{\rm T}$ (left plot)
  and as a function of the charged-lepton-pair  invariant mass.
  The labels ATLAS and CMS refer to the event selections of
  Eq.~(\ref{eq:wwatlascut}) and~(\ref{eq:wwcmscut}), respectively.
  The ratio between the SM predictions at NLO QCD and at LO accuracy
  is also shown (black lines).}
\label{fig:wwqcdlam2}
\efi
On one hand $R^{\rm NLO}_{\rm lin }$ behaves in a similar way to
$R^{\rm LO}_{\rm lin }$, on the other hand the impact of the aTGCs is
in general smaller at NLO QCD in particular at high $p_{\rm T}$ and/or
invariant masses (with the only exception of the $c_{WWW}$ coefficient
that contributes more at NLO QCD\footnote{For on-shell vector
    bosons, it was pointed out in \citere{Azatov:2017kzw} that the
    interference of the ${\cal O}_{WWW}$ operator with the SM
    amplitude is suppressed at LO but not at NLO QCD.}). The situation
  changes for $R^{\rm NLO}_{\rm quad}$, where the sensitivity to the
  aTGCs is strongly reduced with respect to the LO in particular in
  the tails of
the distributions.%
\footnote{This has already been observed in \citere{Baur:1995uv}.} 
At LO the leading contribution to $R^{\rm LO}_{\rm
  quad}$ is given by the positive and growing $\sigma^{\rm LO}_{{\rm
    EFT6}^2}$ terms.  At NLO we have
\begin{equation}
  R^{\rm NLO}_{\rm quad}=R^{\rm NLO}_{\rm lin } + \frac{{\rm d} \sigma_{{\rm EFT6}^2}^{\rm  NLO}}{{\rm d} \sigma_{{\rm SM}^2}^{\rm  NLO}},
  \label{eq:quadratio}
\end{equation}
where
\begin{equation}
  \frac{{\rm d} \sigma_{{\rm EFT6}^2}^{\rm  NLO}}{{\rm d} \sigma_{{\rm SM}^2}^{\rm  NLO}} =
  \frac{{\rm d} \sigma_{{\rm EFT6}^2}^{\rm  LO}}{{\rm d} \sigma_{{\rm SM}^2}^{\rm  LO}} \frac{ \delta^{\rm QCD}_{{\rm EFT6}^2} }{ \delta^{\rm QCD}_{\rm SM} },
  \quad {\rm with } \quad \delta_{{\rm EFT6}^2}^{\rm QCD}=\frac{{\rm d}\sigma^{\rm NLO}_{{\rm EFT6}^2}}{{\rm d}\sigma^{\rm LO}_{{\rm EFT6}^2}},
 \quad \delta_{\rm SM}^{\rm QCD}=\frac{{\rm d}\sigma^{\rm NLO}_{\rm SM}}{{\rm d}\sigma^{\rm LO}_{\rm SM}}.
  \label{eq:quadratio2}
\end{equation}
The NLO QCD corrections suppress the ${\rm EFT6}^2$ terms much
stronger than the SM contributions as shown in \reffi{fig:wwqcdlam2}
for the observables under consideration. Figure~\ref{fig:wwqcdlam2}
also reveals why the ${\rm EFT6}^2$ contribution is more
suppressed for the charged-lepton invariant mass rather than for the
hardest lepton $p_{\rm T}$.

Figures~\ref{fig:wwtgcpth}--\ref{fig:wwtgcmll} also show the relative
EW NLO corrections determined from the ratio between the NLO EW
results and the LO results in the SM. While the introduction of a
jet veto is useful to preserve the sensitivity to the aTGCs, it leads
to large and negative NLO QCD corrections if the jet veto threshold is
small. As a result the effect of the NLO EW corrections is emphasized
and can become larger than the one of the aTGCs.

\subsection{WZ production}
\label{subsect:wzres}

\begin{table}
  \begin{center}
\renewcommand{\arraystretch}{1.4}
    \begin{tabular}{|l|l|l|l|l}
      \hline
       Setup & LO [fb] & NLO QCD [fb] & NLO EW [fb] \\
      \hline
      $\PW^-\PZ$ ATLAS & $12.6455(9)^{+5.5\%}_{-6.8\%}$ & $23.780(4)^{+5.5\%}_{-4.6\%}$ & $11.891(4)^{+5.6\%}_{-6.9\%}$ \\
      \hline
      $\PW^-\PZ$   CMS & $~9.3251(8)^{+5.3\%}_{-6.7\%}$ & $17.215(4)^{+5.4\%}_{-4.3\%}$ & $~8.870(2)^{+5.5\%}_{-6.7\%}$ \\
      \hline
      $\PW^+\PZ$ ATLAS & $18.875(1)^{+5.2\%}_{-6.4\%}$ & $34.253(6)^{+5.3\%}_{-4.3\%}$ & $17.748(8)^{+5.3\%}_{-6.5\%}$ \\
      \hline
      $\PW^+\PZ$   CMS & $14.307(1)^{+5.0\%}_{-6.2\%}$ & $26.357(6)^{+5.4\%}_{-4.3\%}$ & $13.600(4)^{+5.1\%}_{-6.3\%}$ \\
      \hline
    \end{tabular}
  \end{center} 
  \caption{Integrated cross section for $\PW\PZ$ production at $\sqrt{s}=13\TeV$ in the ATLAS and CMS setups
    of Eqs.~(\ref{eq:wzatlascut}) and~(\ref{eq:wzcmscut}), respectively. In the first column $\PW^+\PZ$ ($\PW^-\PZ$)
    is a short-hand notation for the process $\Pp\Pp \to \Pep \nu_{\rm e} \mu^+ \mu^-$
    ($\Pp\Pp \to \Pem \overline{\nu}_{\rm e} \mu^+ \mu^-$). The numbers in parentheses correspond to the statistical
    error on the last digit. The uncertainties are estimated from the scale dependence, as explained in the text.}
  \label{tab:wz-xsec}
\end{table}

The results for the integrated cross sections for WZ production at a
centre-of-mass energy of $13\TeV$ are presented in \refta{tab:wz-xsec}
for the ATLAS and CMS setups of Eqs.~(\ref{eq:wzatlascut})
and~(\ref{eq:wzcmscut}), respectively. In \refta{tab:wz-xsec} and
\reffis{fig:wzptem}--\ref{fig:wztgcmt},$\PW^+\PZ$ ($\PW^-\PZ$) is a
short-hand notation for the process $\Pp\Pp \to \Pep \nu_{\rm e} \mu^+
\mu^-$ ($\Pp\Pp \to \Pem \overline{\nu}_{\rm e} \mu^+ \mu^-$). The LO
predictions are compared to the ones at NLO QCD and NLO EW accuracy.
The numbers in parentheses correspond to the statistical integration
error, while the upper and lower values of the cross sections
correspond to the upper and lower limits from scale
variations~(\ref{eq:scalevar}).  Scale uncertainties are of the same
order at LO and NLO EW and do not decrease significantly at NLO QCD as
a consequence of the large QCD corrections.

The cross sections for the $\PW^+\PZ$ channel are about $50\%$ larger
than the ones for the $\PW^-\PZ$ channel: this can be attributed to
the parton flux within the proton which is larger for the up quark
than for the down quark. The NLO EW corrections are of order $-6\%$
and $-5\%$ in the ATLAS and CMS setups, respectively. The NLO QCD
corrections are positive and reach the value of $+80\%$ and $+90\%$,
depending on the setup. This is due to the fact that diboson
production at LO only proceeds via quark--antiquark annihilation, while
at NLO QCD new channels appear that involve initial-state gluons
(namely $\Pg q \to \PZ \PW^{\pm} q'$ and $\Pg \overline{q} \to \PZ
\PW^{\pm} \overline{q}'$) and are enhanced because of the gluon
luminosity. In principle, the same happens also for $\PW\PW$
production. However, the jet veto in the event selections
~(\ref{eq:wwatlascut}) and~(\ref{eq:wwcmscut}) strongly suppresses the
real QCD corrections and in particular the contributions of the
processes with initial-state gluons: this explains the different
behaviour of NLO QCD corrections for $\PW\PW$ and $\PW\PZ$ shown in
\reftas{tab:ww-xsec} and~\ref{tab:wz-xsec}.

\bfi
\begin{center}
  \begin{minipage}{0.40\textwidth}
    \includegraphics[width=\textwidth]{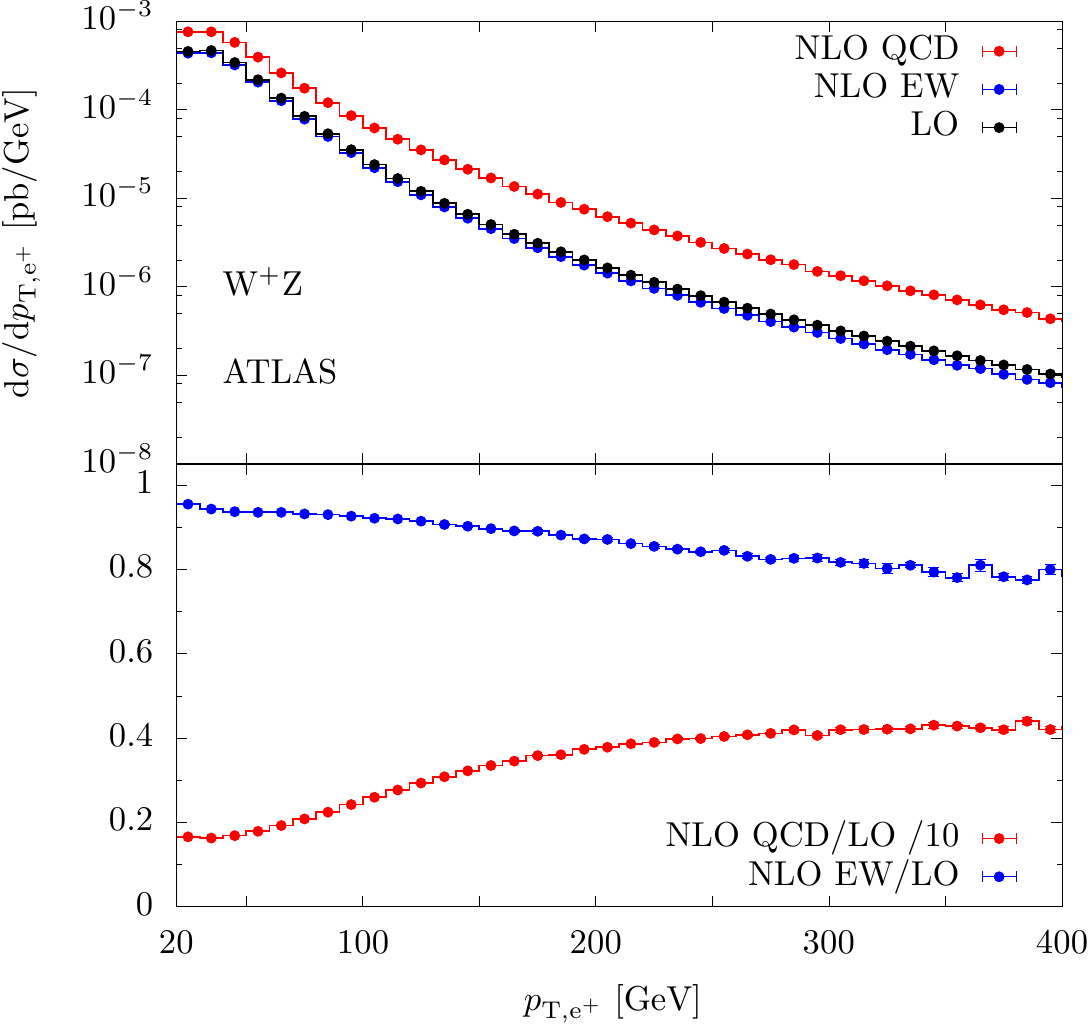}
  \end{minipage}
  \begin{minipage}{0.40\textwidth}
    \includegraphics[width=\textwidth]{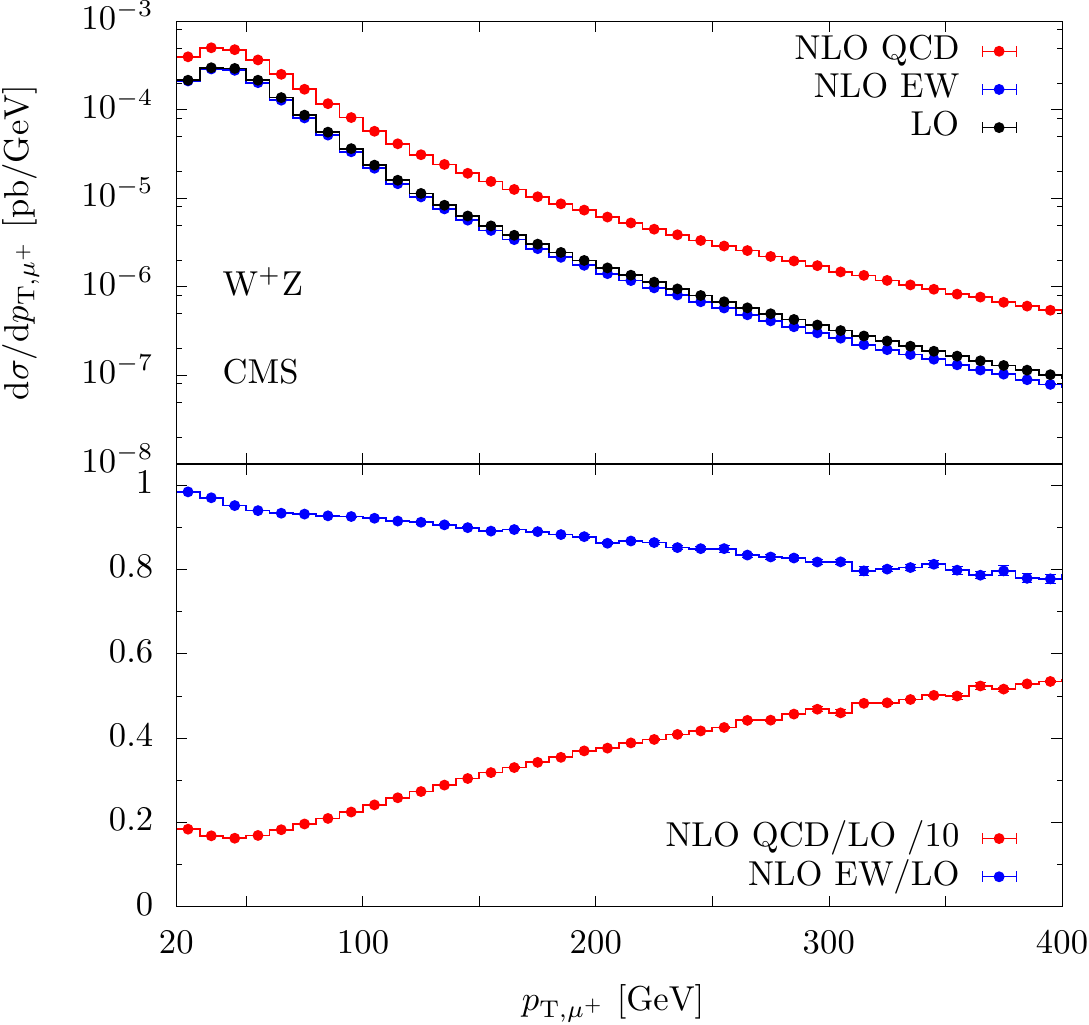}
  \end{minipage}
\end{center}
\caption{Upper panels: differential distributions in the transverse
  momentum of the positron ($p_{{\rm T},\Pe^+}$) and the antimuon
  ($p_{{\rm T},\mu^+}$) for the process $\Pp\Pp \to \Pep \nu_{\rm e}
  \mu^+ \mu^-$ at $\sqrt{s}=13\TeV$ for the ATLAS and CMS event
  selections of Eqs.~(\ref{eq:wzatlascut}) and~(\ref{eq:wzcmscut}),
  respectively. The LO results (black lines) are compared to the ones
  at NLO QCD (red lines) and NLO EW (blue lines). Lower panels: ratio
  of the NLO QCD, and NLO EW contributions with respect to the LO (red
  and blue lines, respectively). In order to improve the plot
  readability the NLO QCD predictions have been divided by a factor 10
  in the ratio ${\rm NLO\,QCD}/{\rm LO}$.  For all curves the central
  value of the factorization and renormalization scales is used and
  the error bars correspond to the statistical integration
  uncertainties. Note that the same PDF set is employed for both the
  LO and NLO predictions.}
\label{fig:wzptem}
\efi
\bfi
\begin{center}
  \begin{minipage}{0.40\textwidth}
    \includegraphics[width=\textwidth]{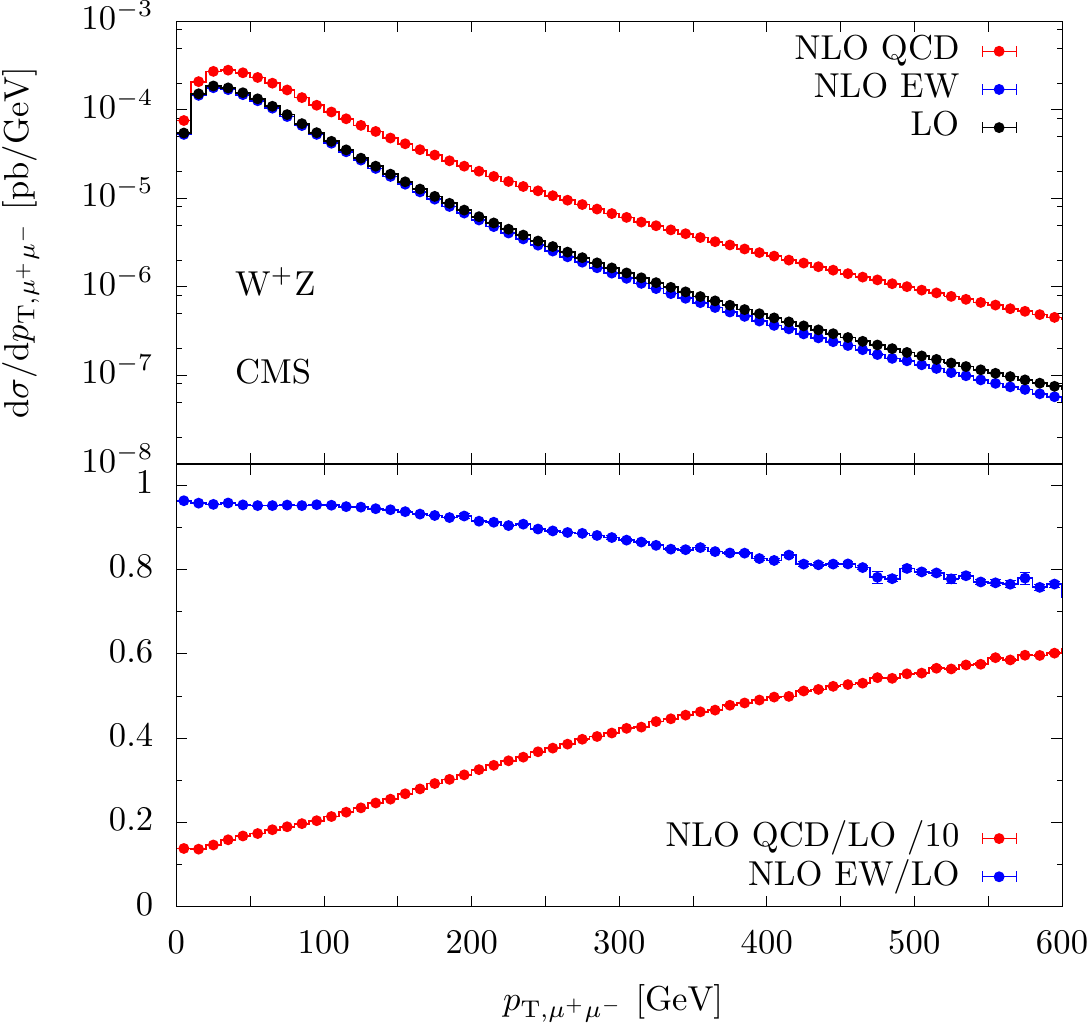}
  \end{minipage}
  \begin{minipage}{0.40\textwidth}
    \includegraphics[width=\textwidth]{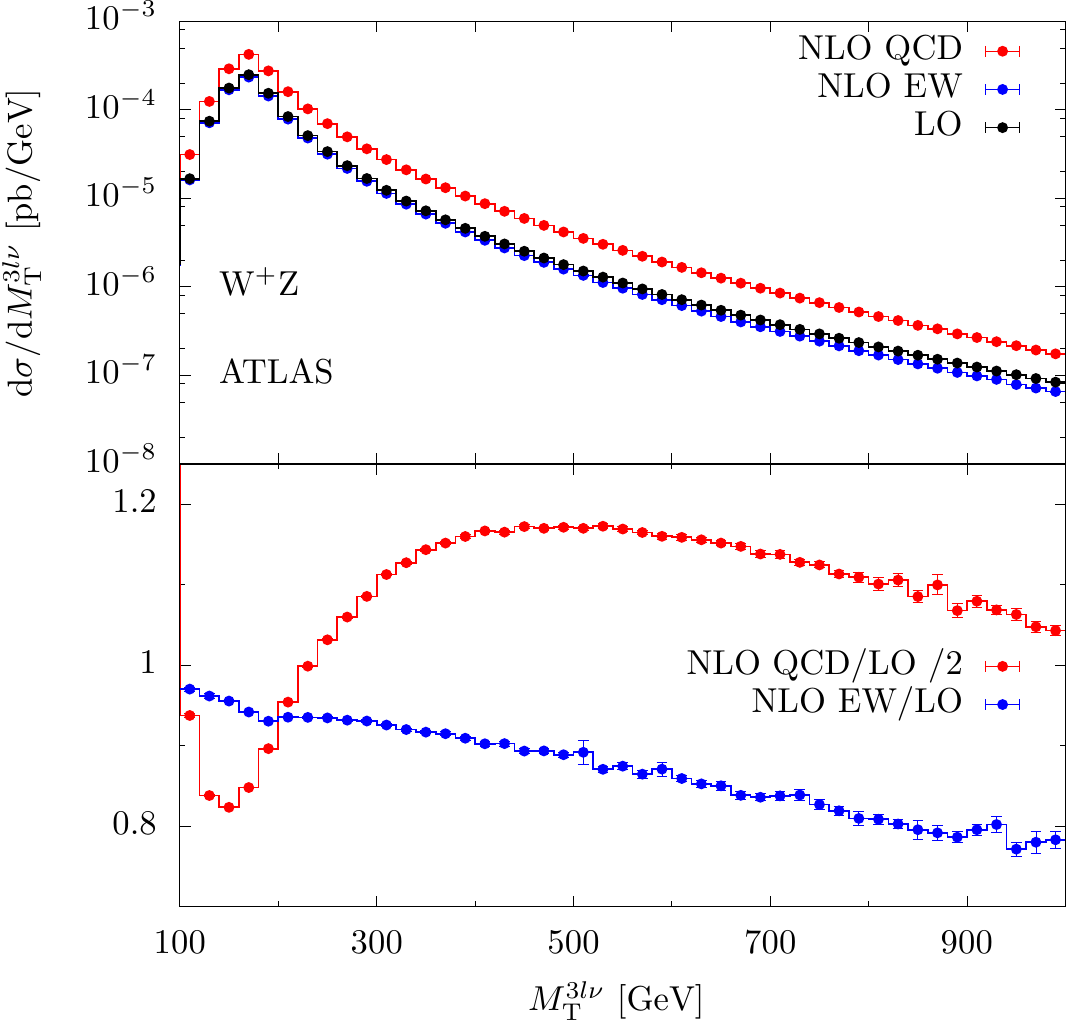}
  \end{minipage}
\end{center}
\caption{Differential distribution in the transverse momentum of the muon--antimuon
  pair ($p_{ {\rm T,}\mu^+ \mu^-}$) and in the $\PW\PZ$ transverse
  mass ($M_{\rm T}^{3l\nu}$) for the process $\Pp\Pp \to \Pep \nu_{\rm
    e} \mu^+ \mu^-$ at $\sqrt{s}=13\TeV$ under the event selections of
  Eqs.~(\ref{eq:wzcmscut}) and~(\ref{eq:wzatlascut}). The NLO QCD
  predictions have been divided by a factor 10 (2) in the ratio ${\rm
    NLO\,QCD}/{\rm LO}$ as a function of $p_{ {\rm T,}\mu^+ \mu^-}$
  ($M_{\rm T}^{3l\nu}$).  Same notations and conventions as in
  \reffi{fig:wzptem}.}
\label{fig:wzmtptz}
\efi
\bfi
\begin{center}
  \begin{minipage}{0.40\textwidth}
    \includegraphics[width=\textwidth]{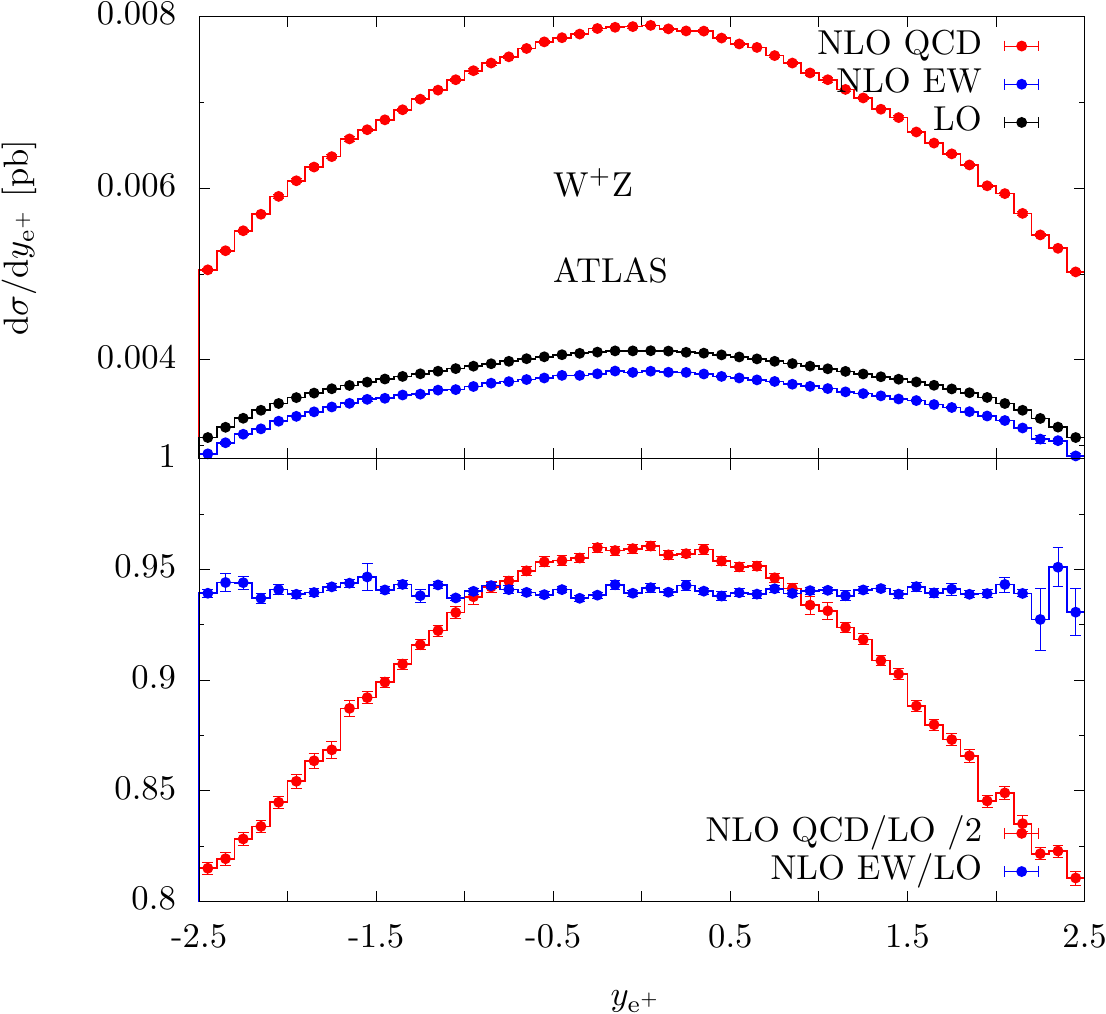}
  \end{minipage}
  \begin{minipage}{0.40\textwidth}
    \includegraphics[width=\textwidth]{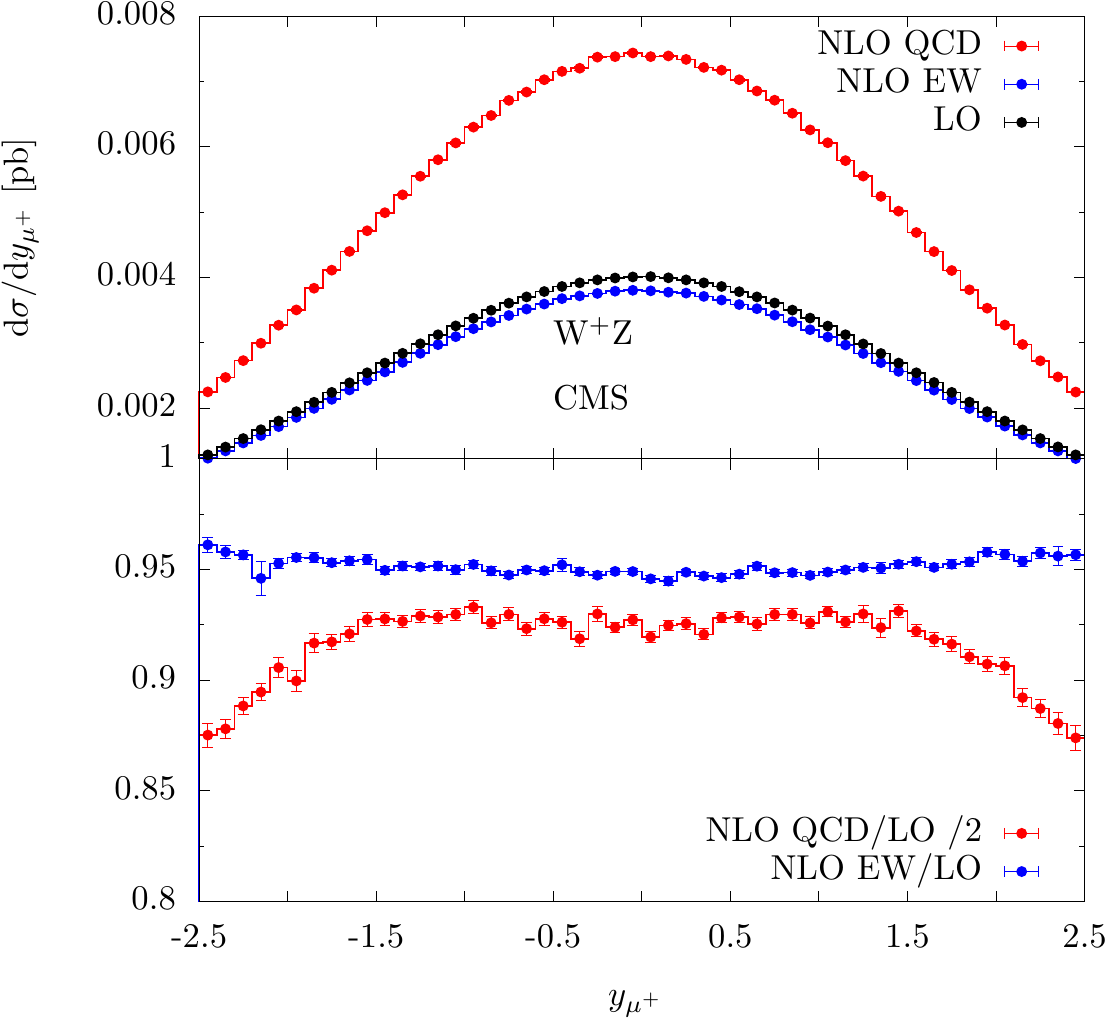}
  \end{minipage}
\end{center}
\caption{Differential distribution in the positron and antimuon rapidities
  ($y_{ {\rm e}^+ }$ and $y_{ \mu^+ }$, respectively) for the process
  $\Pp\Pp \to \Pep \nu_{\rm e} \mu^+ \mu^-$ at $\sqrt{s}=13\TeV$ under
  the event selections of Eqs.~(\ref{eq:wzatlascut})
  and~(\ref{eq:wzcmscut}). Note that the NLO QCD predictions have been
  divided by a factor 2 in the ratio ${\rm NLO\,QCD}/{\rm LO}$. Same
  notations and conventions as in \reffi{fig:wzptem}.}
\label{fig:wzy}
\efi

In \reffis{fig:wzptem}--\ref{fig:wztgcptz} we collect results for
differential distributions for the process $\Pp\Pp \to \Pep \nu_{\rm
  e} \mu^+ \mu^-$.  The distributions in the transverse momentum of
the positron ($p_{{\rm T},\Pe^+}$), the antimuon ($p_{{\rm
    T},\mu^+}$), and the muon--antimuon pair ($p_{ {\rm T,}\mu^+
  \mu^-}$, \ie the $\PZ$-boson transverse momentum) are shown in
\reffis{fig:wzptem} and~\ref{fig:wzmtptz}. For these distributions the
NLO EW corrections are negative and show the typical Sudakov behaviour
above about $100\GeV$ where they start to decrease monotonically and
become of order $-23/{-25}\%$ in the tails of the distributions under
consideration.
The NLO QCD corrections are positive, large, and increasing for large
$p_{\rm T}$. These corrections are dominated by real QCD contributions
as has been verified by playing with jet veto cuts.
In the presence of hard QCD radiation the four-lepton system recoils
against the radiated parton, and the leptons can likely acquire large
transverse momentum. Note that in the plots the NLO QCD corrections
have been divided by a factor~10.

The right plot in Fig.~\ref{fig:wzmtptz} shows the differential distribution in the transverse mass of the $\PW\PZ$ system defined
as:
\begin{align}
  M_{\rT}^{3l\nu}=\sqrt{\left(\sum_{\ell_i=1}^3 p_{\rm
      T,\ell_i}+|\vec{p}_{\rT}^{\,\rm miss}|\right)^2
    -\left[\left(\sum_{\ell_i=1}^3p_{\ell_i,_x}+p_x^{\rm miss}\right)^2
      +\left(\sum_{\ell_i=1}^3p_{\ell_i,_y}+p_y^{\rm
        miss}\right)^2\,\right]}. \label{eq:mtwzdef}
\end{align}
As described in \citere{Biedermann:2017oae}, the NLO EW corrections
are dominated by the real photon radiation below the peak, then show a
plateau between the peak and about $300\GeV$ (where they are of order
$-5\%$), while for larger $M_{\rm T}^{3l\nu}$ values they decrease up
to $-20\%$ for $M_{\rm T}^{3l\nu}=1\TeV$. Compared to the
transverse momentum distributions, the $M_{\rm T}^{3l\nu}$ observable
is less affected by NLO QCD corrections: these contributions are
positive, reach the order of $+135\%$ for $M_{\rm T}^{3l\nu}$
around $500\GeV$ and then start to slowly decrease.

Figure~\ref{fig:wzy} shows the differential distributions in the
positron and the antimuon rapidities ($y_{ {\rm e}^+ }$ and $y_{ \mu^+
  }$, respectively). The NLO EW corrections are basically flat and of
the same order as the NLO EW corrections to the integrated cross
section.  The NLO QCD corrections are positive, slightly more
pronounced in the central region and again of the same order as the
NLO QCD corrections to the integrated cross section.

\bfi
\begin{center}
  \begin{minipage}{0.40\textwidth}
    \includegraphics[width=\textwidth]{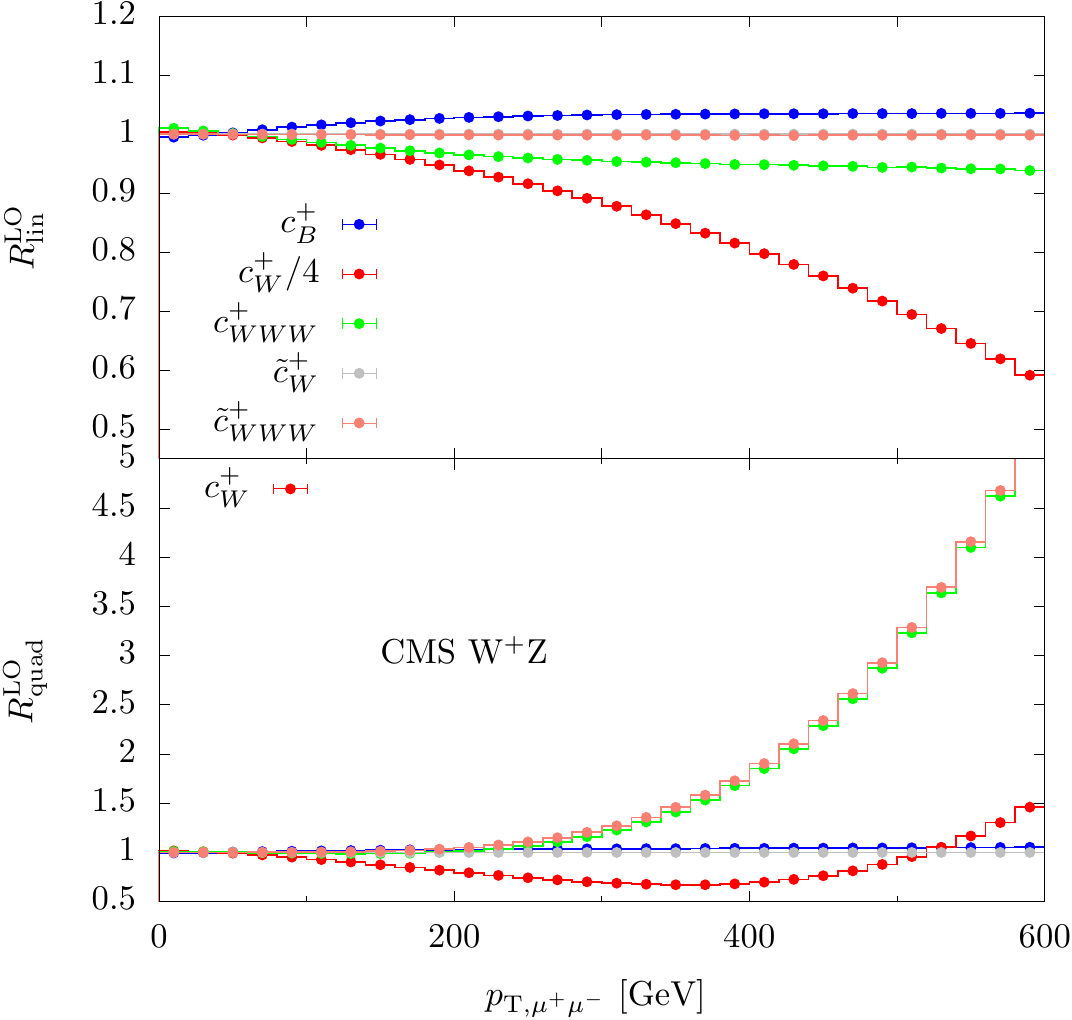}
  \end{minipage}
  \begin{minipage}{0.40\textwidth}
    \includegraphics[width=\textwidth]{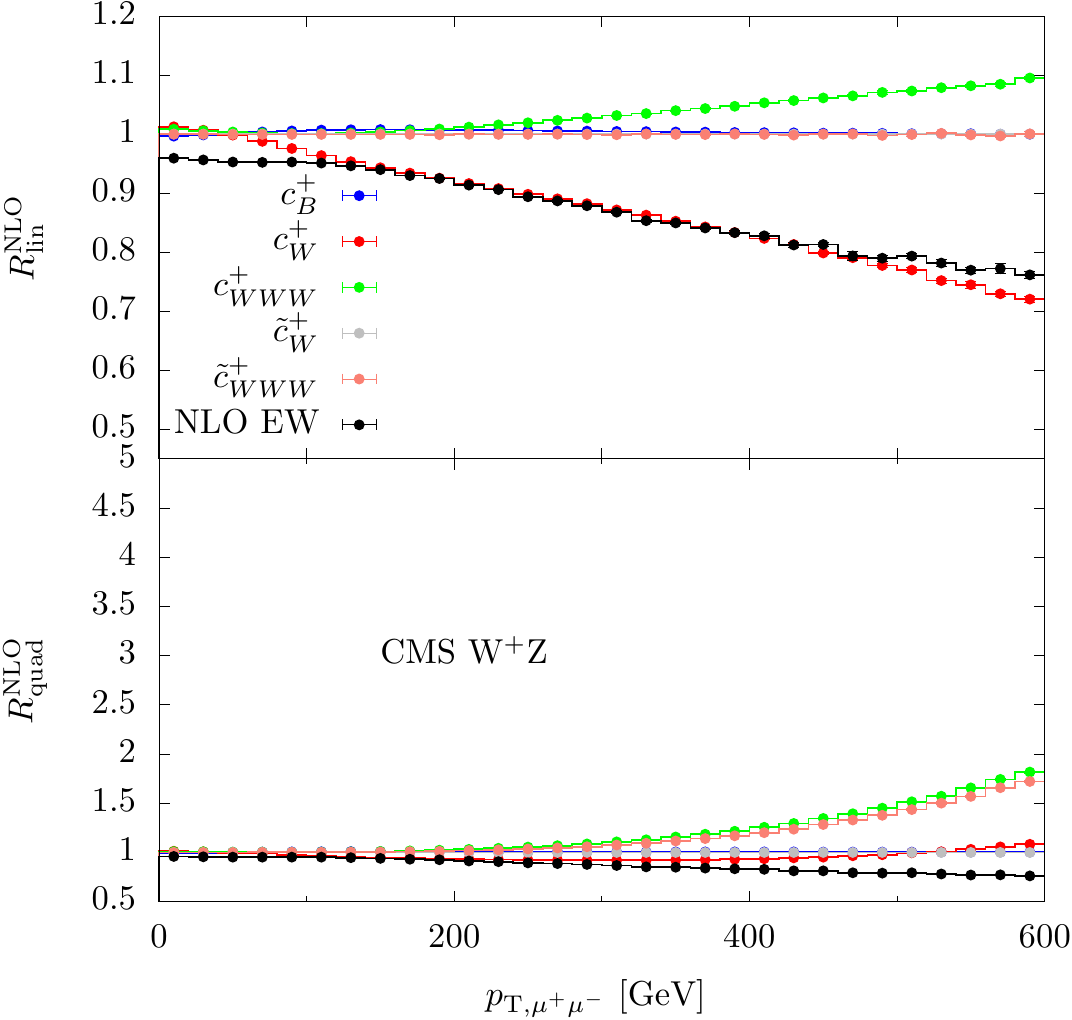}
  \end{minipage}\\[3ex]
  \begin{minipage}{0.40\textwidth}
    \includegraphics[width=\textwidth]{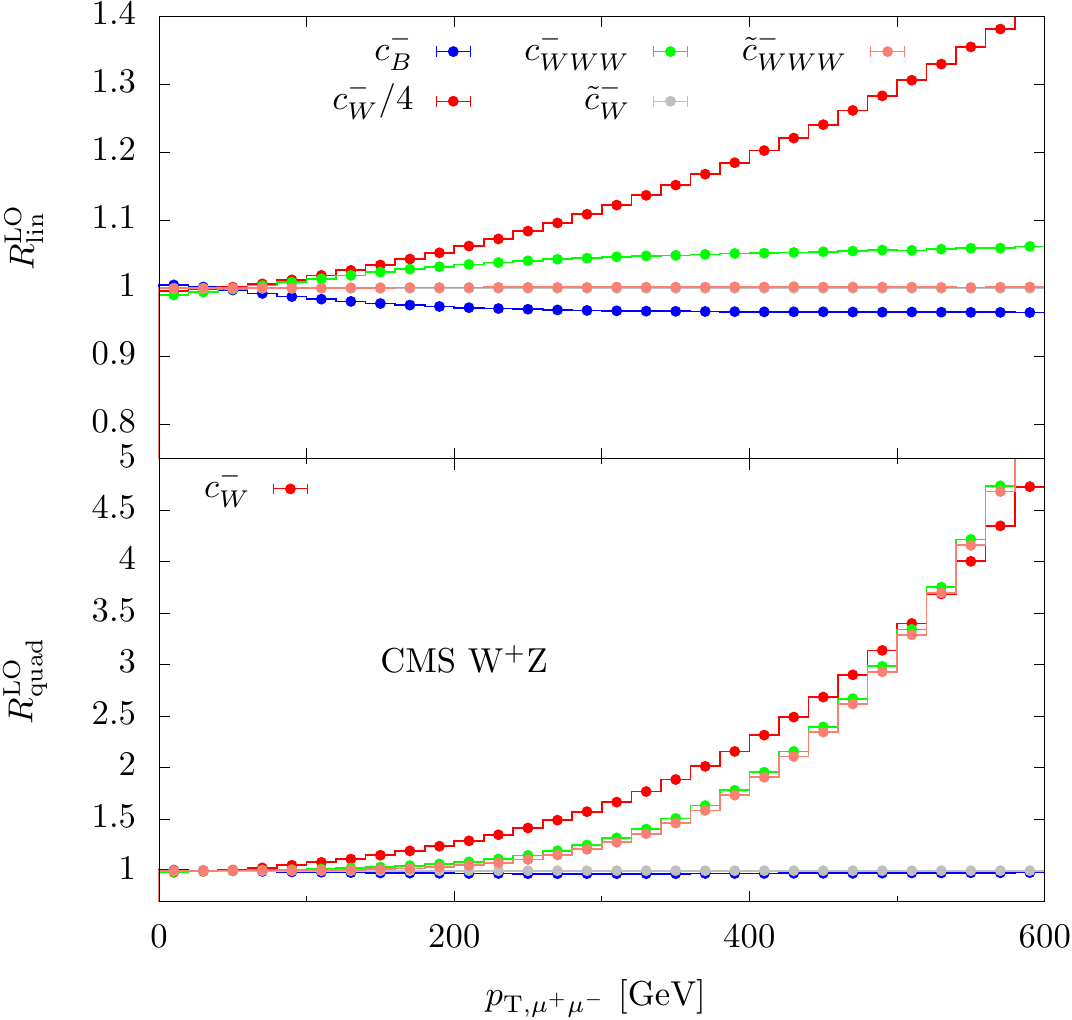}
  \end{minipage}
  \begin{minipage}{0.40\textwidth}
    \includegraphics[width=\textwidth]{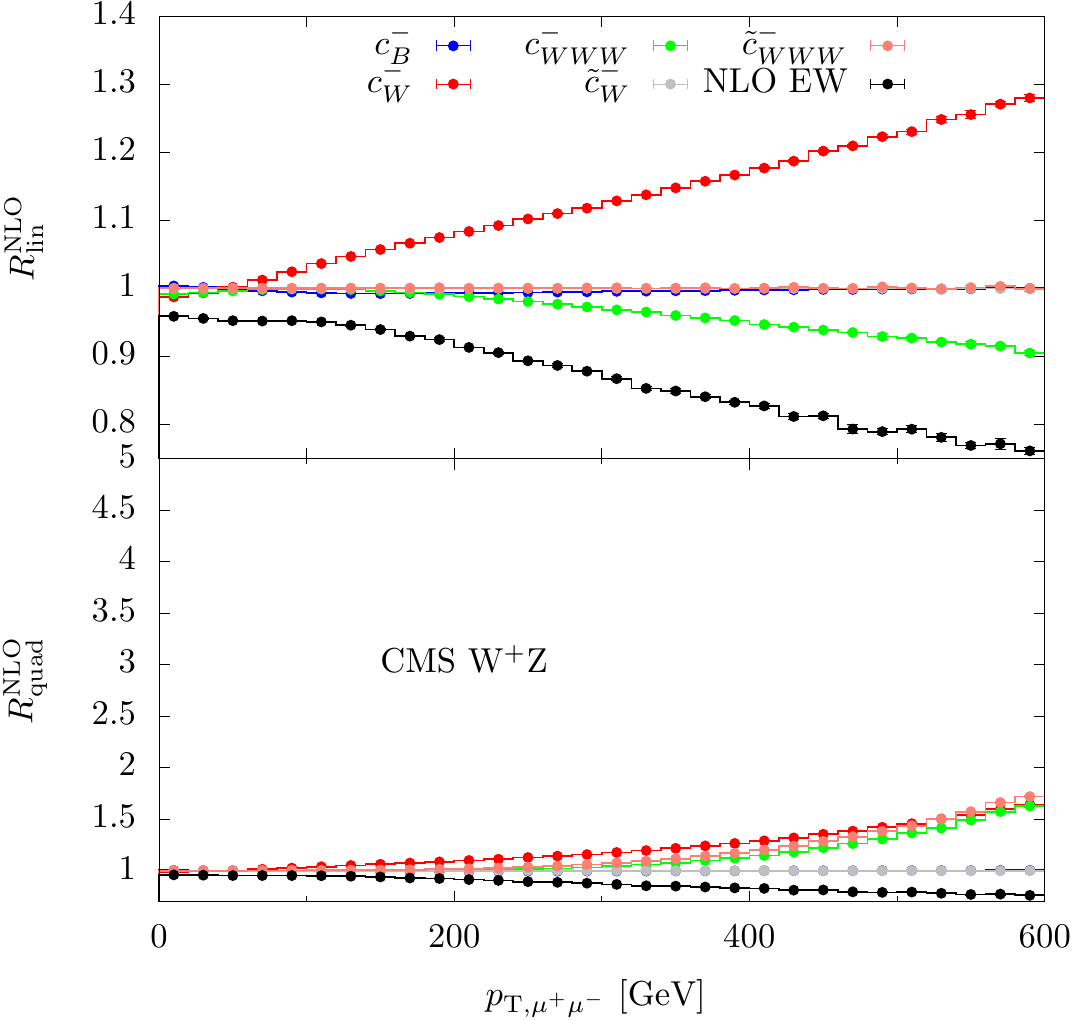}
  \end{minipage}
\end{center}
\caption{Ratio $R^{\rm LO(NLO)}_{\rm lin(quad)}$ as a function of the 
  muon--antimuon transverse momentum for the process $\Pp\Pp \to \Pep \nu_{\rm e} \mu^+ \mu^-$
  in the CMS setup of  Eq.~(\ref{eq:wzcmscut}).
  Same notation and conventions as in \reffi{fig:wwtgcpth}.
  In order to improve the plot readability, in the $R^{\rm LO}_{\rm
    lin}$ ratio (upper panels, left plots) the curves labelled with
  $c_{W}^{\pm}/4$ correspond to our predictions where the
  $c_{W}^{\pm}$ coefficients in Eq.~(\ref{eq:wwwilsoncoeffs})
  have been divided by a factor 4.  }
\label{fig:wztgcptz}
\efi
\bfi
\begin{center}
  \begin{minipage}{0.40\textwidth}
    \includegraphics[width=\textwidth]{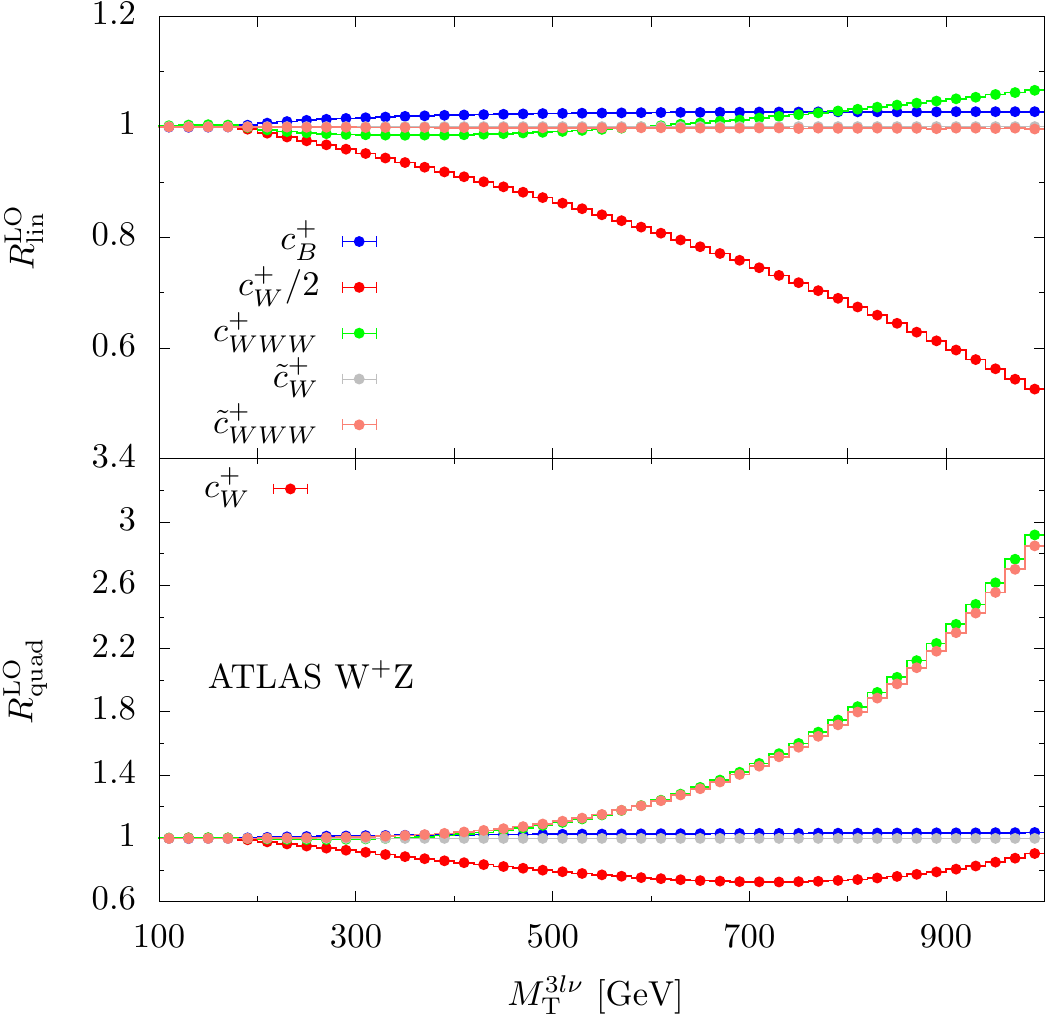}
  \end{minipage}
  \begin{minipage}{0.40\textwidth}
    \includegraphics[width=\textwidth]{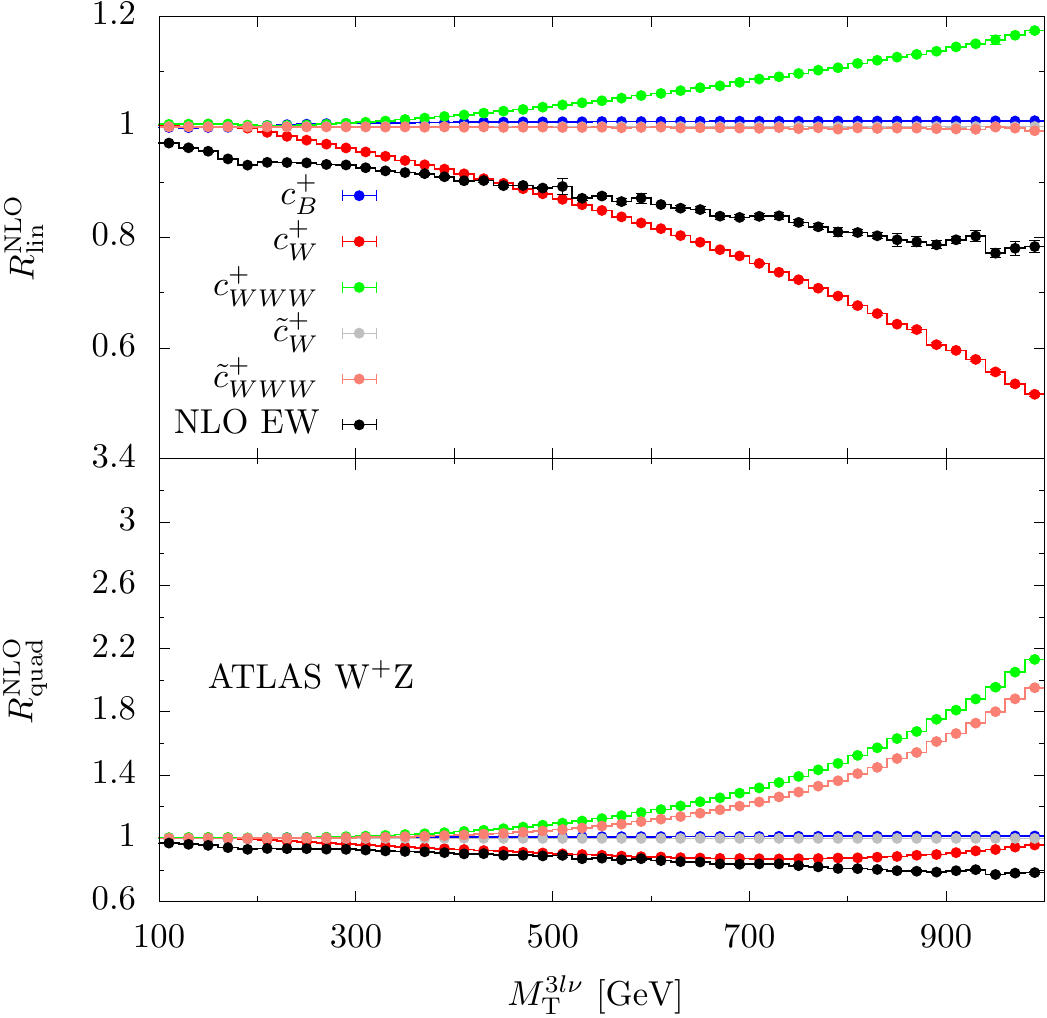}
  \end{minipage}\\[3ex]
  \begin{minipage}{0.40\textwidth}
    \includegraphics[width=\textwidth]{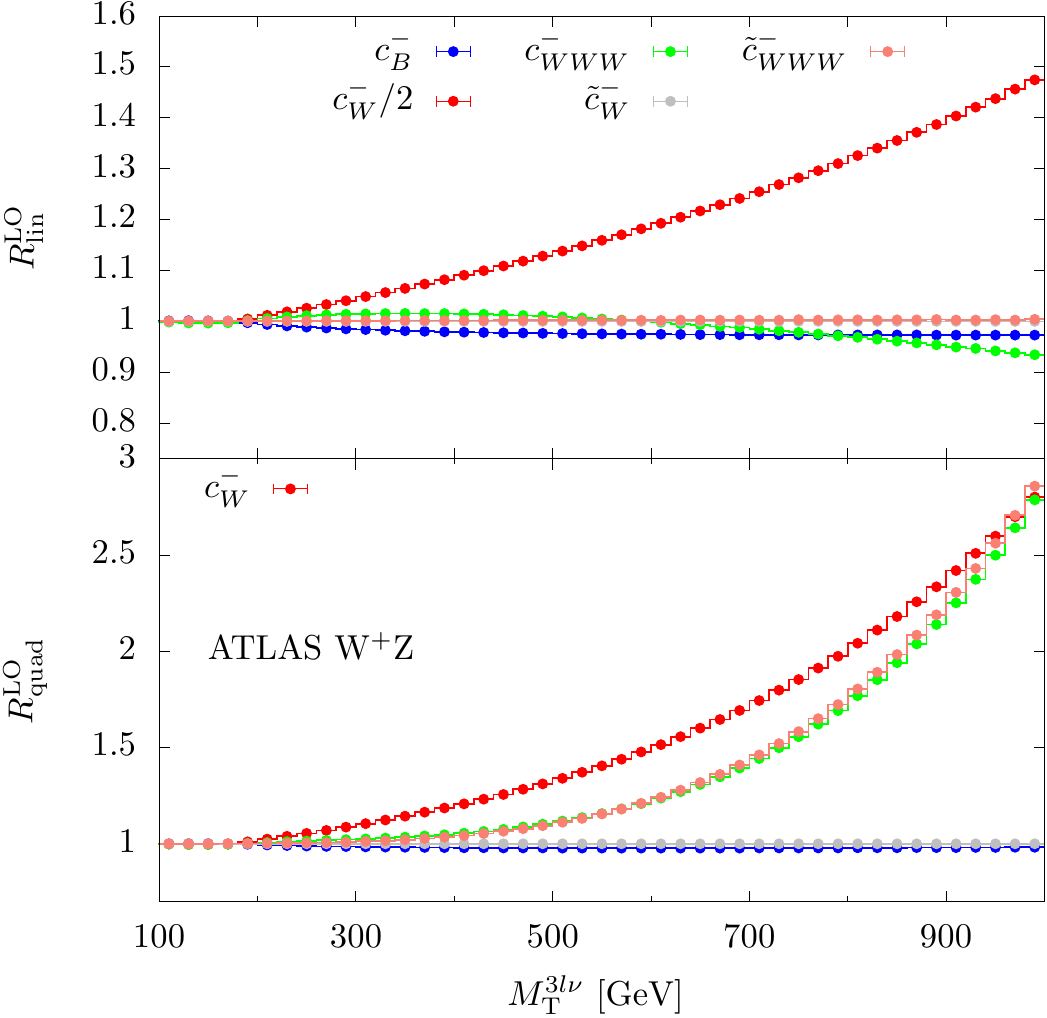}
  \end{minipage}
  \begin{minipage}{0.40\textwidth}
    \includegraphics[width=\textwidth]{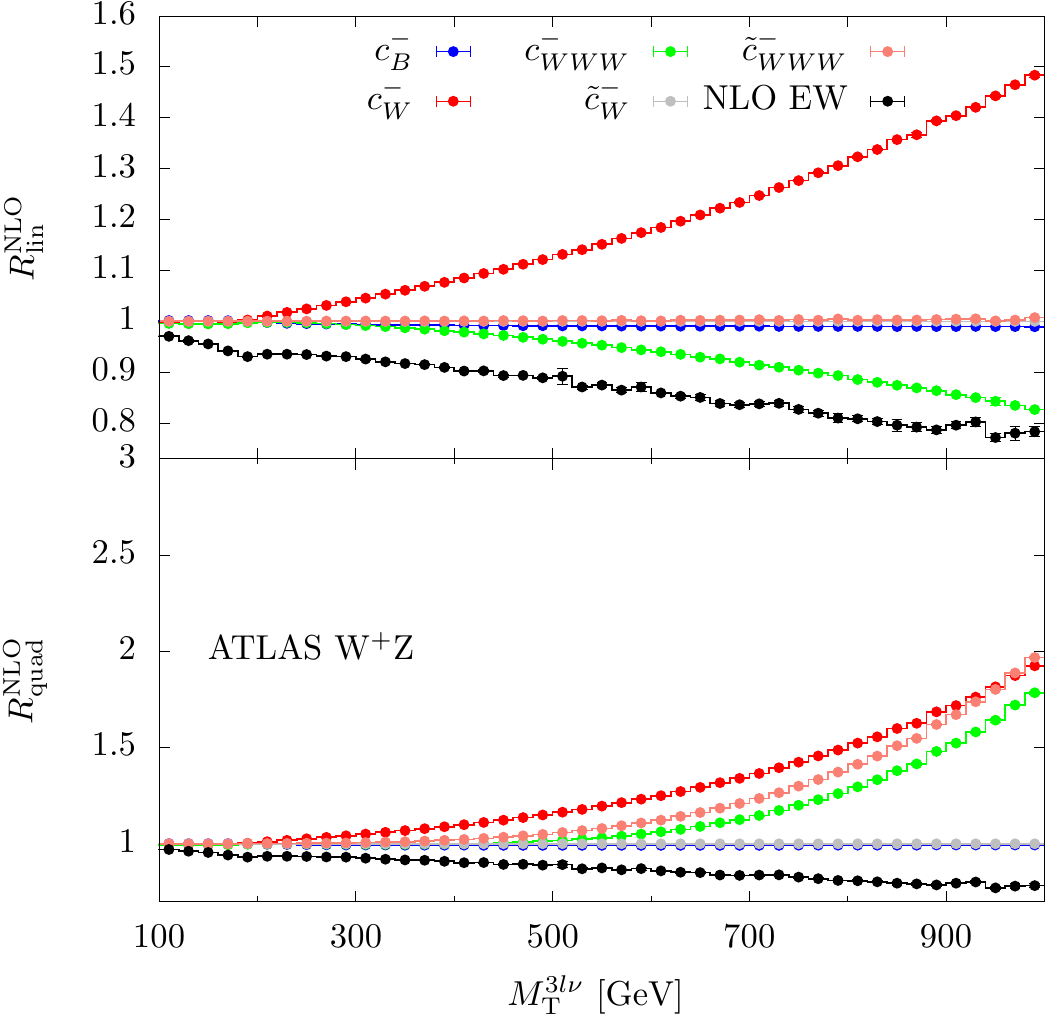}
  \end{minipage}
\end{center}
\caption{Ratio $R^{\rm LO(NLO)}_{\rm lin(quad)}$ as a function of the 
  $\PW\PZ$ transverse mass for the process $\Pp\Pp \to \Pep \nu_{\rm e} \mu^+ \mu^-$
  in the ATLAS setup of  Eq.~(\ref{eq:wzatlascut}).
  Same notation and conventions as in \reffi{fig:wwtgcpth}.
  In order to improve the plot readability, in the $R^{\rm LO}_{\rm
    lin}$ ratio (upper panels, left plots) the curves labelled with
  $c_{W}^{\pm}/2$ correspond to our predictions where the
  $c_{W}^{\pm}$ coefficients in Eq.~(\ref{eq:wwwilsoncoeffs})
  have been divided by a factor 2.}
\label{fig:wztgcmt}
\efi
\bfi
\begin{center}
  \begin{minipage}{0.40\textwidth}
    \includegraphics[width=\textwidth]{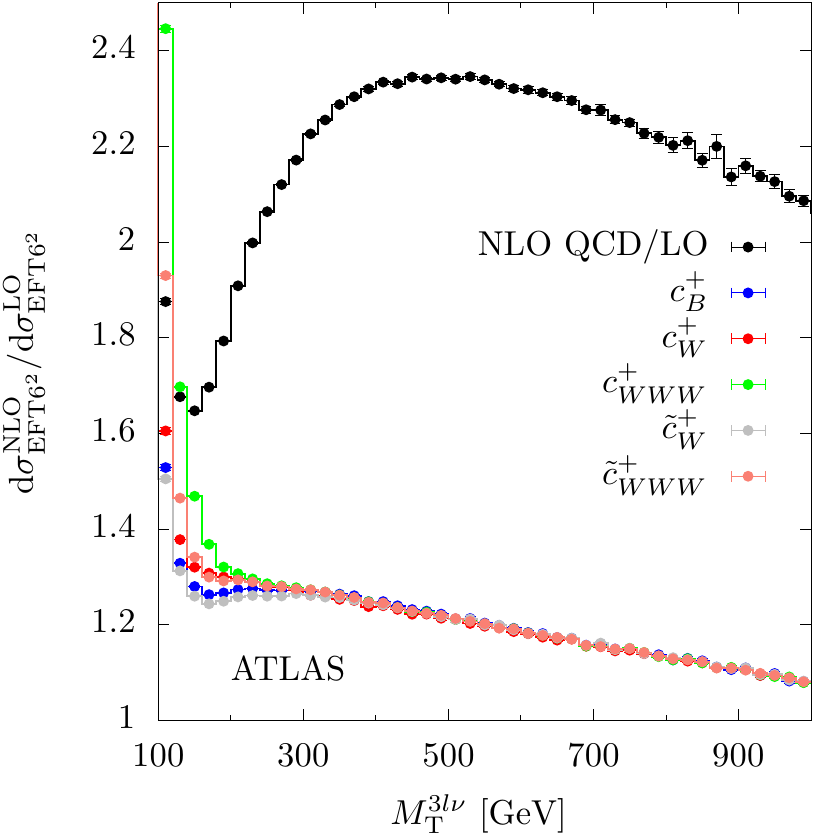}
  \end{minipage}
  \begin{minipage}{0.40\textwidth}
    \includegraphics[width=\textwidth]{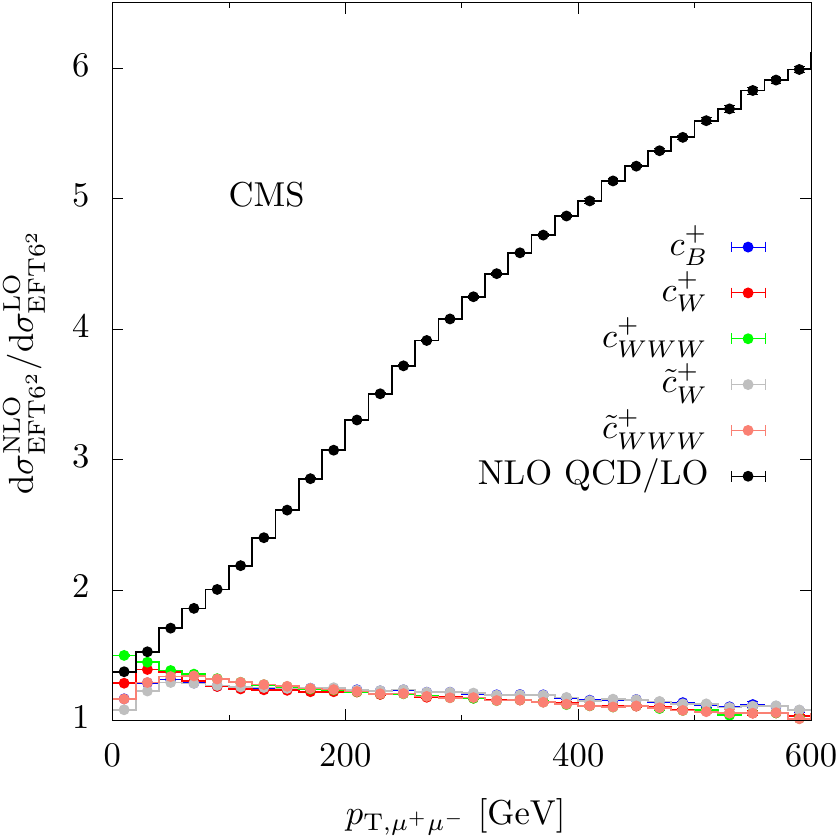}
  \end{minipage}
\end{center}
\caption{Ratio between the ${\rm EFT6}^2$ contribution to the process
  $\Pp\Pp \to \Pep \nu_{\rm e} \mu^+ \mu^-$ computed at NLO QCD
  ($\sigma_{{\rm EFT6}^2}^{\rm NLO}$) and LO ($\sigma_{{\rm EFT6}^2}^{\rm LO}$)
  accuracy  as a function of the $\PW\PZ$ transverse mass (left plot)
  and as a function of the transverse momentum of the muon--antimuon pair.
  The labels ATLAS and CMS refer to the event selections of
  Eq.~(\ref{eq:wzatlascut}) and~(\ref{eq:wzcmscut}), respectively.
  The ratio between the SM predictions at NLO QCD and at LO accuracy
  is also shown (black lines).}
\label{fig:wzqcdlam2}
\efi
The ratios $R^{\rm LO(NLO)}_{\rm lin(quad)}$, defined in
Eq.~(\ref{eq:defratio}), are shown in \reffis{fig:wztgcptz}
and~\ref{fig:wztgcmt} as a function of the $\PZ$-boson transverse
momentum ($p_{ {\rm T,}\mu^+ \mu^-}$) and as a function of the
$\PW\PZ$ transverse mass ($M_{\rm T}^{3l\nu}$).  As in
\reffis{fig:wwtgcpth} and~\ref{fig:wwtgcmll} we use the values in
Eq.~(\ref{eq:wwwilsoncoeffs}) for the Wilson coefficients, and only
one Wilson coefficient is different from zero for each curve.  From a
qualitative point of view, \reffis{fig:wztgcptz} and~\ref{fig:wztgcmt}
show the same behaviour for $R^{\rm LO}_{\rm lin}$, $R^{\rm LO}_{\rm
  quad}$, and $R^{\rm NLO}_{\rm lin}$ as \reffis{fig:wwtgcpth}
and~\ref{fig:wwtgcmll} for $\PW\PW$ production. On one hand, by
comparing the upper and lower panels of the left plots in
\reffis{fig:wztgcptz}--\ref{fig:wztgcmt} we conclude that the largest
contribution comes from the $\sigma^{\rm LO}_{{\rm EFT6}^2}$ terms
(with the only exception of the $c_{W}$ coefficient, for which the
$\sigma^{\rm LO}_{{\rm EFT6}^2}$ terms become larger than the
interference terms only in the tails of the distributions under
consideration). On the other hand, comparing the left and the right
plots in \reffis{fig:wztgcptz}--\ref{fig:wztgcmt} reveals that the NLO
QCD corrections tend to reduce the sensitivity to the aTGCs (with the
exception of the $c_{WWW}$ coefficient in $R^{\rm NLO}_{\rm lin}$) in
particular for the $R^{\rm NLO}_{\rm quad}$ ratio.%
\footnote{For similar results see \citere{Baur:1994aj}.} 
 Even though $R^{\rm
  LO}_{\rm lin(quad)}$ and $R^{\rm NLO}_{\rm lin}$ show the same
qualitative behaviour for $\PW\PW$ and $\PW\PZ$ production, from a
quantitative point of view we notice that $\PW\PZ$ production is more
sensitive to aTGCs and in particular to the $c_{W}$ coefficient.

The shape of the $R^{\rm NLO}_{\rm quad}$ distribution can be
understood by looking at the NLO QCD corrections to the ${\rm EFT6}^2$
contributions (\reffi{fig:wzqcdlam2}) and Eqs.~\refeq{eq:quadratio}
and \refeq{eq:quadratio2}.  At variance with the $\PW\PW$ case, where
the jet veto in the event
selections~(\ref{eq:wwatlascut}), (\ref{eq:wwcmscut}) suppresses real
QCD radiation, for $\PW\PZ$ production the NLO QCD corrections to the
${\rm EFT6}^2$ contributions are positive and large owing to
real-radiation corrections but much
smaller than the corrections to the SM process (this is particularly
evident for the $p_{ {\rm T,}\mu^+ \mu^-}$ distribution). This is due
to the fact that QCD radiation reduces the centre-of-mass energy of
the diboson system with respect to the LO.  Since the aTGCs
contribution increases with the centre-of-mass energy of the diboson
system, at NLO QCD the aTGCs contribution is suppressed.

\subsection{ZZ production}
\label{subsect:zzres}

The results for the fiducial cross sections for the process $\Pp\Pp
\to \Pep \Pem \mu^+ \mu^-$ at $13\TeV$ under the event selection of
Eq.~(\ref{eq:zz4latlascut}) are collected in \refta{tab:zz-xsec}.  The
LO results are compared to the predictions at NLO QCD and NLO EW
accuracy. The contribution of the loop-induced process $\Pg \Pg \to
\PZ \PZ$ is also shown. The NLO EW corrections are of order $-8\%$
while the NLO QCD corrections amount to $+35\%$. The $\Pg\Pg$ channel
contributes about $+17\%$ of the LO prediction. For massless
  quarks the $\Pg\Pg$ channel results only from quark-box diagrams,
  while for the massive top quark also $s$-channel Higgs
  production via a top loop contributes. For a light top quark the
  contribution of the $\Pg\Pg$ channel amounts to $+24\%$ of the LO
  cross section, \ie the large top mass reduces the cross section by
  $7\%$.  The numbers in parentheses represent the integration error
on the last digit, while the upper and lower values for the cross
sections correspond to the uncertainty coming from scale variation
according to Eq.~(\ref{eq:scalevar}).  Scale uncertainties are of the
same order for the LO and the NLO EW predictions and are reduced by a
factor of two at NLO QCD.
\begin{table}
  \begin{center}
\renewcommand{\arraystretch}{1.4}
    \begin{tabular}{|l|l|l|l|l}
      \hline
       LO [fb] & NLO QCD [fb] & NLO EW [fb] & $\Pg\Pg$  [fb] \\
      \hline
       $11.0768(5)^{+6.3\%}_{-7.5\%}$ & $14.993(2)^{+3.1\%}_{-2.4\%}$ & $10.283(2)^{+6.4\%}_{-7.6\%}$ & $1.8584(4)^{+25\%}_{-18\%}$\\
      \hline

    \end{tabular}
  \end{center} 
  \caption{Fiducial cross section for the process $\Pp\Pp \to \Pep \Pem \mu^+ \mu^-$
    at $\sqrt{s}=13\TeV$ in the setup of Eqs.~(\ref{eq:zz4latlascut}). The numbers in
    parentheses correspond to the statistical error on the last digit. The uncertainties
    are estimated from the scale dependence, as explained in the text.}
  \label{tab:zz-xsec}
\end{table}

The differential distribution in the transverse momentum of the
positron ($p_{{\rm T, e}^+}$), the antimuon ($p_{{\rm T,}\mu^+}$), the
muon--antimuon pair ($p_{{\rm T,}\mu^+\mu^-}$), and the hardest $\PZ$
boson [$p_{\rm T, Z}^{\rm max}=\max(p_{{\rm T},\mu^+ \mu^-},\,$
$p_{{\rm T},\Pe^+ \Pe^-})$] are shown in \reffis{fig:zzptem}
and~\ref{fig:zzptz}. For these distributions the NLO EW corrections
are negative and decrease monotonically reaching the value of about
$-40\%$ for $p_{{\rm T, e}^+}$ and $p_{{\rm T,}\mu^+}$ of order $600\GeV$
and $-50\%$ for $p_{{\rm T,}\mu^+\mu^-}$ and $p_{\rm T, Z}^{\rm max}$ of
order $800\GeV$.  The NLO QCD corrections are positive, large, and
increase at high $p_{\rm T}$. As pointed out in
\refse{subsect:wzres}, this is due to the opening of the
gluon-initiated channels that contribute to the real QCD corrections
and enhance the high-$p_{\rm T}$ region.
\bfi
\begin{center}
  \begin{minipage}{0.40\textwidth}
    \includegraphics[width=\textwidth]{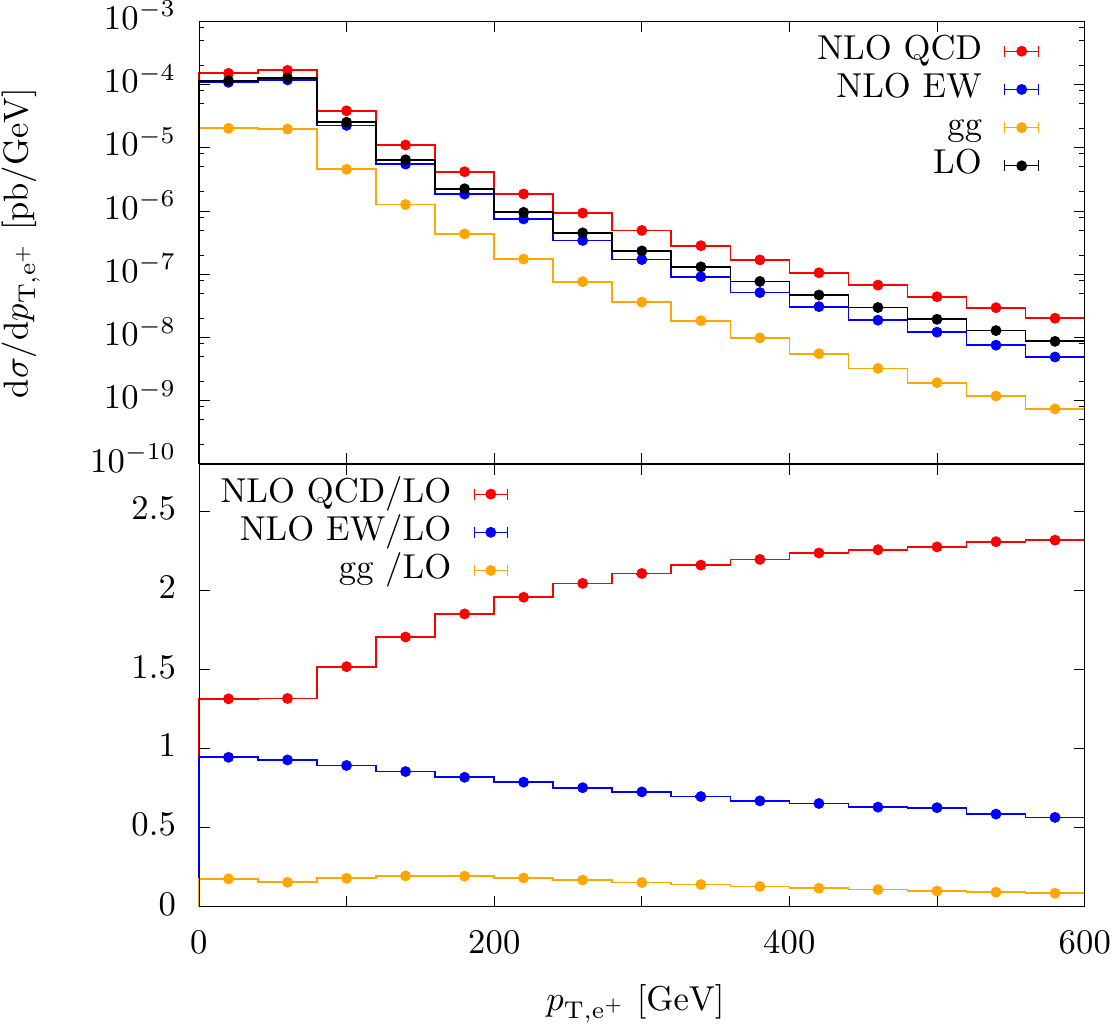}
  \end{minipage}
  \begin{minipage}{0.40\textwidth}
    \includegraphics[width=\textwidth]{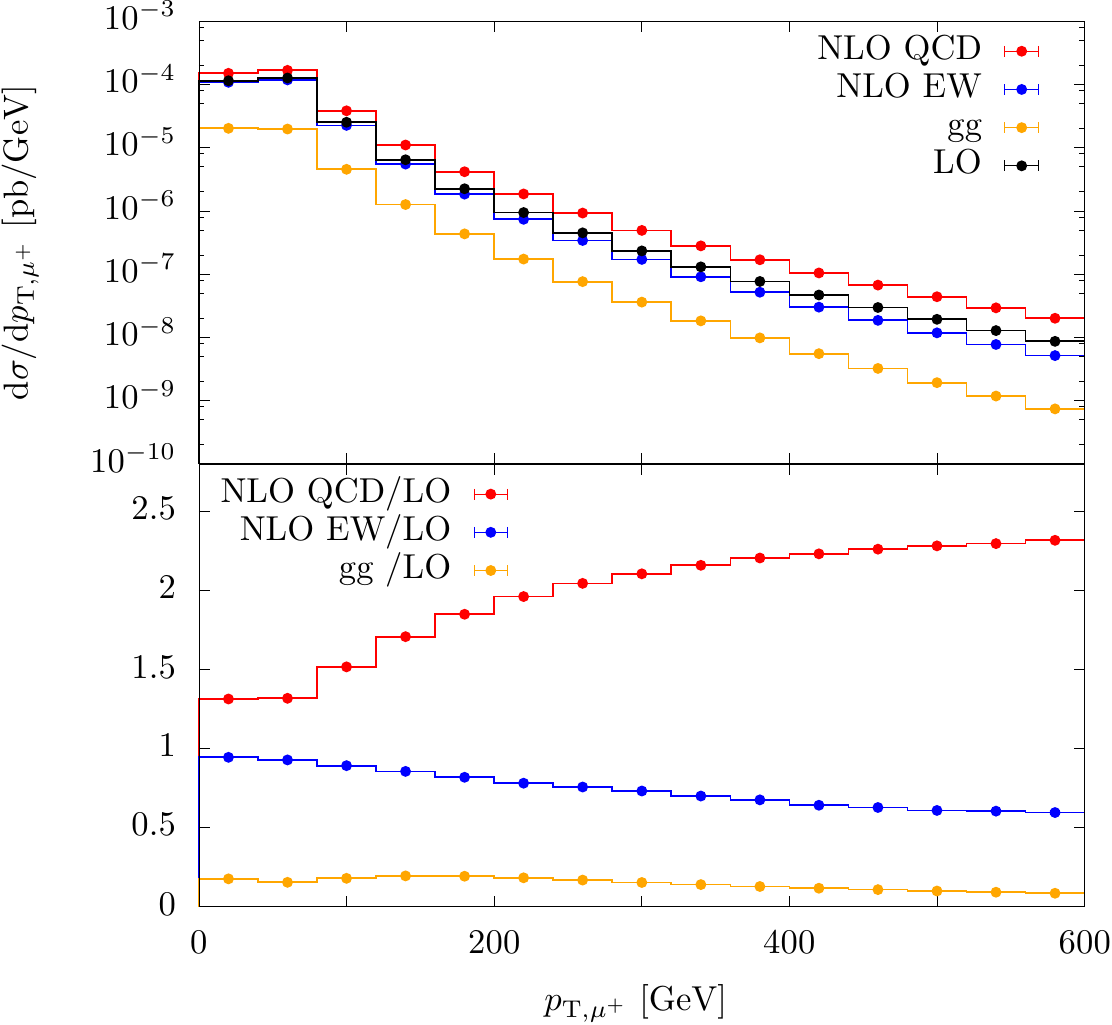}
  \end{minipage}
\end{center}
\caption{Upper panels: differential distributions in the transverse
  momentum of the positron ($p_{{\rm T},\Pe ^{+}}$) and the antimuon
  ($p_{{\rm T},\mu ^{+}}$) for the process $\Pp\Pp \to \Pep \Pem \mu^+ \mu^-$ 
  at $\sqrt{s}=13\TeV$ for the event selection of Eq.~(\ref{eq:zz4latlascut}).
  The LO results (black lines) are compared to the ones
  at NLO QCD (red lines) and NLO EW (blue lines). The $\Pg\Pg$
  contribution is also shown (orange lines).  Lower panels: ratio of
  the NLO QCD, NLO EW and $\Pg\Pg$ contributions with respect to the
  LO (red, blue and orange lines, respectively).
  For all curves the central value of
  the factorization and renormalization scales is used and the error
  bars correspond to the statistical integration uncertainties. Note
  that the same PDF set is employed for both the LO and NLO
  predictions.}
\label{fig:zzptem}
\efi
\bfi
\begin{center}
  \begin{minipage}{0.40\textwidth}
    \includegraphics[width=\textwidth]{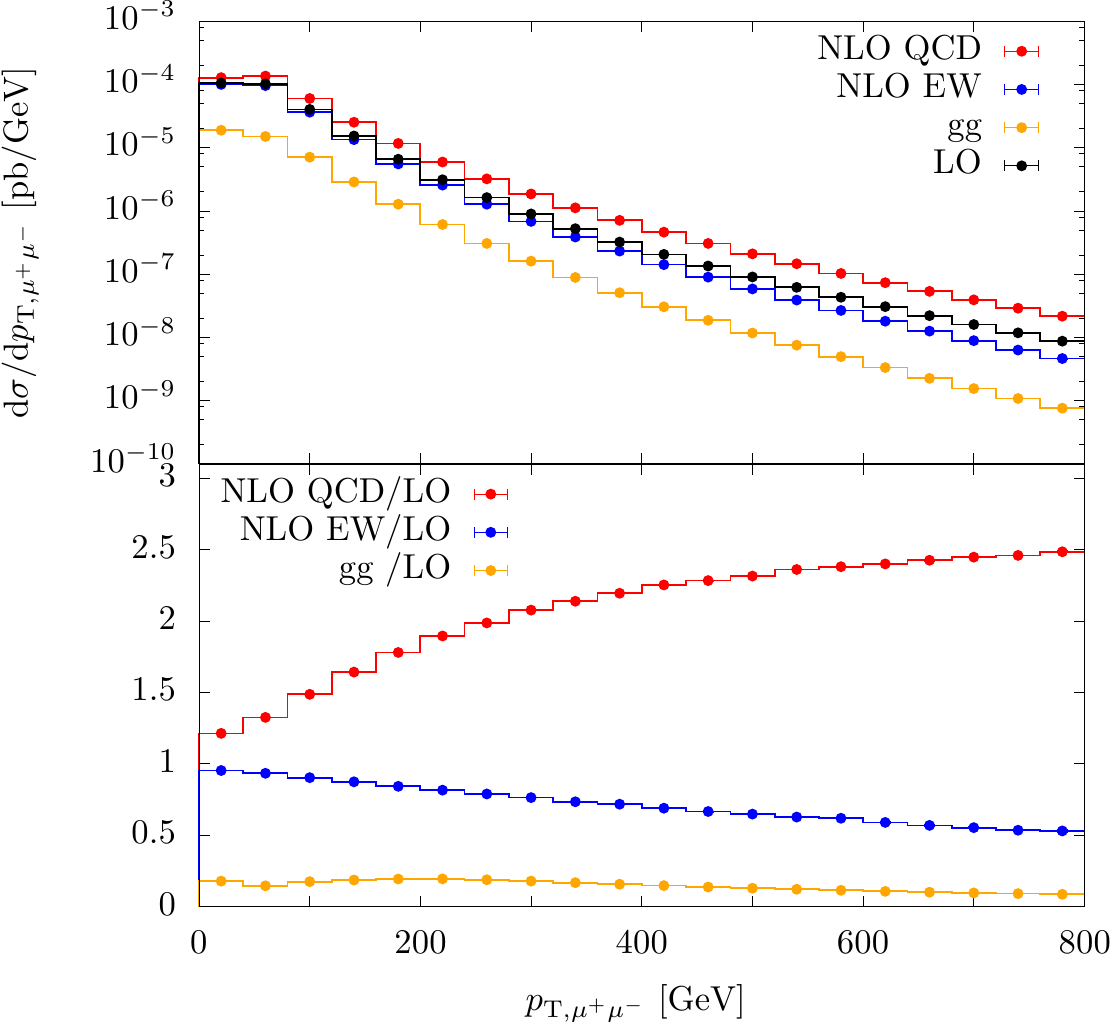}
  \end{minipage}
  \begin{minipage}{0.40\textwidth}
    \includegraphics[width=\textwidth]{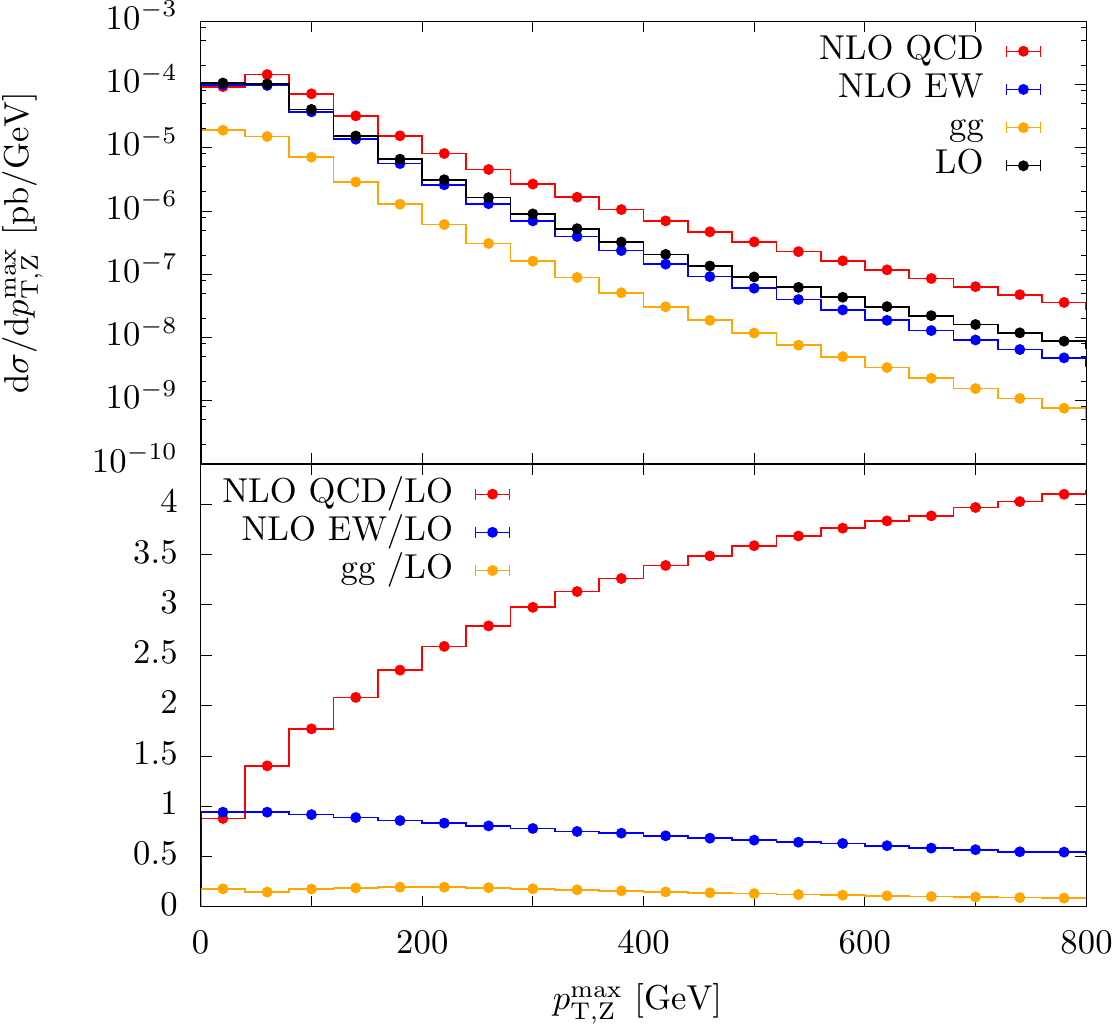}
  \end{minipage}
\end{center}
\caption{Differential distribution in the muon--antimuon-pair transverse
  momentum ($p_{{\rm T},\mu^+ \mu^-}$) and in the hardest $\PZ$-boson
  transverse momentum ($p_{{\rm T,\,Z}}^{\rm max}$) for the process
  $\Pp\Pp \to \Pep \Pem \mu^+ \mu^-$ at $\sqrt{s}=13\TeV$ under the
  event selections of Eq.~(\ref{eq:zz4latlascut}).  Same notations and
  conventions as in \reffi{fig:zzptem}.}
\label{fig:zzptz}
\efi

Figure~\ref{fig:zzdym4l} shows the differential distributions as a
function of the four-lepton invariant mass ($M_{4l}^{\rm inv}$) and as
a function of the rapidity difference of the two $\PZ$ bosons ($\Delta
y_{\rm ZZ}$).  The $M_{4l}^{\rm inv}$ distribution peaks near $2
M_{\PZ}$: below the peak the NLO EW corrections are dominated by real
photon radiation, while above the peak they have the same Sudakov
behaviour found in the $p_{\rm T}$ distributions and reach the value
of $-45\%$ for $M_{4l}^{\rm inv}$ of order $2\TeV$. At variance with
the case of the transverse-momentum distributions, the NLO QCD
corrections to the four-lepton invariant-mass distribution are
relatively flat (they reach the value of $+50\%$ for $M_{4l}^{\rm
  inv}$ between $0.5$ and $1\TeV$ and then they decrease with
$M_{4l}^{\rm inv}$).  Both the NLO EW and the NLO QCD corrections to
the $\PZ$-boson-pair rapidity difference are essentially flat: the NLO
QCD corrections are somewhat larger for large $|\Delta y_{\rm ZZ}|$,
while the contribution of the $\Pg\Pg$ channel is of order $+20\%$ for
$\Delta y_{\rm ZZ}$ between $-2$ and $2$ and decreases for larger
values of $|\Delta y_{\rm ZZ}|$.
\bfi
\begin{center}
  \begin{minipage}{0.40\textwidth}
    \includegraphics[width=\textwidth]{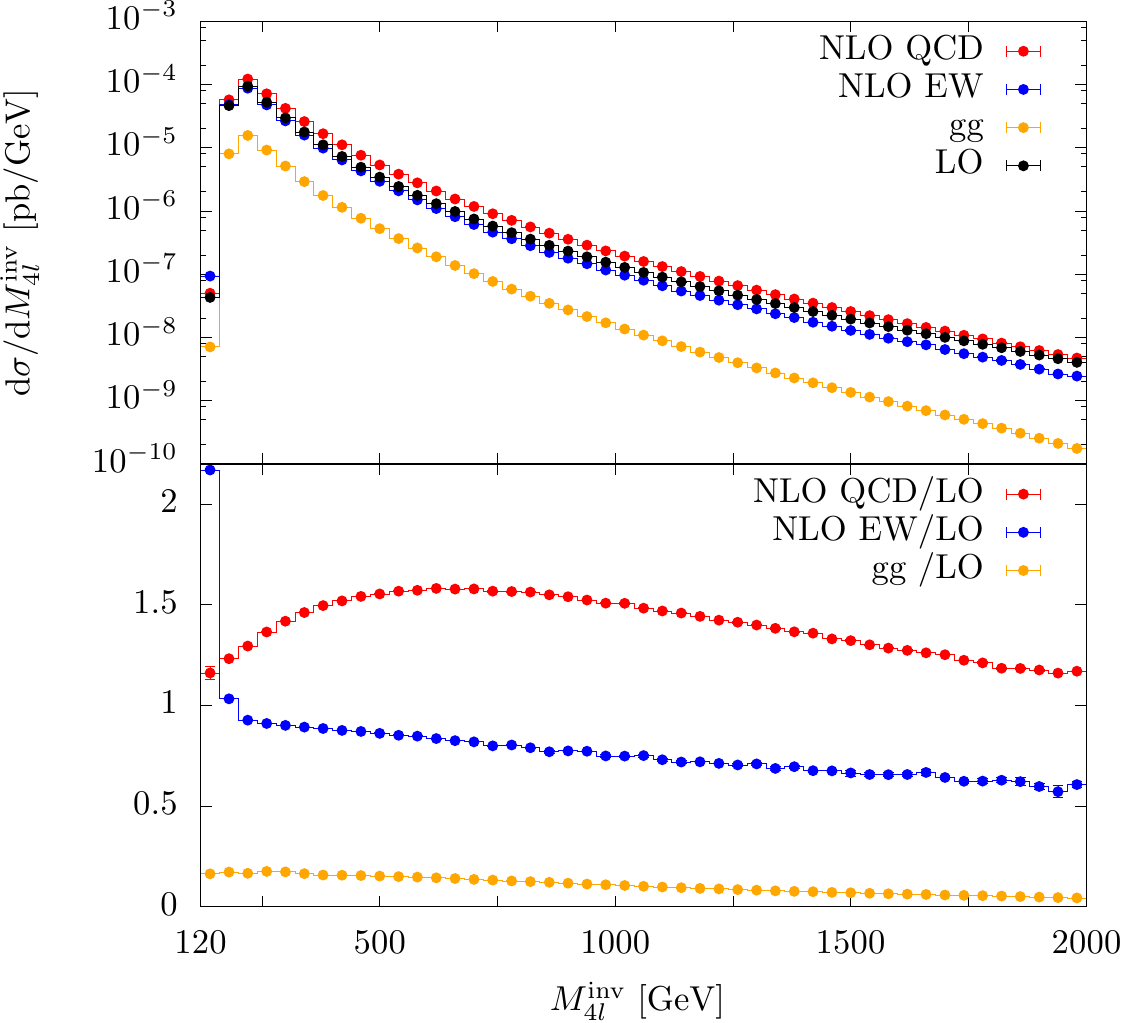}
  \end{minipage}
  \begin{minipage}{0.40\textwidth}
    \includegraphics[width=\textwidth]{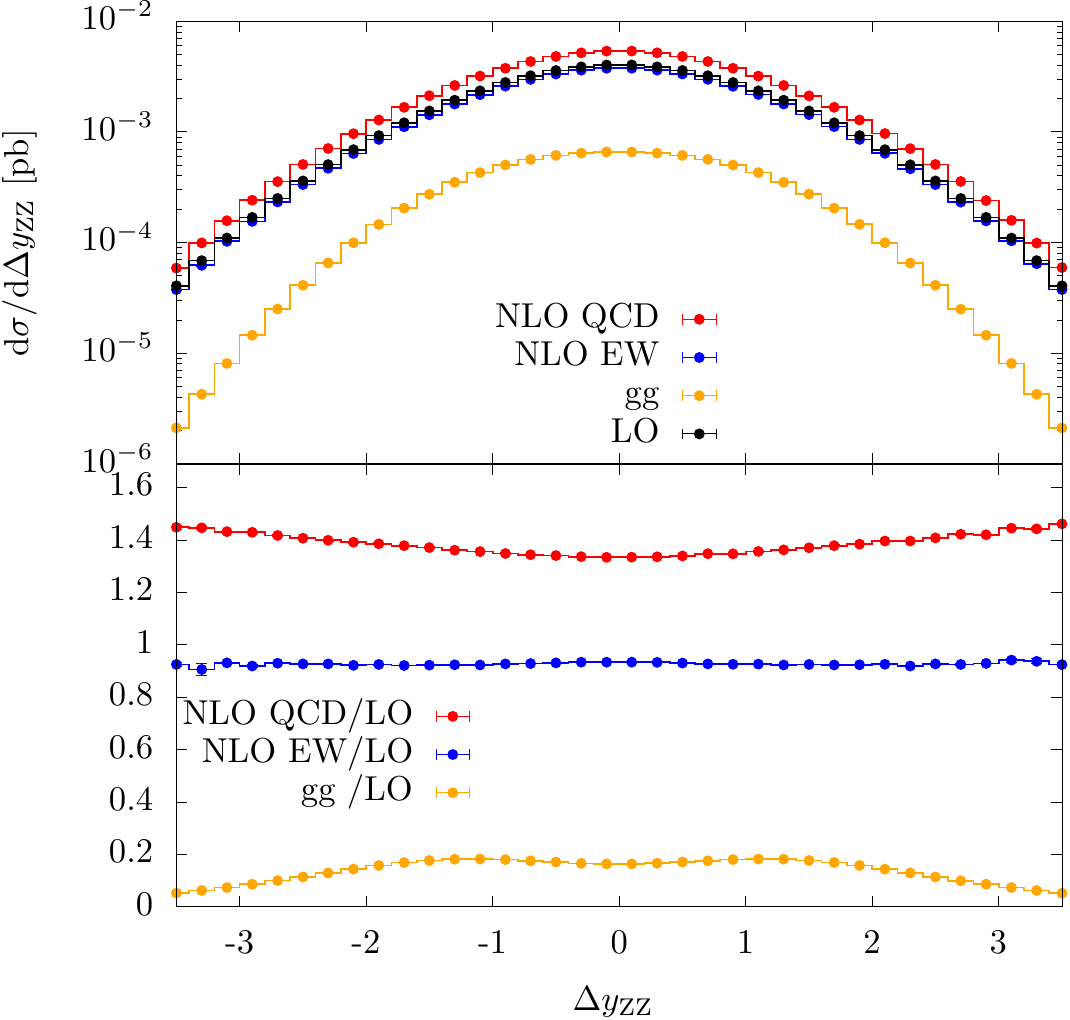}
  \end{minipage}
\end{center}
\caption{Differential distribution in the four-lepton invariant mass
  ($M_{4l}^{\rm inv}$) and in the $\PZ$-pair rapidity difference
  ($\Delta y_{\rm  ZZ}$) for the process
  $\Pp\Pp \to \Pep \Pem \mu^+ \mu^-$  at $\sqrt{s}=13\TeV$
  under the event selections of Eq.~(\ref{eq:zz4latlascut}).
  Same notations and conventions as in \reffi{fig:zzptem}.}
\label{fig:zzdym4l}
\efi

The differential distributions in the positron and antimuon rapidities
($y_{{\rm e}^+}$ and $y_{{\mu}^+}$, respectively) are shown in
\reffi{fig:zzyeym}. Both the NLO EW and the NLO QCD corrections are
basically flat and of the same order as the corrections to the
fiducial cross section.
\bfi
\begin{center}
  \begin{minipage}{0.40\textwidth}
    \includegraphics[width=\textwidth]{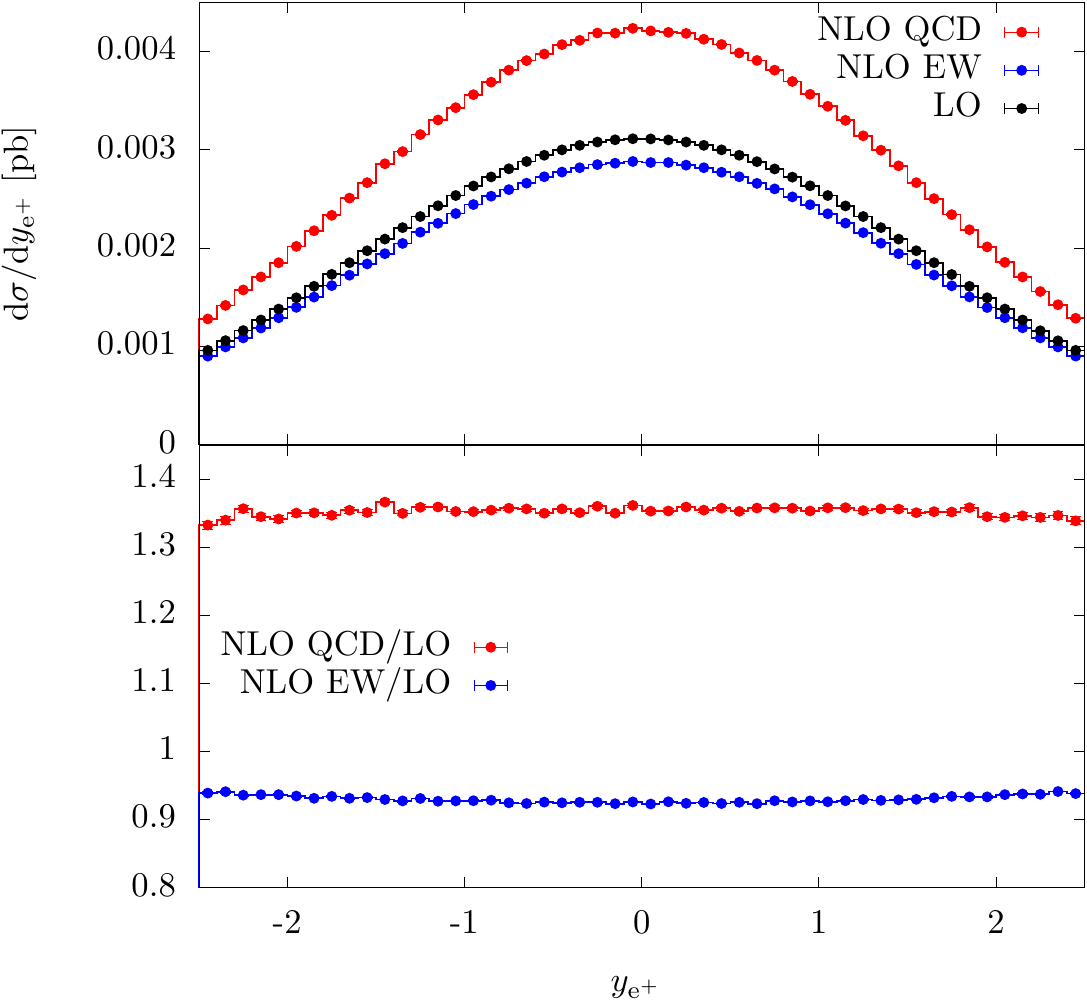}
  \end{minipage}
  \begin{minipage}{0.40\textwidth}
    \includegraphics[width=\textwidth]{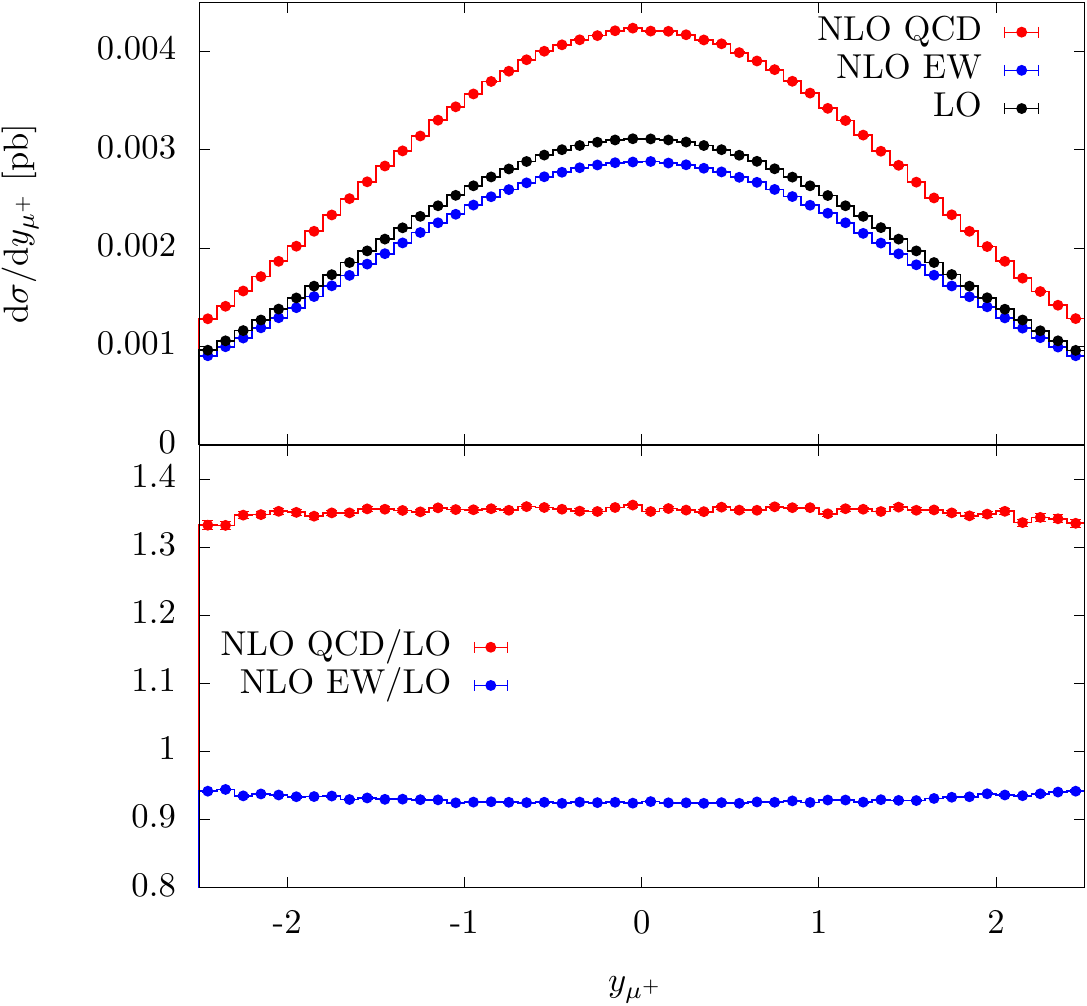}
  \end{minipage}
\end{center}
\caption{Differential distribution in the positron and antimuon  rapidities
  ($y_{{\rm e}^+}$ and $y_{\mu^+}$) for the process
  $\Pp\Pp \to \Pep \Pem \mu^+ \mu^-$  at $\sqrt{s}=13\TeV$
  under the event selections of Eq.~(\ref{eq:zz4latlascut}).
  Same notations and conventions as in \reffi{fig:zzptem}.}
\label{fig:zzyeym}
\efi

Concerning the sensitivity to the neutral aTGCs,
\reffis{fig:zztgcptmax} and~\ref{fig:zztgcm4l} show the differential
distribution of the ratios $R^{\rm LO(NLO)}_{\rm lin(quad)}$, defined
in Eq.~(\ref{eq:defratio}), as a function of the transverse momentum
of the hardest $\PZ$~boson and as a function of the four-lepton
invariant mass. For the Wilson coefficients of the dimension-8
operators involved in $\PZ\PZ$ production we use the values listed in
Eq.~(\ref{eq:zzwilsoncoeffs}). As in \reffis{fig:wwtgcpth}
and~\ref{fig:wwtgcmll}, for each curve in the plot only one Wilson
coefficient is different from zero.  Figures~\ref{fig:zztgcptmax}
and~\ref{fig:zztgcm4l} confirm the same pattern already described for
the dimension-6 operators in $\PW\PW$ and $\PW\PZ$ production.  First
of all, by comparing $R^{\rm LO(NLO)}_{\rm lin}$ and $R^{\rm
  LO(NLO)}_{\rm quad}$ we notice that the leading effect comes from
the ${\rm EFT}8^2$ contributions: this feature is much more evident
than in the $\PW\PW$ and $\PW\PZ$ case, since $R^{\rm LO(NLO)}_{\rm
  lin}$ turns out to be sensitive only to the $c_{\tilde{B}W}$
coefficient, while for $R^{\rm LO(NLO)}_{\rm quad}$ there is a
dependence on all four possible Wilson coefficients.  Even for
$c_{\tilde{B}W}$, the ${\rm EFT}8^2$ contributions always dominate
over the ${\rm SM\times EFT8}$ contributions.  By comparing $R^{\rm
  LO}_{\rm lin(quad)}$ and $R^{\rm NLO}_{\rm lin(quad)}$ we conclude
that the NLO QCD corrections reduce the dependence on the Wilson
coefficients of the dimension-8 operators. For $R^{\rm NLO}_{\rm
  quad}$ the reduction in the sensitivity to the aTGCs is more
pronounced for the $p_{\rm T, Z}^{\rm max}$ observable rather than for
the four-lepton invariant-mass distribution.  This can be understood
by comparing the NLO QCD corrections to the ${\rm EFT}8^2$ terms,
which furnish the leading contribution to $R^{\rm LO}_{\rm quad}$,
with the NLO QCD corrections to the SM results, using equations
analogous to Eqs.~\refeq{eq:quadratio} and \refeq{eq:quadratio2}.
\bfi
\begin{center}
  \begin{minipage}{0.40\textwidth}
    \includegraphics[width=\textwidth]{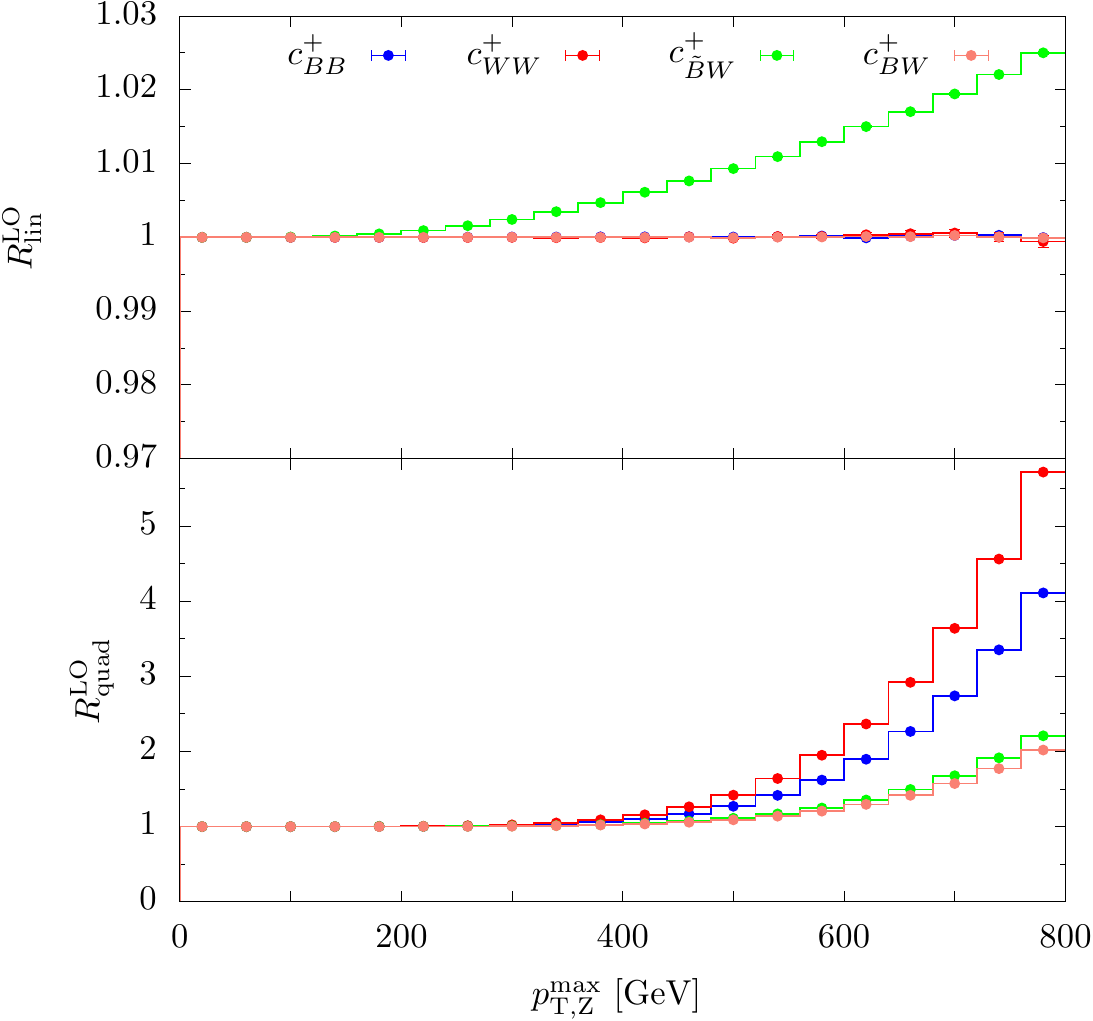}
  \end{minipage}
  \begin{minipage}{0.40\textwidth}
    \includegraphics[width=\textwidth]{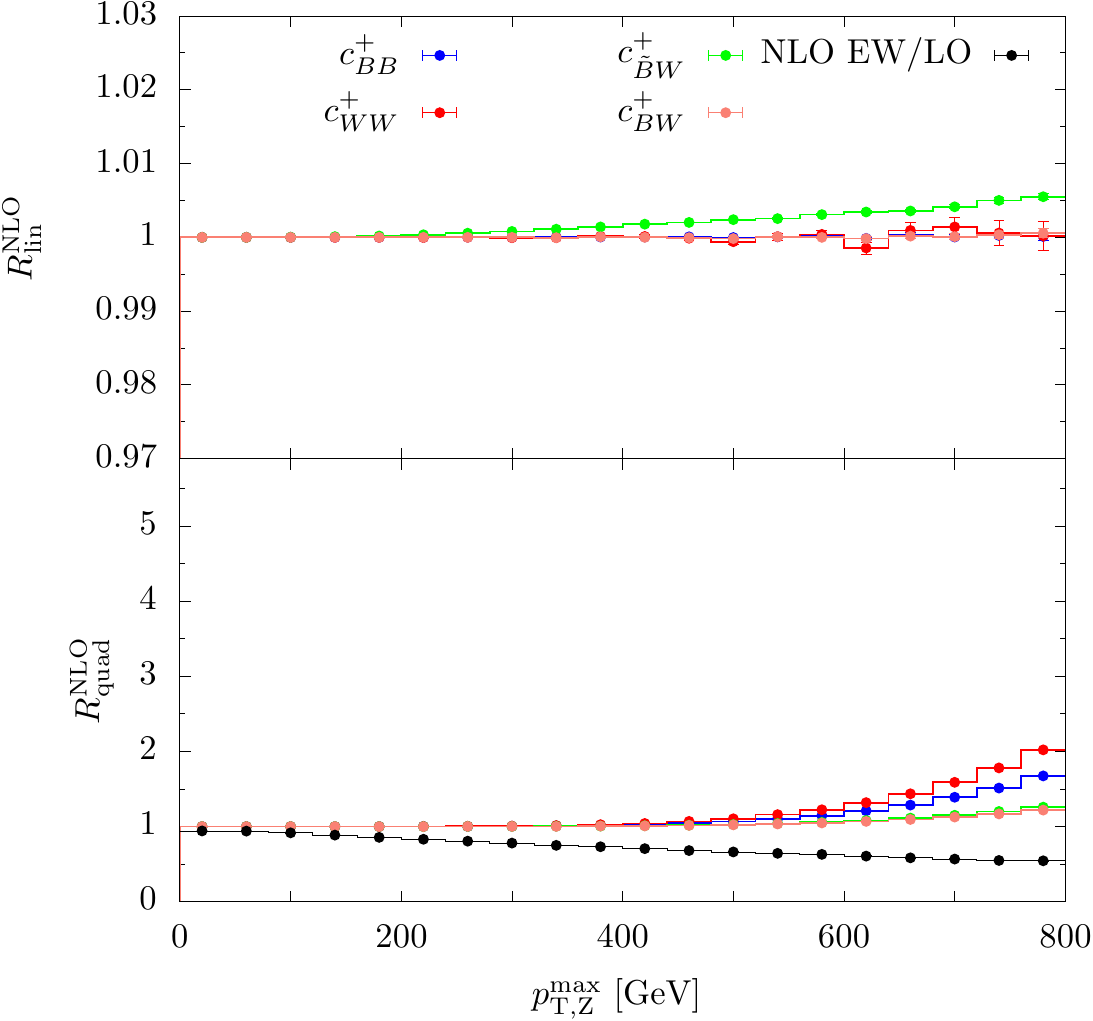}
  \end{minipage}\\[3ex]
  \begin{minipage}{0.40\textwidth}
    \includegraphics[width=\textwidth]{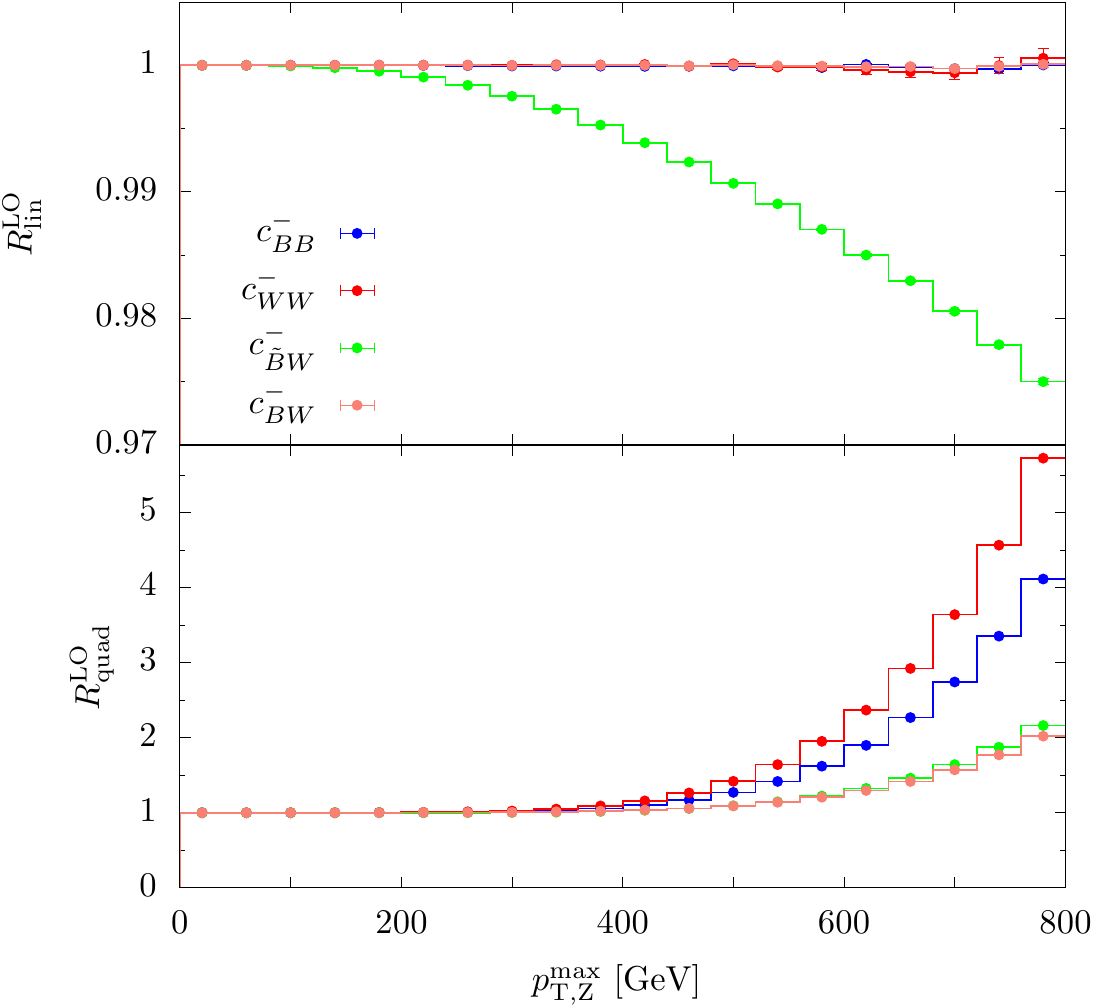}
  \end{minipage}
  \begin{minipage}{0.40\textwidth}
    \includegraphics[width=\textwidth]{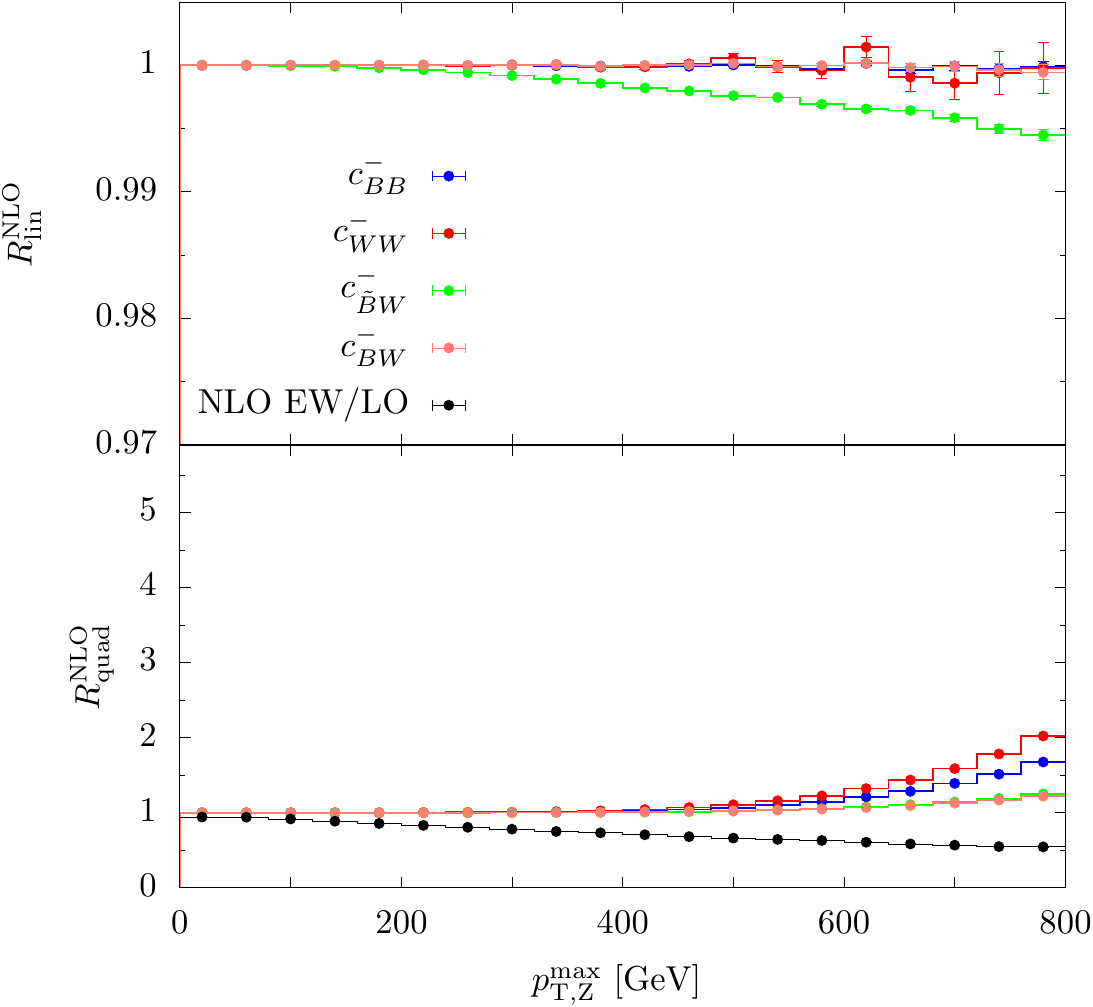}
  \end{minipage}
\end{center}
\caption{Ratio $R^{\rm LO(NLO)}_{\rm lin(quad)}$ as a function of the
   transverse momentum of the hardest $\PZ$ boson for the process
  $\Pp\Pp \to \Pep \Pem \mu^+ \mu^-$  at $\sqrt{s}=13\TeV$
  under the event selections of Eq.~(\ref{eq:zz4latlascut}).
  Same notation and conventions as in \reffi{fig:wwtgcpth}.}
\label{fig:zztgcptmax}
\efi
\bfi
\begin{center}
  \begin{minipage}{0.40\textwidth}
    \includegraphics[width=\textwidth]{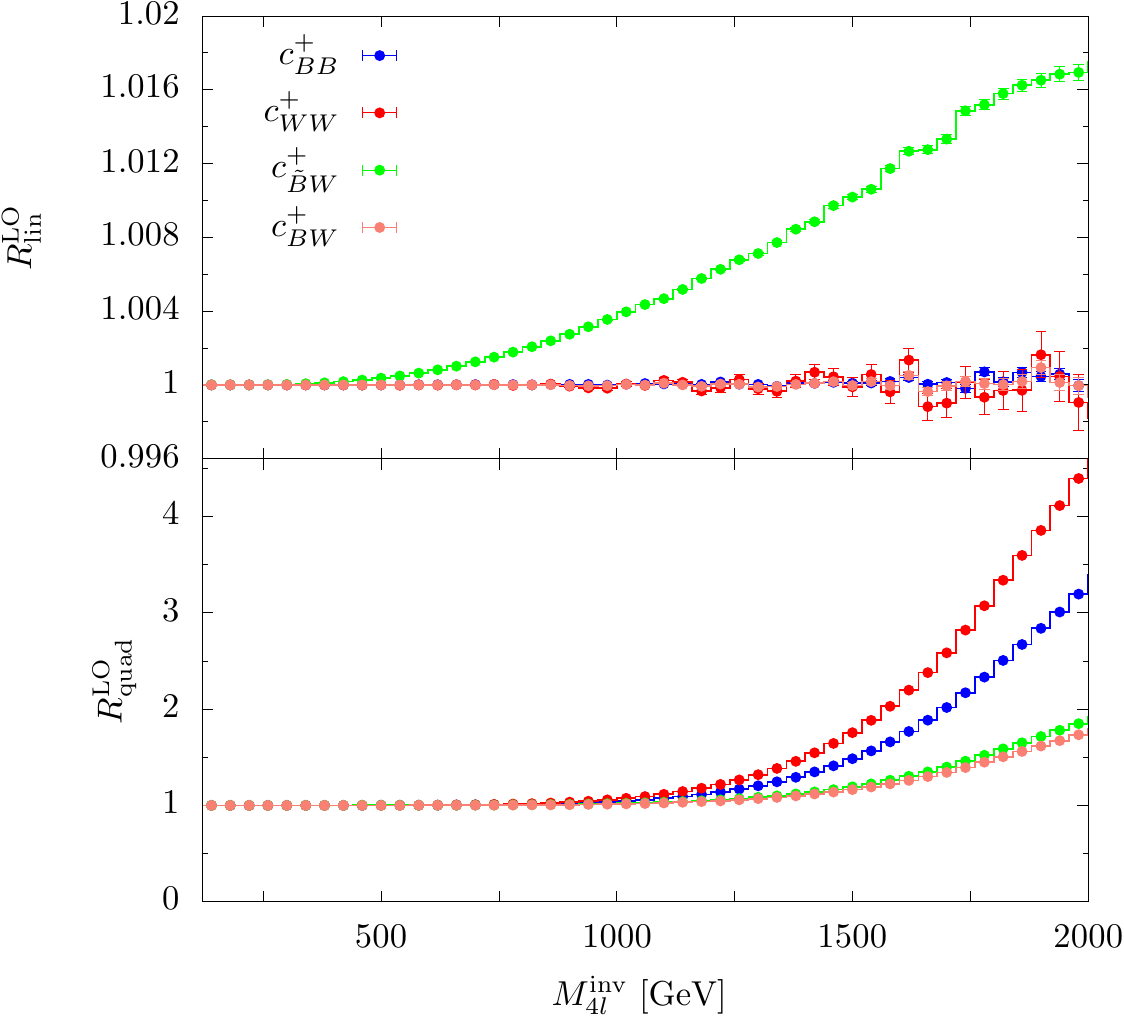}
  \end{minipage}
  \begin{minipage}{0.40\textwidth}
    \includegraphics[width=\textwidth]{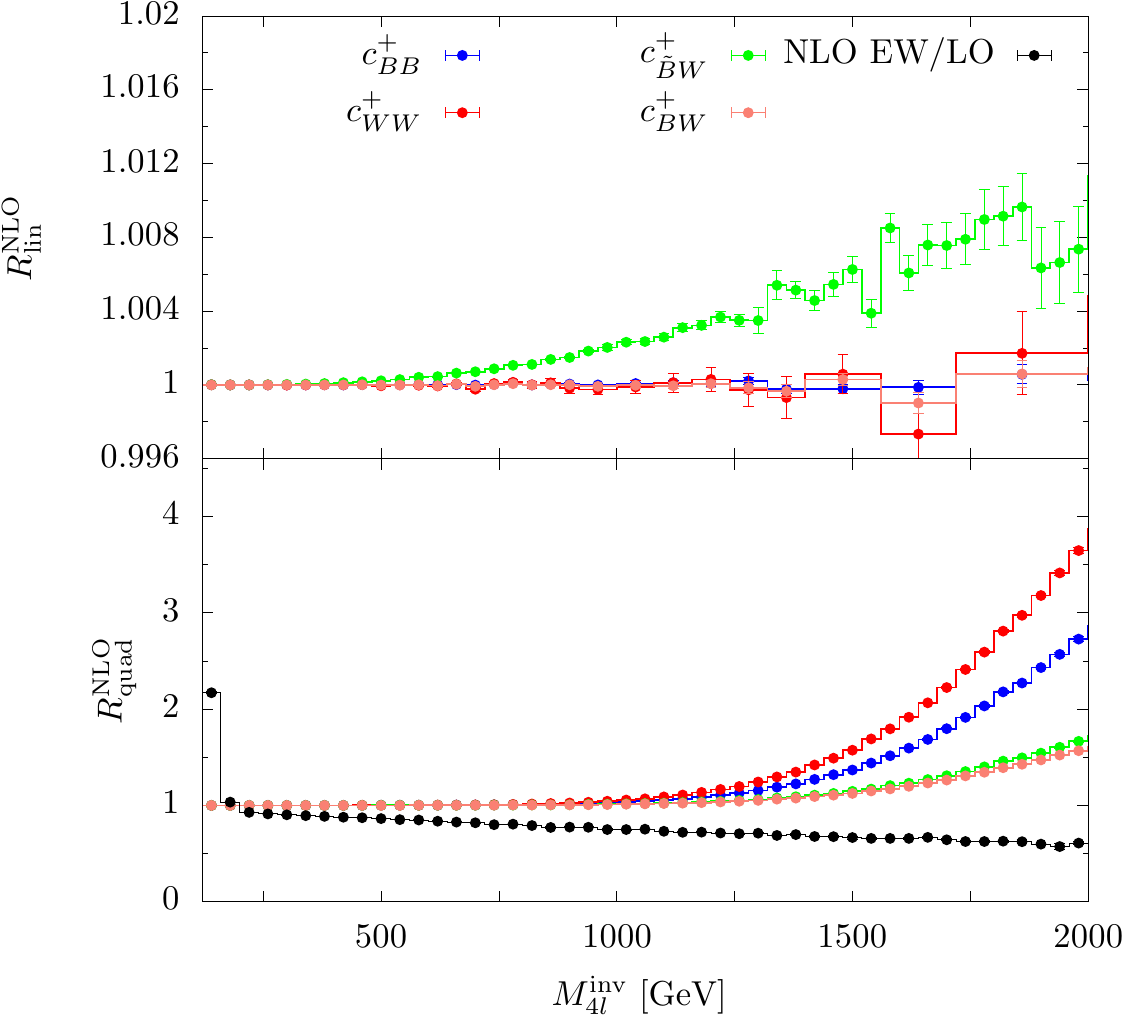}
  \end{minipage}\\[3ex]
  \begin{minipage}{0.40\textwidth}
    \includegraphics[width=\textwidth]{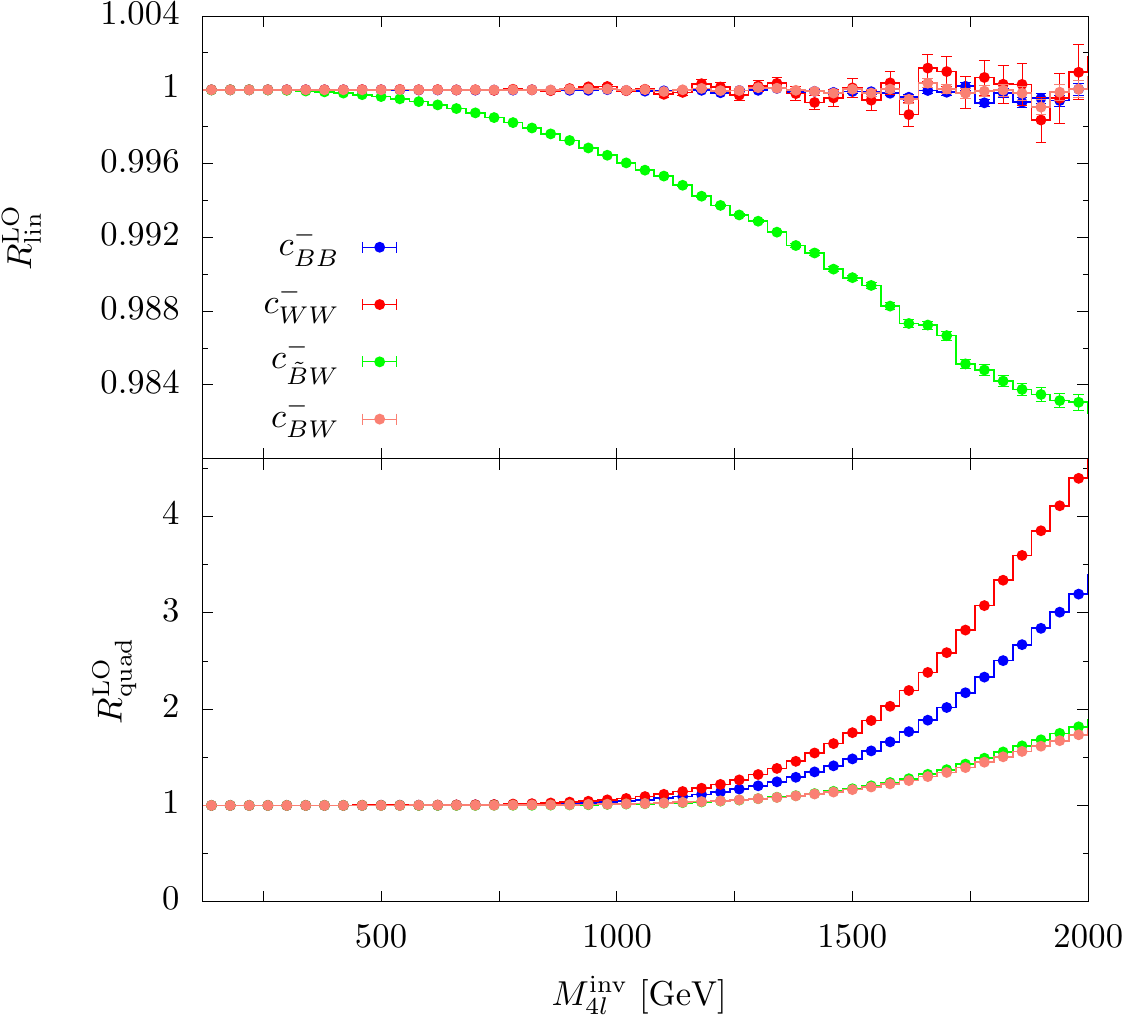}
  \end{minipage}
  \begin{minipage}{0.40\textwidth}
    \includegraphics[width=\textwidth]{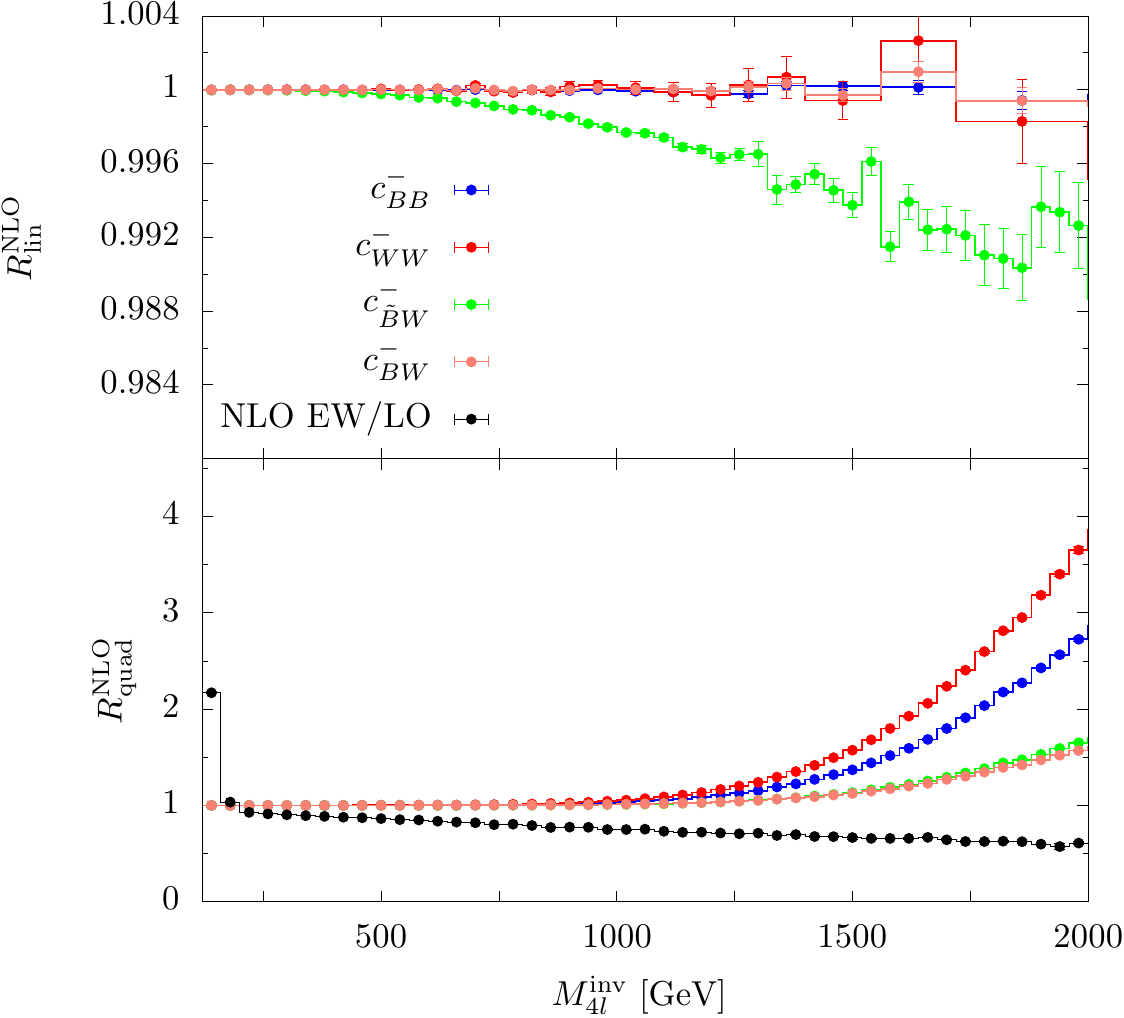}
  \end{minipage}
\end{center}
\caption{Ratio $R^{\rm LO(NLO)}_{\rm lin(quad)}$ as a function of the
  four-lepton invariant mass  for the process
  $\Pp\Pp \to \Pep \Pem \mu^+ \mu^-$  at $\sqrt{s}=13\TeV$
  under the event selections of Eq.~(\ref{eq:zz4latlascut}).
  Same notation and conventions as in \reffi{fig:wwtgcpth}.}
\label{fig:zztgcm4l}
\efi
\bfi
\begin{center}
  \begin{minipage}{0.40\textwidth}
    \includegraphics[width=\textwidth]{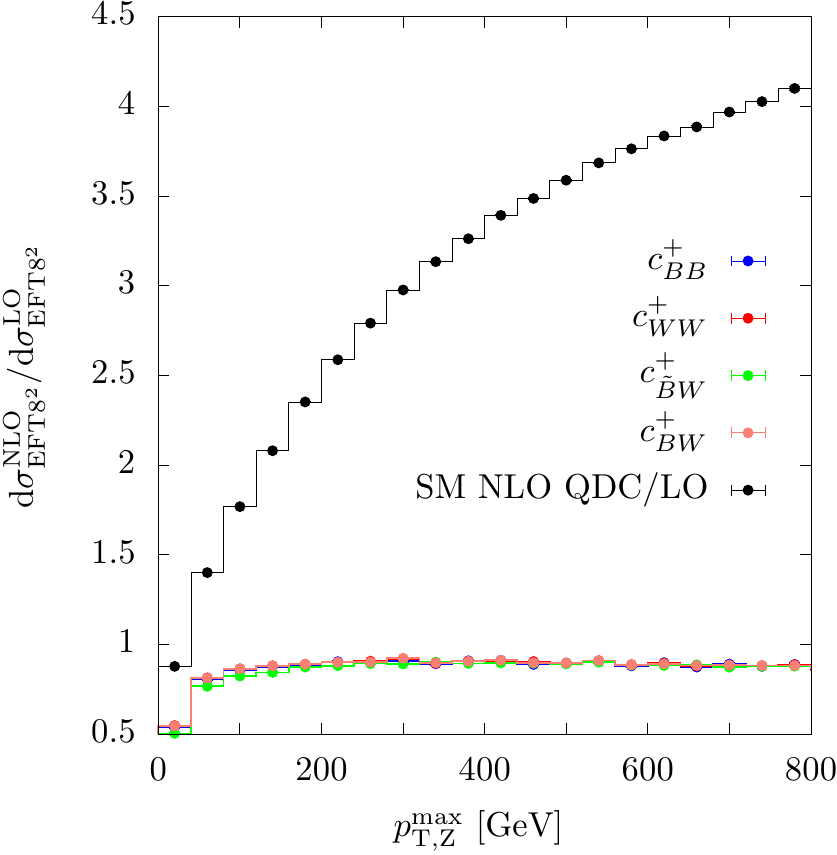}
  \end{minipage}
  \begin{minipage}{0.40\textwidth}
    \includegraphics[width=\textwidth]{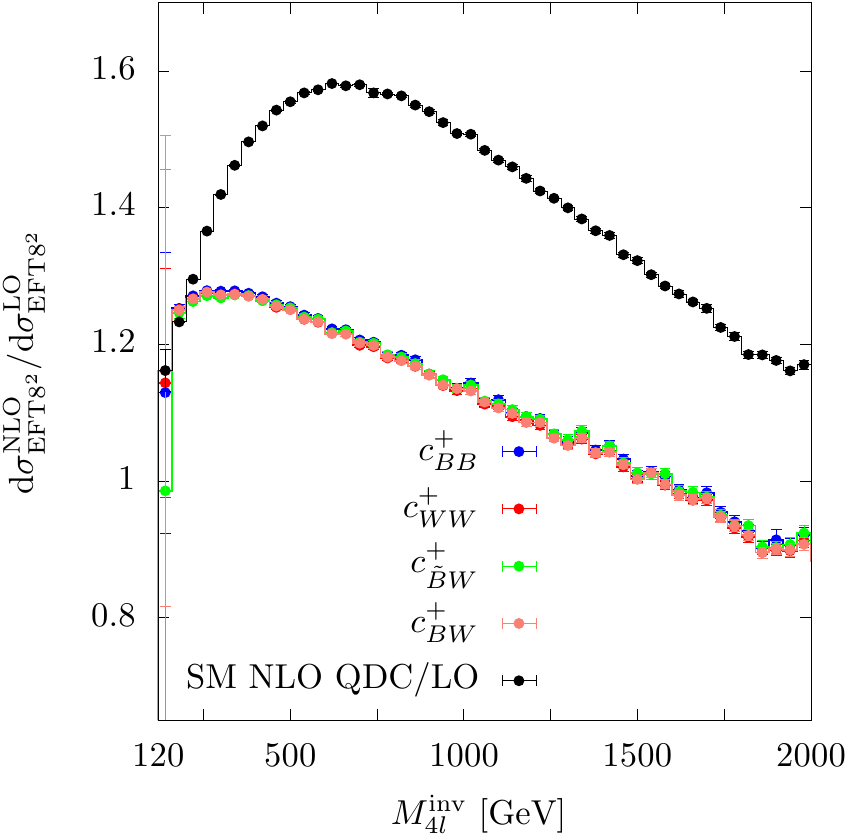}
  \end{minipage}
\end{center}
\caption{Ratio between the ${\rm EFT8}^2$ contribution to the process
  $\Pp\Pp \to \Pep \nu_{\rm e} \mu^+ \mu^-$ computed at NLO QCD
  ($\sigma_{{\rm EFT8}^2}^{\rm NLO}$) and LO ($\sigma_{{\rm EFT8}^2}^{\rm LO}$)
  accuracy  as a function of the transverse momentum of the hardest $\PZ$
  boson (left plot)
  and as a function of the four-lepton invariant mass (right plot)
  under the event selections of Eq.~(\ref{eq:zz4latlascut}).
  The ratio between the SM predictions at NLO QCD and at LO accuracy
  is also shown (black lines).}
\label{fig:zzqcdlam2}
\efi
The distributions of $\delta_{{\rm EFT8}^2}^{\rm QCD}$ and
$\delta_{\rm SM}^{\rm QCD}$ are shown in \reffi{fig:zzqcdlam2}.  While
for the distribution in the transverse momentum of the leading
$\PZ$~boson we find the same behaviour as already described in
\refse{subsect:wzres} for $\PW\PZ$ production, for the distribution in
$M_{4l}^{\mathrm{inv}}$ the NLO QCD corrections to the ${\rm EFT}8^2$
contribution and the ones to the SM prediction are similar (they
differ only by up to $40\%$).

\section{Conclusions}
\label{sect:conclusion}

A precise theoretical understanding of diboson production processes at
the LHC is crucial both in the context of tests of the SM and in
the one of the direct searches for anomalous triple-gauge-boson
interactions.

In this paper we presented a phenomenological study of $\PW\PW$~($\to
\Pep \Pne \Pmum \bar{\nu}_{\mu}$), $\PW\PZ$~($\to \Pem
\overline{\nu}_{\rm e} \mu^+ \mu^-$), and $\PZ\PZ$~($\to \Pep \Pem
\mu^+ \mu^-$) production considering event selections of interest for
the aTGCs searches at the LHC.  For $\PW\PW$ and $\PZ\PZ$ production
we included the impact of the loop-induced $\Pg\Pg \to VV$ processes
at LO.

The calculation described in this paper is the first application of
\recolatwo in the EFT context: a \UFO model file including the SM
Lagrangian as well as the dimension-6 (-8) operators relevant for
$\PW\PW$ and $\PW\PZ$ ($\PZ\PZ$) production have been implemented
using the {\sc Mathematica} package {\FeynRules}.  The model file has
been converted to a \recolatwo model file by means of the {\Python}
library \Reptil.  All  NLO QCD and NLO EW corrections in this paper
have been computed with \recolatwo.

The code has been used to study the effect of the aTGCs in the EFT
framework at LO and at NLO QCD for some observables of experimental
interest.  We found that the sensitivity to the aTGCs is in general
reduced at NLO QCD because of real radiation contributions, like the
opening $\Pg q$/$\Pg \overline{q}$ channels, which are less sensitive
to the aTGCs.  From a quantitative point of view, the reduction in the
sensitivity to aTGCs  depends on the analysis setup and on the observables
under consideration. If the terms involving squared anomalous
couplings (${{\rm EFT}^2}$ terms) are taken into account, this effect
is proportional to the ratio of the NLO QCD corrections to the ${{\rm
    EFT}^2}$ terms and the NLO QCD corrections to the SM predictions.
We also disentangled the effect of the interference terms linear in
the anomalous couplings (${{\rm SM}\times {\rm EFT}}$) and the ${{\rm
    EFT}^2}$ terms and we showed how the latter dominate over the
interference terms almost everywhere in the distributions of interest
for the aTGCs searches at the LHC.

\subsection*{Acknowledgements}
The work of M.C. and A.D. was supported by the German Science
Foundation (DFG) under reference number DE 623/5-1.  J.-N. Lang
acknowledges support from the Swiss National Science Foundation (SNF)
under contract BSCGI0-157722. A.D. and J.-N.L are grateful to the
Mainz Institute for Theoretical Physics (MITP) for its hospitality and
partial support during the completion of this work.

\bibliographystyle{JHEPmod}
\bibliography{dibosonbibfile}

\end{document}